\documentclass{aa}  

\usepackage{graphicx}
\usepackage{hyperref}
\usepackage{txfonts}
\usepackage{lscape}

\newcommand{\artemis}{ArT\'eMiS} 
\newcommand{\degree}{^\circ}
\newcommand{\ccol}{$^{13}$CO(2--1) }
\newcommand{\cco}{$^{13}$CO }
\newcommand{\PhA}[1]{{#1}}

\begin{document}

\title{Understanding the star formation efficiency in dense gas: Initial~results from the CAFFEINE survey with ArT\'eMiS 
\thanks{This publication is based on data acquired with the Atacama Pathfinder Experiment (APEX) under projects 1102.C-0745 and 106.21MS. APEX is a collaboration between the Max-Planck-Institut für Radioastronomie, the European Southern Observatory, and the Onsala Space Observatory.}}

\titlerunning{Star formation efficiency in CAFFEINE molecular clouds}

   \author{M. Mattern\inst{1}
          \and
          Ph. Andr{\'e}\inst{1}
          \and 
          A. Zavagno\inst{2,3}
          \and
          D. Russeil\inst{2}
          \and
          H. Roussel\inst{4}
          \and
          N. Peretto\inst{5}
          \and
          F. Schuller\inst{6}
          \and
          Y. Shimajiri\inst{7}
          \and\\
          J. Di Francesco\inst{8,9}
          \and
          D. Arzoumanian\inst{10}
          \and
          V. Rev{\'e}ret\inst{1}
          \and
          C. De Breuck\inst{11}
          }

   \institute{Laboratoire d’Astrophysique (AIM), Université Paris-Saclay, Université Paris Cité, CEA, CNRS, AIM, 91191 Gif-sur-Yvette,
France\\
              \email{michael.mattern@cea.fr}
              \and    
              Aix Marseille Univ, CNRS, CNES, LAM, Marseille, France 
              \and 
              Institut Universitaire de France, Paris, France
              \and 
              IAP, Sorbonne Université, CNRS (UMR 7095), 75014 Paris, France
              \and 
              School of Physics and Astronomy, Cardiff University, The Parade,Cardiff CF24 3AA, UK    
              \and
              Leibniz-Institut für Astrophysik Potsdam (AIP), An der Sternwarte 16, 14482 Potsdam, Germany
              \and
              Kyushu Kyoritsu University, 1-8, Jiyugaoka, Yahatanishi-ku, Kitakyushu-shi, Fukuoka 807-8585, Japan
              \and
              National Research Council of Canada, Herzberg, Astronomy and Astrophysics Research Centre, 5071 West Saanich Road, Victoria, BC V9E 2E7, Canada
              \and
              Department of Physics and Astronomy, University of Victoria, Victoria, BC V8P 5C2, Canada
              \and
              National Astronomical Observatory of Japan, Osawa 2-21-1, Mitaka, Tokyo 181-8588, Japan
              \and
              European Southern Observatory, Karl Schwarzschild Straße 2, 85748 Garching, Germany
              }

   \date{Received ...; accepted 18 May 2024}

 
  \abstract
   {Despite recent progress, the question of what regulates the star formation efficiency (SFE) in galaxies remains one of the most debated problems in astrophysics. 
   According to the dominant picture, star formation (SF) is regulated by turbulence and feedback, and the SFE is $\sim$\,1--2\% or less per local free-fall time 
   on 
   all scales from  
   Galactic clouds to high-redshift galaxies. In an alternate scenario, the star formation rate (SFR) in galactic disks is linearly proportional to the mass of dense gas above some critical density threshold 
 $\sim$\,$10^4\, {\rm cm}^{-3}$.}
   {We aim to discriminate between these two pictures thanks to high-resolution submillimeter and mid-infrared 
    imaging observations, which trace both dense gas and young stellar objects (YSOs) for a comprehensive sample of 49 nearby massive SF complexes out 
    to a distance of $d \sim 3$~kpc in the Galactic disk.}
   {We used data from CAFFEINE, a complete 350/450\,$\mu$m survey with APEX/ArT\'eMiS of the densest portions of all southern molecular clouds at $d \lesssim 3$~kpc, 
   in combination with {\it Herschel} data to produce column density maps at a factor of $\sim$\,4 higher resolution ($8\arcsec$) than 
   standard {\it Herschel} column density maps ($36\arcsec$). Our maps are free of any saturation effect around luminous high-mass protostellar objects and 
   resolve the structure of dense gas and the typical $\sim$\,0.1\,pc 
   width of molecular filaments out to 3~kpc, which is the most important asset of 
   the present study and is impossible to achieve with {\it Herschel} data alone. 
   Coupled with SFR estimates derived from {\it Spitzer} mid-infrared  observations of the YSO content of the same clouds, this allowed us 
   to study the dependence of the SFE on density in the CAFFEINE clouds. 
   We also combine our findings with existing SF efficiency measurements in nearby clouds to extend our analysis down to lower column densities.
   }
   {Our results suggest that the SFE does not increase with density above the critical threshold and support a scenario in which the SFE in dense gas 
   is approximately constant (independent of free-fall time). However, the SF efficiency measurements traced by Class~I YSOs in nearby clouds are more inconclusive, 
   since they are consistent with both the presence of a density threshold 
   and a dependence on density above the threshold. 
   Overall, we suggest that the SF efficiency in dense gas is 
   primarily governed by the physics of filament fragmentation into protostellar cores.
   }
   {}
               

   \maketitle
%
\section{Introduction}

Understanding what regulates the star formation efficiency (SFE) in the giant molecular clouds (GMCs) of galaxies is a fundamental 
open question in star formation (SF) research. 
The star formation rate (SFR) on multiple scales in galaxies is known to be strongly correlated with the mass of available molecular 
gas \citep[e.g.,][]{Kennicutt1998,Bigiel2011}. 
Overall, SF is observed to be a very inefficient process. It has been argued that the SF per free-fall time, defined 
as the fraction of molecular gas that is converted into stars per free-fall time [$\epsilon_{\rm ff} \equiv ({\rm SFR}/M_{\rm gas}) \times t_{\rm ff}$], 
is a quasi-universal quantity with a typical value of $\sim$~1--2\% on a wide range of spatial scales from local Galactic clouds 
to high-redshift galaxies \citep[][]{Krumholz2007,Krumholz2014}, in agreement with the turbulence-regulated SF model of \citet{Krumholz2005}. 
If $\epsilon_{\rm ff} $ is indeed constant, this implies that the non-normalized SFE, 
${\rm SFE} \equiv {\rm SFR}/M_{\rm gas} \propto t_{\rm ff}^{-1} $, should increase with gas density $\rho$, since $ t_{\rm ff} \propto \rho^{-1/2} $. 
In agreement with this scenario, observational evidence that $\epsilon_{\rm ff} $ may be roughly constant $\sim$2.6\% {\it within} nearby molecular clouds 
has been recently reported \citep{Pokhrel2021}. 
However, resolved observations of nearby disk galaxies indicate a roughly constant depletion time $t_{\rm dep} \equiv {\rm SFE}^{-1} \sim 1$--2\,Gyr \citep{Bigiel2011,Schruba2011} 
on kpc scales, which is at variance with a strictly constant $\epsilon_{\rm ff} $ value on all scales. Indeed, adopting an average volume density $n_{\rm H_2} \sim 10^2\, {\rm cm}^{-3} $ 
and thus an average free-fall time $t_{\rm ff} \sim 3$\,Myr for CO molecular clouds leads to $\epsilon_{\rm ff} = t_{\rm ff}/t_{\rm dep} \sim 0.15$--0.3\% on galactic scales, which is 
an order of magnitude lower than the $\sim$~1--2\% value quoted above.
Moreover,
direct observations of the connection between dense gas and SF in nearby resolved molecular clouds suggest 
a picture in which the SFR is directly proportional to the mass of dense molecular gas, $M_{\rm dense}$. 
Here, dense gas is understood as gas with surface density $\Sigma_{\rm H_2} > \Sigma_{\rm th}  \approx 100$--$200\, {\rm M_\odot \, pc^{-2}}$, 
which translates to volume densities $n_{\rm H_2} \ga n_{\rm th}  \approx1$--$2 \times 10^4\,{\rm cm}^{-3} $ 
assuming clouds with a typical radial density profile $n_{\rm H_2} \propto r^{-2}$ \citep[][]{Lada2010,Lada2012,Evans2014,Shimajiri2017}.
In this alternate picture, SF occurs mostly (if not only) in dense cloud gas above $\Sigma_{\rm th}$  in surface density or $n_{\rm th}$ in volume density, 
and ${\rm SFE_{dense}} \equiv {\rm SFR}/M_{\rm dense}$ is roughly constant at $\sim 5 \times 10^{-8}\, {\rm yr^{-1}} $, corresponding 
to a typical depletion time for dense gas of $\sim 20\,$Myr. 
Accordingly, $\epsilon_{\rm ff} = {\rm SFE} \times t_{\rm ff}$ is expected to {\it decrease} with density above $\sim 1$--$2 \times 10^4\,{\rm cm}^{-3} $. 
Such a threshold or transition picture for the SFE fits well within the filament paradigm of SF \citep[e.g.,][]{Andre2014}, 
in which prestellar cores and protostars form primarily 
in dense, thermally supercritical filaments 
with masses per unit length of $> 16\, M_\odot$/pc, typical widths of $\sim$\,0.1 pc \citep[e.g.,][]{Arzoumanian2019}, 
and a typical core formation efficacy of $\sim$\,15-20\% \citep[e.g.,][]{Konyves2015}. 
The latter scenario leads to a natural transition surface density $\Sigma_{\rm th} \sim 160\, M_\odot\, {\rm pc}^{-2} $, 
equivalent to a transition volume density $n_{\rm th} \sim 2 \times 10^4\,{\rm cm}^{-3} $, set by the critical mass per unit length of molecular gas filaments.

In this paper, we present results from CAFFEINE, a complete 350/450 $\mu$m survey with APEX/ArT\'eMiS of the densest portions 
of all molecular clouds at distances of $d \la 3$\,kpc (Section \ref{sec:CAFFEINE_obs}), which help us discriminate between the scenarios described above for the SFE in dense gas. 
Combining the ArT\'eMiS observations with {\it Herschel} data, 
we produced column density maps with a factor of $\sim$\,4 higher resolution than standard {\it Herschel} column density maps (Section \ref{sec:colden_maps}). 
This improvement allows us to resolve the structure of dense gas and the typical $\sim$\,0.1 pc width of molecular filaments/cores out to 3~kpc, 
which is impossible with {\it Herschel} data alone. 
After correcting for line-of-sight contamination (Section \ref{sec:contamination}), we used our high-resolution column density maps in conjunction with SFR estimates from {\it Spitzer} young stellar object (YSO) and protostellar clump catalogs (Section \ref{sec:SFR_CAF}) to study the dependence of the SFE on density in a sample of CAFFEINE clouds. We present our results and a comparison with nearby clouds in Section \ref{sec:Results} and discuss their meaning in the context of two theoretical models in Section \ref{sec:Discussion}, before drawing our conclusions in Section \ref{sec:Conclusion}.

\section{Submillimeter and infrared data}
In the present study, we introduce and focus on data from the Core And Filament Formation/Evolution In Natal Environments (CAFFEINE) survey with ArT\'eMiS. 
We complement these data with additional submillimeter 
dust continuum observations from the {\it Herschel} space observatory and the {\it Planck} all-sky survey, and molecular line observations of the $^{13}$CO molecule. Furthermore, to derive estimates of the SFR, we used catalogs of YSOs compiled from several infrared surveys and of protostellar clumps identified on \textit{Herschel} maps.

\subsection{The CAFFEINE imaging survey with \artemis}
The CAFFEINE survey was carried out between August 2018 and 
August 2022 as a large program of the European Southern Observatory (ESO) 
under projects IDs E-1102.C-0745A-2018 and 106.21MS.
The data were taken with the \artemis ~bolometer-array camera \citep{Reveret2014,Andre2016,Talvard2018}
at the Atacama Pathfinder EXperiment (APEX) telescope \citep{Guesten2006} 
in Chile. \artemis  ~produces images of the continuum emission at both $\rm 350~\mu m$ and $\rm 450~\mu m$ simultaneously. 
At these wavelengths, the half-power beam width (HPBW) angular resolution achieved by \artemis  ~on 
the $\rm 12~m$ APEX antenna 
is $\sim$$8\arcsec$ and $\sim$$10\arcsec$, respectively.
The pointing accuracy was checked  every $\sim$\,1--2~h and found to be better than $\lesssim 3\arcsec$. 
To ensure high sensitivity and data quality, observations were carried out only in submillimeter weather conditions 
with a precipitable water vapor content (PWV) of 
$\rm < 0.7~mm$.

\begin{figure}
    \centering
    \includegraphics[width=0.49\textwidth,trim={0.5cm 2cm 0.5cm 2cm},clip]{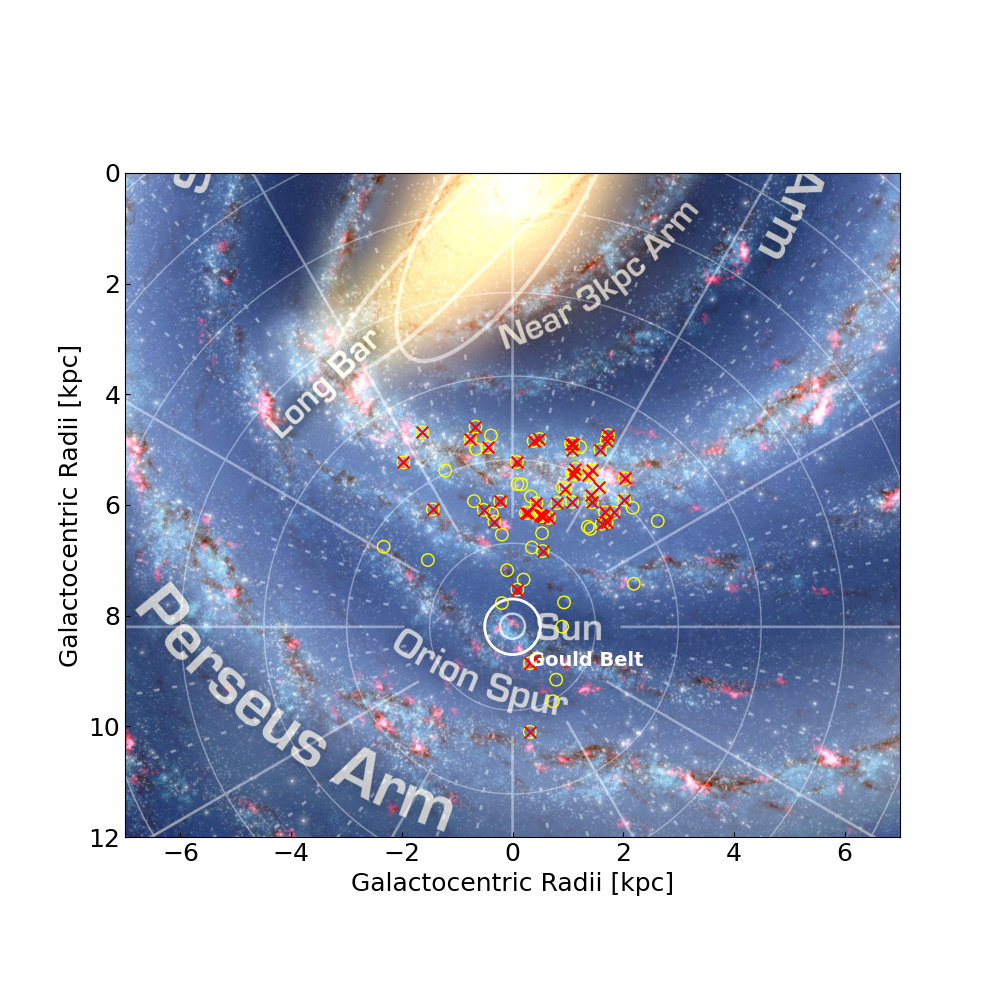}
    \caption{Locations of the CAFFEINE targets marked as yellow circles in an artist's impression of the Milky Way by R. Hurt (NASA; JPL-Caltech). The targets selected for this study are marked by red crosses.}
    \label{fig:target_locations}
\end{figure}

The CAFFEINE survey targets the densest  ($A_V > 40$ mag) portions of all dense star-forming complexes visible from APEX as seen by the ATLASGAL and Hi-GAL surveys \citep{Schuller2009, Molinari2010}
out to a distance $d \sim \rm 3~kpc$ (Fig. \ref{fig:target_locations}). This distance limit allows us to achieve a physical resolution of $\rm < 0.1~pc$ in all regions. 
The significance of the CAFFEINE cloud sample is that it contains two to three orders of magnitude more dense gas mass at high column density ($A_V > 50$ -- 200 mag)  
than the nearby clouds of Gould's Belt and therefore allows us to probe the high-mass SF regime and the dependence of the SFE with density.

The CAFFEINE survey encompasses observations of 80 dense molecular clouds within a distance range from $\rm \sim 0.6~kpc$ to $ \rm \sim 3.0~kpc$, following the estimates of associated Hi-GAL clumps \citep{Russeil2011, Elia2021}. In particular, we assign the average distance of the largest group of clumps with similar distances (about $\pm 0.1~\rm kpc$) and additionally compare with the velocities of additional molecular line data (see Sect. \ref{sec:contamination}). As the Scutum-Centaurus spiral-arm of the Milky Way is located just outside of the $\rm 3.0~kpc$ radius a few prominent and potentially interesting sources were also included with a strict limit at $\rm 4.0~kpc$ (Fig. \ref{fig:target_locations}).

\begin{figure*}[t]
    \centering
    \begin{minipage}{0.33\textwidth}
        \includegraphics[width=\textwidth,trim={3cm 0 3cm 0},clip]{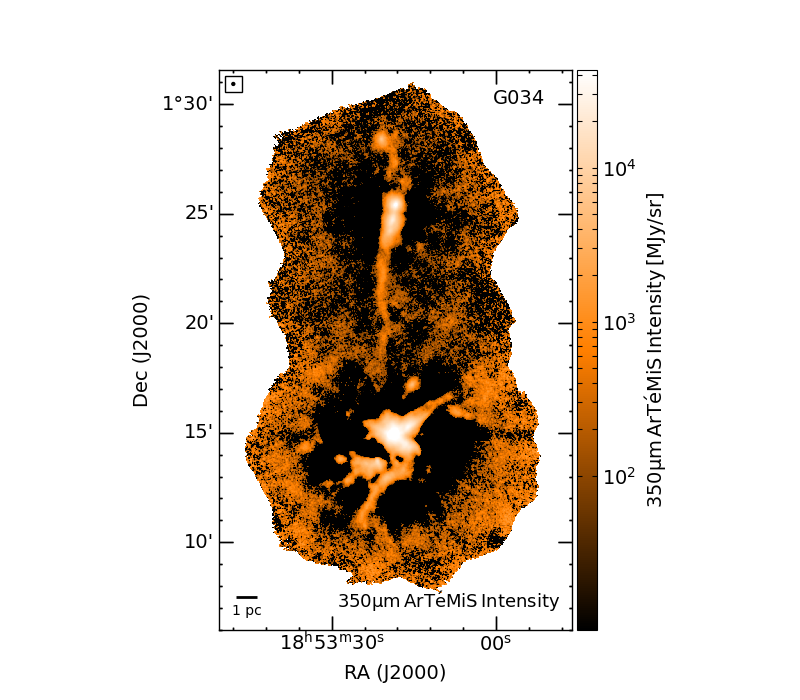}
    \end{minipage}
    \begin{minipage}{0.33\textwidth}
        \includegraphics[width=\textwidth,trim={3cm 0 3cm 0},clip]{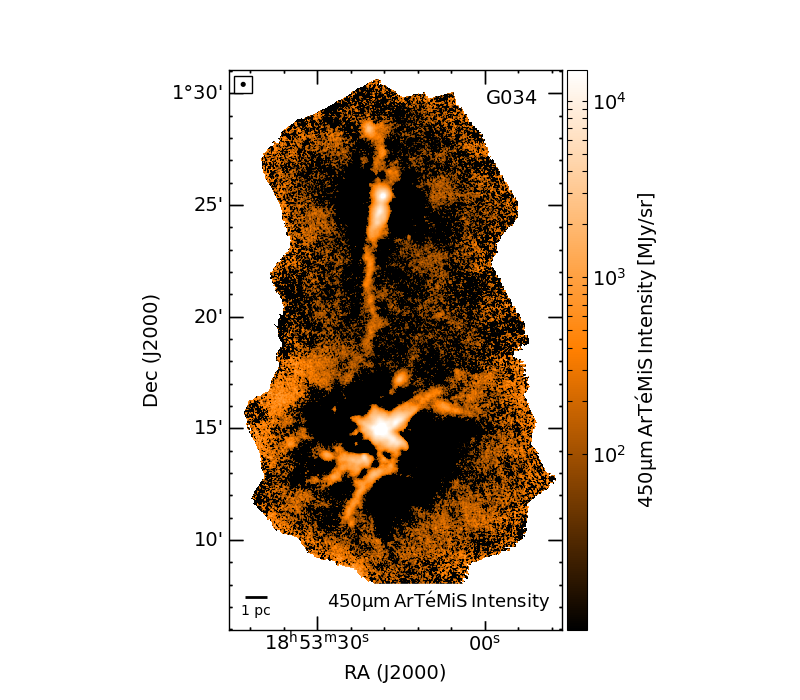}
    \end{minipage}
    \begin{minipage}{0.33\textwidth}
        \includegraphics[width=\textwidth,trim={3cm 0 3cm 0},clip]{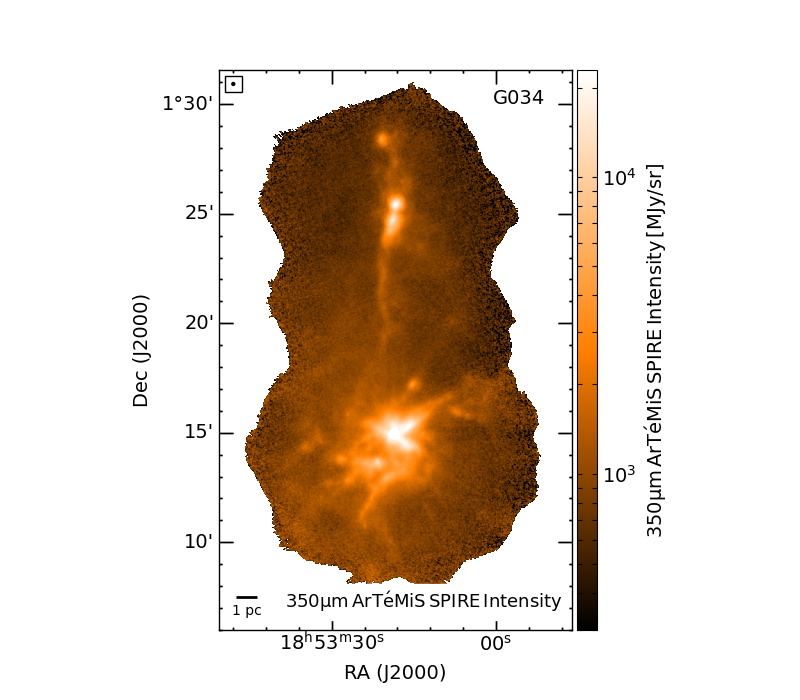}
    \end{minipage}
    \caption{Examples of data products derived from the CAFFEINE survey. \textbf{Left} Dust continuum emission of the G034 region as observed at $\rm 350~\mu m$ by \artemis ~at $8\arcsec$ resolution. \textbf{Center:} Dust continuum emission of the G034 region as observed at $\rm 450~\mu m$ 
    by \artemis ~at $10\arcsec$ resolution. \textbf{Right:} Combined $\rm 350~\mu m$ \artemis--SPIRE dust continuum emission map with $8\arcsec$ resolution and high dynamic range. }
    \label{fig:map_example}
\end{figure*}

The \artemis ~data were processed with the dedicated  \artemis ~Pipeline using IDL and Scanamorphos
(APIS)\footnote{See \url{https://www.apex-telescope.org/ns/data-reduction-for-bolometer-arrays/\#ArtemisPipeline}}, 
which takes care of converting the raw data to IDL-friendly data structures, 
applying flux calibration and atmospheric opacity correction, rejecting unusable pixels,
removing atmospheric emission fluctuations and instrumental noise, and making maps for astrophysical use. 
The final intensity and  weight maps were saved in \textsc{FITS} format. 
The average rms noise is $0.46\rm ~Jy/8\arcsec\,beam$ at $\rm 350~\mu m$ and $0.37\rm ~Jy/10\arcsec\,beam$ at $\rm 450~\mu m$.
We present in Fig.~\ref{fig:map_example} (left and center) the G034 region as an example of the \artemis~ data products created.

\subsubsection{Field selection}
 
For the present analysis of the SFE at high densities, 
we selected a subsample of 49 clouds that have a surface area larger than $\rm 1~pc^2$ above a column density of $\rm 2 \times 10^{22}~cm^{-2}$. 
We also included the MonR2 region to compare with the results of \cite{Pokhrel2021}. The selected clouds are marked in Fig.~\ref{fig:target_locations} and listed in Table \ref{tab:CAFFEINE_sources} by order of right ascension.
The selected fields were observed within $\rm \sim 215~h$ with \artemis/APEX, which corresponds to $\sim 75$\% of the total CAFFEINE observing time.

\begin{table*}
\caption{CAFFEINE cloud sample \PhA{analyzed in the present study}}
\label{tab:CAFFEINE_sources}
\begin{tabular}{lccccccrrrr}
\hline\hline
Cloud & $d$ & RA$_{\rm J2000}$ & DEC$_{\rm J2000}$ & $l$ & $b$ & $N({\rm H_2})_\text{back}$\tablefootmark{a} & $A_{22.15}$\tablefootmark{b} & $A_{22.90}$\tablefootmark{b} & $M_{22.15}$\tablefootmark{c} & $M_{22.90}$\tablefootmark{c} \\
 & kpc & hms & dms & $\degree$ & $\degree$ & $\rm 10^{22}~cm^{-2}$ & $\rm pc^{-2}$ & $\rm pc^{-2}$ & $\rm M_\odot$ & $\rm M_\odot$ \\
\hline
MonR2 & 0.8 & 06:07:48 & -06:23:24 & 213.71 & -12.6 & 0.0 & 1.34 & 0.06 & 964 & 147 \\
R5180 & 2.0 & 06:08:48 & 21:34:48 & 188.99 & 0.83 & 0.0 & 3.87 & 0.31 & 6474 & 4256 \\
NGC2264 & 0.8 & 06:41:02.4 & 09:31:12 & 203.28 & 2.04 & 0.0 & 2.49 & 0.16 & 1952 & 415 \\
GAL316 & 2.5 & 14:44:52.8 & -59:49:48 & 316.75 & -0.03 & 1.33 & 12.38 & 1.16 & 10755 & 2900 \\
I14453 & 2.7 & 14:48:57.6 & -59:25:48 & 317.38 & 0.1 & 1.02 & 3.23 & 0.15 & 2366 & 423 \\
SDC317p7 & 3.0 & 14:51:12 & -59:16:48 & 317.7 & 0.12 & 1.19 & 6.83 & 0.19 & 4371 & 485 \\
SDC317p9 & 2.4 & 14:53:50.4 & -59:32:24 & 317.89 & -0.28 & 1.14 & 3.92 & 0.26 & 2637 & 672 \\
RCW87 & 2.6 & 15:04:24 & -57:33:36 & 320.03 & 0.83 & 0.46 & 9.52 & 0.33 & 6498 & 818 \\
G322 & 3.3 & 15:18:33.6 & -56:38:24 & 322.16 & 0.64 & 0.51 & 7.31 & 0.9 & 7314 & 3246 \\
SDC326 & 2.6 & 15:43:33.6 & -54:03:00 & 326.55 & 0.74 & 0.32 & 75.35 & 5.41 & 56927 & 16364 \\
G327 & 2.9 & 15:52:57.6 & -54:36:36 & 327.28 & -0.55 & 0.77 & 17.49 & 1.89 & 19994 & 9824 \\
I15541 & 2.7 & 15:58:07.2 & -53:58:12 & 328.27 & -0.54 & 1.39 & 21.85 & 2.65 & 22164 & 8112 \\
GAL332 & 3.0 & 16:15:07.2 & -49:49:12 & 332.99 & 0.78 & 0.72 & 24.05 & 1.12 & 16195 & 3738 \\
G332 & 3.1 & 16:16:19.2 & -50:52:12 & 332.4 & -0.11 & 1.7 & 17.94 & 0.3 & 10629 & 790 \\
RCW106 & 3.7 & 16:19:04.8 & -51:01:12 & 332.61 & -0.52 & 0.0 & 116.26 & 2.46 & 63738 & 7062 \\
G333 & 2.4 & 16:20:09.6 & -49:36:00 & 333.73 & 0.37 & 1.08 & 4.26 & 0.17 & 3191 & 343 \\
G333.2 & 3.5 & 16:20:55.2 & -50:38:24 & 333.09 & -0.45 & 1.76 & 73.8 & 8.0 & 71624 & 26371 \\
I16175 & 3.8 & 16:21:16.8 & -50:51:36 & 332.97 & -0.65 & 0.47 & 6.89 & 0.69 & 5926 & 2414 \\
G337.1 & 1.4 & 16:37:45.6 & -47:39:00 & 337.15 & -0.39 & 1.49 & 1.69 & 0.03 & 1058 & 72 \\
I16351 & 2.9 & 16:38:52.8 & -47:27:00 & 337.42 & -0.39 & 1.39 & 6.34 & 0.28 & 4596 & 1103 \\
I16367 & 3.0 & 16:40:09.6 & -47:09:36 & 337.78 & -0.36 & 1.39 & 5.53 & 0.16 & 3268 & 413 \\
SDC338 & 2.6 & 16:40:40.8 & -46:19:12 & 338.47 & 0.12 & 1.38 & 29.46 & 1.63 & 21064 & 4237 \\
G337.9 & 2.9 & 16:41:16.8 & -47:06:36 & 337.95 & -0.47 & 1.39 & 15.69 & 0.81 & 11613 & 2647 \\
G339 & 2.3 & 16:46:04.8 & -45:38:24 & 339.6 & -0.14 & 1.41 & 5.36 & 0.14 & 2877 & 397 \\
G340.8 & 3.3 & 16:51:09.6 & -44:45:00 & 340.86 & -0.25 & 1.98 & 14.59 & 0.45 & 8866 & 1286 \\
OH341 & 3.4 & 16:52:38.4 & -44:30:36 & 341.2 & -0.3 & 1.6 & 20.96 & 1.55 & 16250 & 4109 \\
SDC340 & 2.0 & 16:54:28.8 & -45:14:24 & 340.85 & -1.01 & 0.73 & 6.58 & 0.44 & 5206 & 1356 \\
G341.9 & 3.4 & 16:54:38.4 & -43:52:12 & 341.94 & -0.17 & 1.34 & 7.88 & 1.36 & 8719 & 3988 \\
RCW116B & 2.0 & 17:00:40.8 & -40:30:36 & 345.26 & 1.04 & 0.76 & 7.3 & 0.47 & 6062 & 1651 \\
I16572 & 2.0 & 17:00:55.2 & -42:24:00 & 343.8 & -0.15 & 1.55 & 5.66 & 0.15 & 3349 & 469 \\
SDC344 & 2.0 & 17:03:57.6 & -42:28:12 & 344.09 & -0.64 & 0.42 & 39.82 & 1.6 & 28516 & 5014 \\
SDC345.4 & 2.0 & 17:04:16.8 & -40:45:00 & 345.49 & 0.36 & 1.25 & 5.13 & 0.82 & 5458 & 2508 \\
SDC345.0 & 2.0 & 17:05:24 & -41:27:00 & 345.06 & -0.23 & 1.31 & 20.06 & 0.52 & 12949 & 1126 \\
SDC348 & 2.2 & 17:19:40.8 & -38:57:36 & 348.67 & -0.97 & 0.52 & 21.24 & 2.37 & 19393 & 6717 \\
NGC6334N & 2.0 & 17:22:33.6 & -35:11:24 & 352.1 & 0.7 & 0.57 & 14.58 & 0.15 & 7962 & 321 \\
I17233a & 0.6 & 17:26:04.8 & -36:15:00 & 351.63 & -0.48 & 0.99 & 0.74 & 0.09 & 683 & 260 \\
I17233b & 3.4 & 17:26:04.8 & -36:15:00 & 351.63 & -0.48 & 1.03 & 5.49 & 0 & 2531 & 0 \\
NGC6357SE & 2.0 & 17:26:09.6 & -34:31:48 & 353.06 & 0.46 & 1.71 & 8.71 & 0.2 & 5489 & 437 \\
OH353 & 3.3 & 17:30:12 & -34:42:00 & 353.38 & -0.32 & 1.44 & 24.57 & 3.76 & 25425 & 10412 \\
G358.5 & 2.9 & 17:43:33.6 & -30:29:24 & 358.44 & -0.43 & 1.18 & 12.63 & 0.62 & 8601 & 1726 \\
G005 & 2.2 & 18:00:31.2 & -24:07:12 & 5.84 & -0.42 & 1.38 & 5.93 & 0.39 & 4801 & 1330 \\
G008 & 3.2 & 18:02:33.6 & -21:48:00 & 8.09 & 0.32 & 0.93 & 10.0 & 0.47 & 6672 & 1319 \\
SDC010 & 1.9 & 18:09:04.8 & -20:12:00 & 10.23 & -0.23 & 1.76 & 30.95 & 1.6 & 21461 & 4315 \\
G010 & 3.6 & 18:09:28.8 & -19:42:36 & 10.71 & -0.07 & 1.76 & 44.05 & 0.06 & 19927 & 133 \\
W33 & 3.4 & 18:14:04.8 & -17:43:12 & 12.97 & -0.07 & 2.43 & 218.75 & 11.26 & 156765 & 31851 \\
SDC014 & 2.1 & 18:18:36 & -16:43:48 & 14.35 & -0.55 & 1.51 & 23.15 & 0.65 & 14677 & 2089 \\
W42 & 3.8 & 18:38:19.2 & -06:47:24 & 25.39 & -0.19 & 1.03 & 15.44 & 0.58 & 10333 & 1624 \\
G034 & 3.5 & 18:53:19.2 & 01:18:36 & 34.31 & 0.18 & 0.53 & 169.28 & 4.63 & 96118 & 14203 \\
W48B & 2.5 & 18:58:09.6 & 01:37:12 & 35.14 & -0.76 & 0.63 & 12.86 & 1.27 & 11538 & 3862 \\
\hline
\end{tabular}
\tablefoot{
\tablefoottext{a}{Spatial average column density estimate for the foreground plus background along the line of sight toward the cloud.}
\tablefoottext{b}{Areas within column density contours of $N_{\rm H_2}=10^{22.15}~\rm cm^{-2}$ and $N_{\rm H_2}=10^{22.90}~\rm cm^{-2}$, respectively.}
\tablefoottext{c}{Gas masses within column density contours of $N_{\rm H_2}=10^{22.15}~\rm cm^{-2}$ and $N_{\rm H_2}=10^{22.90}~\rm cm^{-2}$, respectively.}
}
\end{table*}

\subsection{\textit{Herschel} imaging data }
While the \artemis ~observations provide high angular resolution to identify small-scale structures, they are not tracing emission from scales larger 
than 2.5\arcmin--5\arcmin, because such extended emission is difficult to distinguish from sky noise. 
To recover extended emission on these scales, we utilized data from \textit{Herschel}  published 
as part of the Hi-GAL and HOBYS surveys \citep{Molinari2010,Motte2010}. 
In particular, we used the PACS $\rm 160~\mu m$ and SPIRE $\rm 250~\mu m$, $\rm 350~\mu m$ and $\rm 500~\mu m$ emission data 
which provide FWHM resolutions of $13.5\arcsec$(Hi-GAL)/$11.7\arcsec$(HOBYS), $18.2\arcsec$, $24.9\arcsec$, $36.3\arcsec$, respectively.

For comparison with the results obtained here on the CAFFEINE sample of massive star-forming complexes, we also used column density maps 
of lower-mass, nearby star-forming clouds (Table~\ref{tab:GB_clouds}), 
constructed from {\it Herschel} Gould Belt survey (HGBS) data\footnote{See \url{http://gouldbelt-herschel.cea.fr/archives}} at a FWHM resolution of $18.2\arcsec$ 
\citep[][]{Andre2010, Palmeirim2013}.

\section{Constructing column density maps}
\label{sec:colden_maps}
\subsection{Combining \artemis ~and \textit{Herschel} data}

To recover extended emission at scales  
larger than $2.5 \arcmin$, we combined the \artemis ~maps with \textit{Herschel}-SPIRE maps at similar wavelengths using the \textit{immerge} task in MIRIAD \citep{Sault1995} and 
the same approach as  \cite{Andre2016} and \cite{Schuller2021b}. 
The \textit{immerge} algorithm combines the available data in the Fourier domain after determining an optimal calibration factor to align the flux scales of the input images (here from \artemis ~and SPIRE) in a common annulus of the UV plane. Here, we adopted an annulus corresponding to the range of baselines from $\rm 0.5~m$ (the baseline $b$ sensitive to angular scales $\lambda/b \sim 2.4\arcmin$ at $\lambda = \rm 350~\mu m$) to $\rm 3.5~m$ (the diameter of the \textit{Herschel} telescope) to align the flux scale of the \artemis ~$\rm 350~\mu m$ map to the flux scale of the SPIRE $\rm 350~\mu m$ map. We used the same method to combine the \artemis ~$\rm 450~\mu m$ data with \textit{Herschel}-SPIRE maps interpolated from $\rm 350~\mu m$ and $\rm 500~\mu m$ to $\rm 450~\mu m$, assuming a linear relation in log space. 

In the combination process, we could identify several areas that suffer from saturation effects in the \textit{Herschel} data. 
While clearly saturated pixels are masked in the published data, we found more pixels in bright areas that deviate from the typical relation between the \textit{Herschel} and \artemis ~pixel intensities. 
Therefore, we used this relation to correct the saturated \textit{Herschel} data 
before performing the combination.

An example of a combined \artemis--SPIRE map is shown in Fig.~\ref{fig:map_example} (right) for the G034 region. The effective resolution is $\sim$$8\arcsec$ and $\sim$$10\arcsec$ (HPBW) at $\rm 350~\mu m$ and $\rm 450~\mu m$, respectively. 
For the clouds GAL316, G337.9, and W33, our method is not sufficient to recover all of the extended emission and leaves areas of ``negative emission''. 
Therefore, we used a more sophisticated
method employing the \textit{Scanamorphos} software for ArTéMiS \citep{Roussel2013, Roussel2018}, 
as described in Appendix~\ref{sect:alt_combination}.
Given the complexity of this alternative method, we applied it only in cases for which it was really necessary.
All of the discussion that follows is based on these combined \artemis ~and SPIRE maps.

\subsection{Constructing high-resolution column density maps}
We used the combined \artemis--SPIRE maps at $350~\mu$m ($8\arcsec $ resolution) and 
$450~\mu$m
($10\arcsec $ resolution)
to construct high-resolution column density maps, employing a pixel by pixel SED (spectral energy distribution) fitting approach. 
To cover a broader range of wavelengths and improve the robustness of the fitting, we also included \textit{Herschel} SPIRE 250$~\mu$m  ($18.2\arcsec $ resolution)
and PACS 160$~\mu$m  ($13.5\arcsec $ resolution) data. 

To derive appropriate zero-level offsets for the \artemis$+$\textit{Herschel} maps and \textit{Herschel-only} maps, 
we used the all-sky high sensitivity maps provided by the \textit{Planck} collaboration. 
The {\it Planck} maps have an angular resolution of $330\arcsec$ and an absolute calibration uncertainty of 7\% \citep{Planck2011}. 
We made sufficiently large cut-outs ($\sim 1\degree \times 1\degree$) around the targeted clouds using the \textit{healpix} library. 
These data were then used to derive expected intensity maps at the \textit{Herschel} wavelengths. 
After smoothing and adapting the \textit{Herschel} maps to the \textit{Planck} reference the median differences at each wavelength were used as zero-level offsets \citep[cf.][]{Bracco2020}.

\begin{figure*}[t]
    \centering
    \begin{minipage}{0.49\textwidth}
        \includegraphics[width=\textwidth,trim={3cm 0 3cm 0},clip]{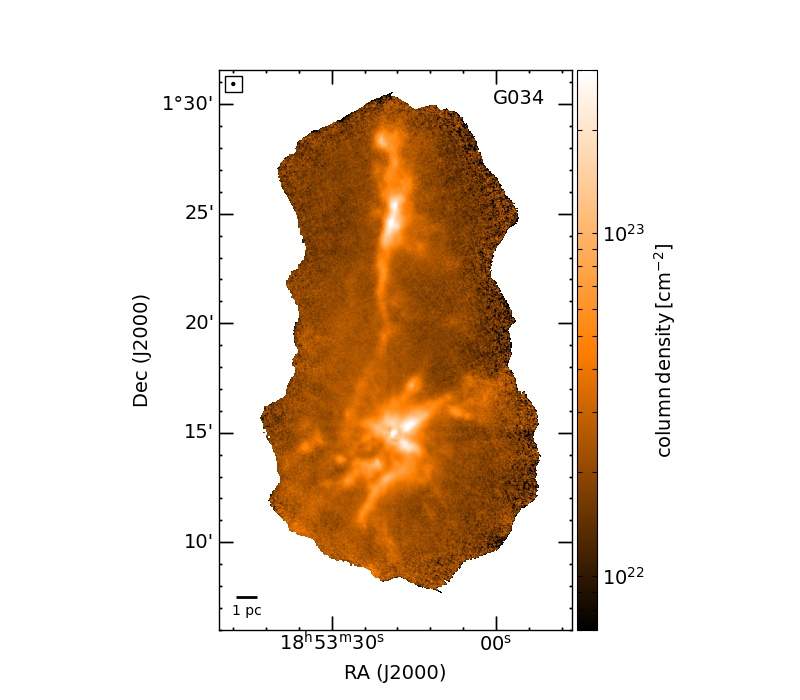}
    \end{minipage}
    \begin{minipage}{0.49\textwidth}
        \includegraphics[width=\textwidth,trim={3cm 0 3cm 0},clip]{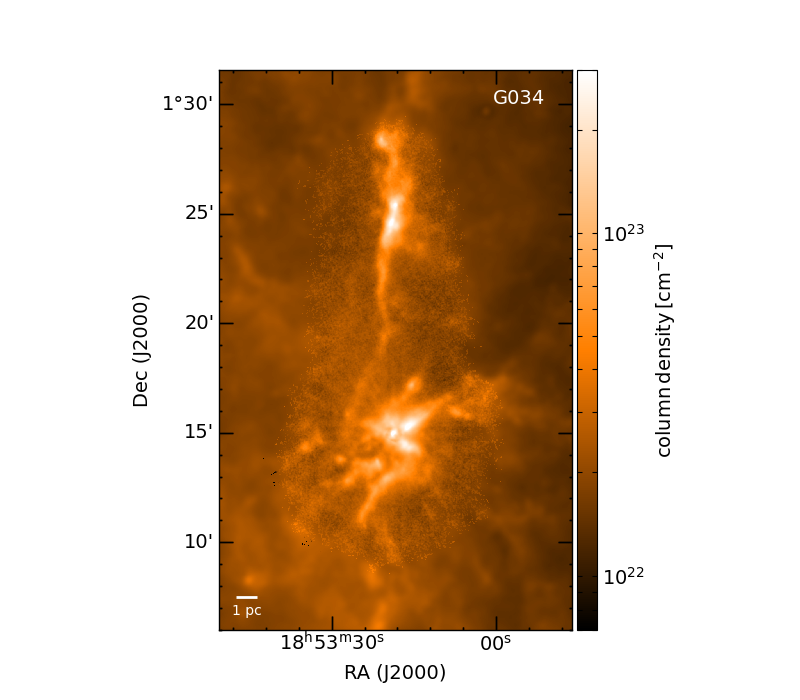}
    \end{minipage}
    \caption{Examples of high level data products from the CAFFEINE survey. \textbf{Left:} Column density map of the G034 region at $8\arcsec$ resolution derived from combined \artemis\ and \textit{Herschel} data. 
    \textbf{Right:} Final multiresolution column density map of the G034 region, with $8\arcsec$ resolution above $N_{H_2}  \sim 4 \times 10^{22}\, {\rm cm}^{-2} $ and $18.2\arcsec$ resolution at lower column densities.}
    \label{fig:column_density_maps}
\end{figure*}

Column density maps were then derived from the offset-calibrated data. 
To achieve the highest possible resolution ($\sim  8\arcsec$), we made use of the methods described in \cite{Schuller2021b}  
and Appendix~A of \cite{Palmeirim2013} for \artemis ~and {\it Herschel} data, respectively. 
The method introduced by \cite{Schuller2021b} can be summarized in the following steps: 
The images in all four bands from $160~\mu$m to 
$450~\mu$m were first smoothed to the lowest resolution of  $18.2\arcsec$, set by the SPIRE $\rm 250\, \mu m$ data. 
A modified blackbody function $\propto B_\nu(T_d) \, \kappa_{\nu}\, N(\rm{H_2}) $ 
was then fitted to the specific intensities in the four bands at every position of the field to derive 
a dust temperature map (and an auxiliary column density map) at $18.2\arcsec$ resolution.
Here, $ B_\nu(T_d) $ denotes the Planck function at frequency $\nu$ (or wavelength $\lambda$) 
for temperature $T_d$ and $ \kappa_{\nu} $ (or $ \kappa_{\lambda} $) is the dust opacity at that frequency (or wavelength).  
The dust temperature map was subsequently used to convert the combined  \artemis--SPIRE map at $350\, \mu$m  
to a $8\arcsec $-resolution column density map (Fig.~\ref{fig:column_density_maps}) based on the following relation:
\begin{equation}
    N(\rm{H_2}) = \frac{I_{\rm 350\mu m}}{B_{350}(T_d) \, \kappa_{350} \, \mu_{H2} m_p}
,\end{equation}
where $I_{\rm 350\mu m}$ is the specific intensity (in MJy/sr)  in the 
\artemis--SPIRE map, $\mu_{H2}=2.8$ is the mean molecular weight 
per $H_2$ molecule \citep{Kauffmann2008}, $m_p$ is the proton mass, and $\kappa_{350} $ is the dust opacity at $\lambda = {\rm 350\, \mu m} $.
For consistency with earlier {\it Herschel} (e.g. HGBS) work, we adopted the following simple dust opacity law at submillimeter wavelengths:
$ \kappa_{\lambda} = 0.1 \times (\lambda/300\, {\rm \mu  m})^{-\beta}\,\rm cm^2 g^{-1}$ for gas $+$ dust \citep{Hildebrand1983}.
(We note that this implicitly assumes a standard gas-to-dust mass ratio of 100.) 
This simple opacity law, very similar to that modeled by \cite{Preibisch1993} and 
\cite{Ossenkopf1994} for dust grains with thin ice mantles 
but no coagulation (often referred to as OH4), 
has been shown to yield column density estimates agreeing within $\sim$50\% with independent dust extinction measurements 
for $ 3 \times 10^{21}\, {\rm cm}^{-2} \la N_{H_2} \la 10^{23}\, {\rm cm}^{-2} $ \citep[see][]{Roy2014}. 
At higher column densities, $\kappa_{300\, {\rm \mu  m}}$ may exceed our default value of $0.1\, \rm cm^2 g^{-1}$ 
by a factor of $\sim$2, as in the dust model(s) with grain coagulation (OH5 or OH6) advocated by \cite{Ossenkopf1994} for dense protostellar cores. 
In turn, our column density maps may overestimate actual column densities by a factor of $\sim$2 in the densest areas, 
with very little influence on the discussion presented in Sect.~\ref{sec:Discussion}.

Following the multiscale decomposition approach introduced by \cite{Palmeirim2013}, 
we also created column density maps at $18.2\arcsec$ resolution based purely on \textit{Herschel} data,   
which extend beyond the dense ($A_V > 40$~mag) areas observed with good sensitivity with \artemis ~and cover the whole extents of the target star-forming complexes. 
In a final step, we combined the $18.2\arcsec$-resolution column density maps from \textit{Herschel} with the $8\arcsec$-resolution column density maps 
derived from \artemis$+$\textit{Herschel} data as explained above, to produce multiresolution column density maps of the target fields (Fig.~\ref{fig:column_density_maps}). 
These multiresolution column density maps coincide with the \artemis$+$\textit{Herschel} column density maps and thus have a resolution 
of $\sim \, 8\arcsec$ in the dense ($A_V > 40$~mag) inner portions of the fields and a resolution of $18.2\arcsec$ in the lower-density outer parts. 

The combined 
\artemis--SPIRE maps at $\rm 350~\mu m$ and $\rm 450~\mu m$, the \artemis ~weight maps, and 
the final multiresolution column density maps, 
are made publicly available as data products through the ESO archive\footnote{\url{https://archive.eso.org/scienceportal/home}} and the CAFFEINE web page\footnote{\url{https://sites.google.com/view/caffeine-2024/fits-maps}}.

\section{Methods and analysis}

\subsection{Evaluating and subtracting the line-of-sight contamination}
\label{sec:contamination}
The submillimeter dust continuum emission traced by ArT\'eMiS, \textit{Herschel}, and \textit{Planck} 
data covers a wide range of column densities 
and is therefore ideal to derive reliable cloud mass estimates. Because of the otherwise desirable sensitivity, these measurements, however, also trace 
emission from dust along the line-of-sight which is not associated with the targeted molecular cloud. 
This contribution is especially critical within the inner Galaxy where the number density of molecular clouds is high.

To address this problem, we used \cco observations from several surveys covering the Galactic Plane. We preferably used 
the SEDIGISM \ccol survey \citep{Schuller2021a} and the catalog of identified molecular clouds \citep{Duarte-Cabral2021}. 
Identifying the CAFFEINE clouds within the SEDIGISM catalog ensures that the clouds are truly velocity coherent structures and that the SEDIGISM cloud masks define their spatial extents in the plane of the sky.

We found three regions where the emission detected by CAFFEINE is assigned to multiple SEDIGISM clouds. In the case of SDC338 and SDC010, 
the high column density areas are located within a common larger area of low column densities and 
the distances to the different SEDIGISM clouds are comparable within the uncertainties. 
Therefore, we decided to treat them as single clouds, as they were observed with CAFFEINE. 
In the case of I17233, three SEDIGISM clouds toward its direction exhibit large differences in distance from the Sun. 
We therefore separated them into two components I17233A and I17233B. 
The third cloud has a distance of $\rm 6.5~kpc$ and was thus excluded from the present analysis.

In regions for which the SEDIGISM survey is not available, we used $^{13}$CO(1--0) data from the GRS and THRUMMS surveys \citep{Jackson2006, Barnes2015}. 
As there are no cloud masks available for these surveys, we identified clouds directly from the corresponding data. 
To do so, we first selected the dominant velocity component in the line-of-sight to each region and then integrated over this component down to a signal-to-noise ratio (S/N) limit of $\rm S/N \ge 5$. 
The resulting integrated intensity map gave us the extent of the region similar to those of the SEDIGISM clouds. 

From the molecular line data we also derived an estimate of the diffuse ISM along the line-of-sight in the Milky Way \citep[see, e.g.,][]{Mattern2018b, Peretto2023}. 
The above-mentioned CO surveys trace any significant amount of molecular gas within the covered areas. 
Therefore, we derived integrated intensity maps  over the full velocity range and over the velocity range of the cloud of interest, respectively. 
We then created a ratio map to determine the contribution of the cloud compared to the total line-of-sight. 
We found that all CAFFEINE clouds are the main component in their respective directions. 
To estimate the contribution of the diffuse gas to the total column density, we smoothed the ratio map to the resolution of the \textit{Planck} column density map and subtracted this fraction. 
The average of the 
subtracted column densities over the area of each target cloud was used as an estimate of the diffuse gas column density, 
which we refer to as background (see $N({\rm H_2})_\text{back}$ in Table~\ref{tab:CAFFEINE_sources}).

Notably, 
the clouds  SDC340, RCW116B, SDC348, and SDC014 are located slightly off the Galactic Plane ($|b| \approx 0.5$ -- $1.0\degree$) and are not covered by any of the 
above CO surveys. For these clouds the impact of diffuse line-of-sight emission is probably not negligible, but we did not find significant confusion with other clouds by comparing the 
distances of compact Hi-GAL sources \citep{Elia2021} toward these clouds.  
We thus defined the background as a spatially constant value derived as the mean over low column density areas within each map. 
The clouds MonR2, R5180, and NGC2264 are even further away from the Galactic Plane or toward the outer Galaxy where the diffuse background emission is negligible.

\subsection{Star formation rates}
\label{sec:SFR_CAF}
\subsubsection{Star formation rates from young stellar objects}
\label{sec:SFR-CAF-YSO}
To evaluate the SFRs of the CAFFEINE clouds we first need to identify the YSOs present in each. 
For that, we combined the YSOs catalogs from \cite{Marton2016, Marton2019} and \cite{Kuhn2021}. \cite{Marton2016} produced a Class I/II YSO catalog from a training method analyzing the reliable 2MASS and WISE photometric data. Adding Gaia-DR2 information to the WISE data \cite{Marton2019} produced a more recent catalog that included the probability for an object to be a YSO. 
From this catalog, we selected objects with a probability higher than 80\% of being a YSO. 
In addition, the YSO class (class I, II, III or flat SED) was established from the infrared 
spectral index $\alpha_{\rm IR}$ \citep[calculated from the WISE magnitudes following][]{Kang2017} using the ranges of $\alpha_{\rm IR}$ values 
recommended by \cite{Greene1994} for each class. 
An alternative YSO catalog is the {\it Spitzer}/IRAC candidate YSO (SPICY) catalog produced by \cite{Kuhn2021}. 
This catalog is mainly based on near-IR data and is more sensitive to Class II YSOs (as illustrated by their Figure 15). 
Kuhn et al. identified candidate YSOs from infrared excesses consistent with the SEDs 
of pre-main sequence stars with disks or envelopes. Further, they assigned a class (Class I, II, III or flat SED) to every object from the infrared spectral index \citep[also using the $\alpha_{\rm IR}$ ranges from][]{Greene1994}. 
For each catalog we selected Class I, II and flat-spectrum YSOs, and then concatenated the results (taking into account duplicates) to produce a global YSO catalog. 
For YSO classification, these catalogs used a machine learning method which is designed to minimize contamination from galaxies and evolved stars efficiently. 
A fraction of 6 $\%$ of false YSO identification 
is estimated by \cite{Marton2019}.  
For the MonR2 region, we adopted the publicly available YSO catalog compiled by Gutermuth et al. (in preparation) from the {\it Spitzer} Extended Solar Neighborhood Archive (SESNA). 
 
\begin{figure}
    \centering
    \includegraphics[width=0.49\textwidth,trim={0.5cm 0.5cm 0.0cm 2.0cm},clip]{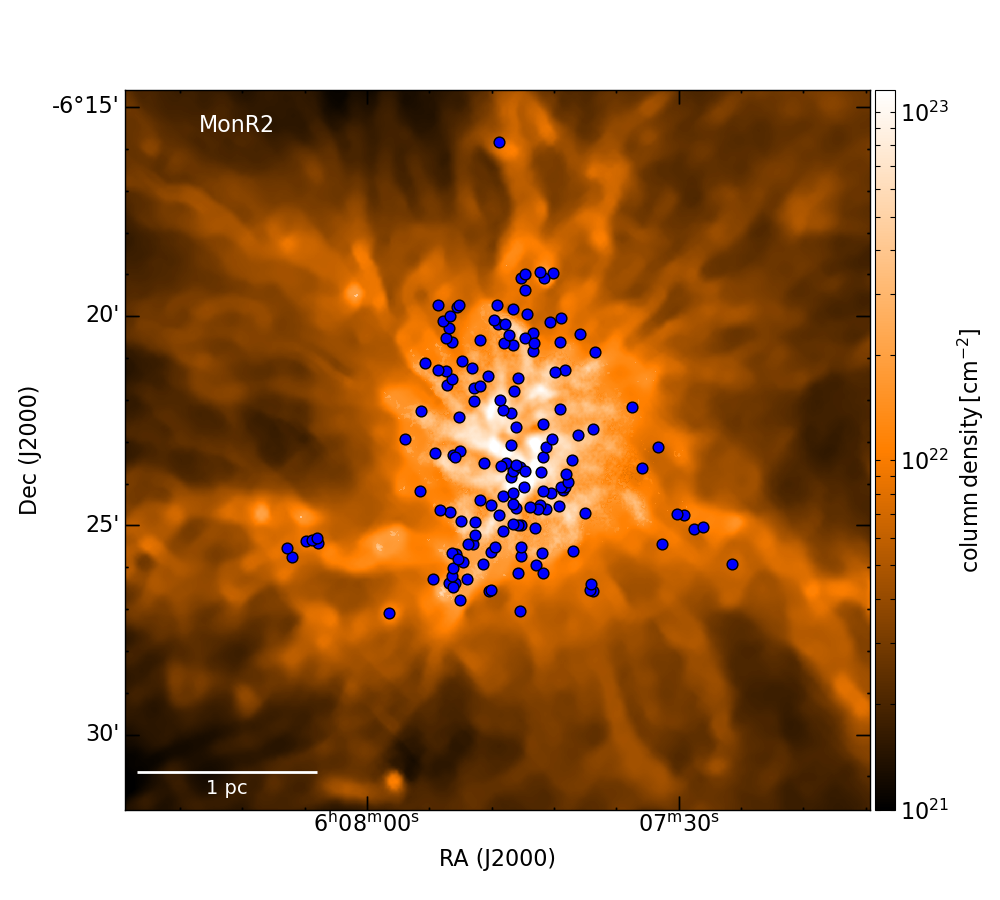}
    \caption{Multiresolution column density map of the MonR2 region based on ArT\'eMiS $+$ {\it Herschel} data, with blue dots marking the locations of identified YSOs within the cloud area. 
    The resolution of the map ranges from $\sim$$8\arcsec$ at $N_{H_2}  > 4 \times 10^{22}\, {\rm cm}^{-2} $ to $18.2\arcsec$ resolution at lower column densities.
    }
    \label{fig:map_YSO_example}
\end{figure}

We then selected all of the YSOs falling in the cloud mask (see Fig.~\ref{fig:map_YSO_example} for the MonR2 example) and for each YSO we ran the SED fitting tool from \cite{Robitaille2017}. 
Following \cite{Sewilo2019} after fitting the models, we counted the number of good fits provided by each model set. Of these,  the one with the largest number of good fits was selected 
as the best model set from which the adopted best model was that with the smallest $\chi^{2}$. The selected model then gave the modeled radius and temperature of the source from which we evaluated the mass using 
the zero-age main sequence (ZAMS) luminosity-mass relation\footnote{The ZAMS luminosity-mass relation is a reasonable assumption here, since 
most of the {\it Spitzer}/WISE YSOs selected in the CAFFEINE regions are relatively high luminosity objects ($> 80\%$ are more luminous than $10\ \rm L_\odot $, and $> 50\%$ more luminous than $50\ \rm L_\odot$)
and thus unlikely to be dominated by accretion. The detailed {\it Spitzer} study of the W51 
complex by \citet[][]{Kang2009} 
confirms that the ZAMS relation is generally 
appropriate for YSOs with $L_{\rm tot} > $\ 10--50\ $ \rm L_\odot$. Moreover, the luminosity used here to derive $M_\star$ is the stellar luminosity $L_\star$ estimated from the model fitting, not the observed $L_{\rm tot} = L_\star + L_{\rm acc}$. } 
as in \cite{Immer2012b}. We then followed the method described in \cite{Immer2012b} by fitting the mass distribution histogram with a power-law curve $\psi = A M^{-2.3}$ (where $A$ is the normalization factor),  
in accordance with the form of the \cite{Kroupa2002} initial mass function (IMF) for $M \ge \rm 0.5~M_\odot$. 
The fit allowed us to solve for the value of the normalization factor $A$. Then, assuming a continuous IMF at $ M = \rm 0.5~M_\odot$, we deduced the normalization factors for the low-mass end of the \cite{Kroupa2002} IMF 
in the range 0.01$\rm~M_\odot$ to 0.5~$\rm M_\odot$, and estimated the total mass of YSOs in the mass range 0.01 to 120 $\rm M_\odot$. 
The SFR was estimated by dividing this YSO mass by a SF timescale of $\rm 0.5 + 2 = 2.5~Myr$
corresponding to the estimated total duration 
of the Class~I $+$ Class~II YSO evolutionary phases \cite[see][]{Evans2009}. 
This timescale is probably uncertain by factor of $\sim$2 on either side. 
On one hand, the median lifetime of Class~II YSOs with significant infrared excess may be slightly longer than previously thought \citep[i.e., $\sim$3--4\,Myr,][and references therein]{Dunham2015}.
On the other hand, theoretical work on pre-main sequence (PMS) evolutionary tracks suggests that episodic accretion during the embedded protostellar phase 
can make a $\sim$1\,Myr old PMS star appear as old as $\la$10\,Myr if standard, non-accreting isochrones are used in the Hertzsprung–Russell diagram to derive its age \citep{Baraffe2009}. 
In view of these uncertainties, for better consistency and easier comparison with recent literature on SFRs and SFEs, we adopted the same timescale values as \citet[][]{Evans2009}.

To evaluate the completeness of the YSO population 
we retrieved the WISE 4.6 $\mu$m and the IRAC 4.5 $\mu$m magnitude for PMS stars of age 0.5 Myr (typical lifetime for a Class I YSO) and 2 Myr (typical lifetime for a Class II YSO) from PARSEC\footnote{http://stev.oapd.inaf.it/cgi-bin/cmd} isochrones \citep{Bressan2012}. We adopted a sensitivity limit of 15.7 mag and 14.5 mag for WISE 4.6 $\mu$m and the IRAC 4.5 $\mu$m \citep{Koenig2014}. 
In this way, we estimated that YSOs more massive than 3 $\rm M_\odot$, at 2 kpc, should be detected at least up to a column density of $1.2 \times 10^{23}\rm \, cm^{-2}$. 
For $1 \rm M_\odot$ YSOs, the detection limits are at column densities of $2 \times 10^{22}\, \rm cm^{-2}$ and  $7.5\times 10^{22}\, \rm cm^{-2}$ for Class~II and Class~I objects, respectively. 
These estimates are conservative as they do not take into account the infrared excess emission expected from accretion in Class~I and Class~II objects.

On the other hand, the YSOs population can be underestimated due to confusion with extended nebulosity, neighboring bright sources and/or saturation \citep[e.g.][]{Koenig2014}. 
For example \cite{Megeath2016} estimated that the fraction of detected stars dropped to about 10$\%$ in the brightest parts of the Orion Nebula compared to regions with only faint nebulosity. 

In practice, for the studied regions, the histograms of YSO masses suggests that good YSO completeness is reached between 2 and 4~M$_\odot$ on average.

The uncertainty of the SFR in each cloud is given by
\begin{equation}
    \frac{\Delta{\rm SFR}_\text{YSO}}{{\rm SFR}_\text{YSO}}=\sqrt{ \left(\frac{\Delta M_\star}{M_\star} \right)^2+\left(\frac{\Delta t_\star}{t_\star} \right)^2}
    \label{equ:dSFR-YSO}
,\end{equation}
where $\Delta M_\star$ and $\Delta t_\star$ are the stellar mass and evolutionary timescale uncertainties, respectively. 
We adopted a relative uncertainty of $\frac{\Delta t_\star}{t_\star} = 100\%$ for the duration of the combined Class I plus Class II YSO evolutionary phases \citep[cf.][]{Evans2009}.
We evaluated the typical value of $\frac{\Delta M_\star}{M_\star}$ from the quadratic sum of the relative mass uncertainties coming from the SED fitting (45\%), the IMF binning and fitting algorithm (40\%), 
and the (in)completeness impact (30\%).   

\subsubsection{Star formation rates from protostellar clumps}
\label{sec:Meth_clump_relation}
\begin{figure}
    \centering
    \includegraphics[width=0.49\textwidth,trim={0.5cm 0.5cm 0.0cm 2.0cm},clip]{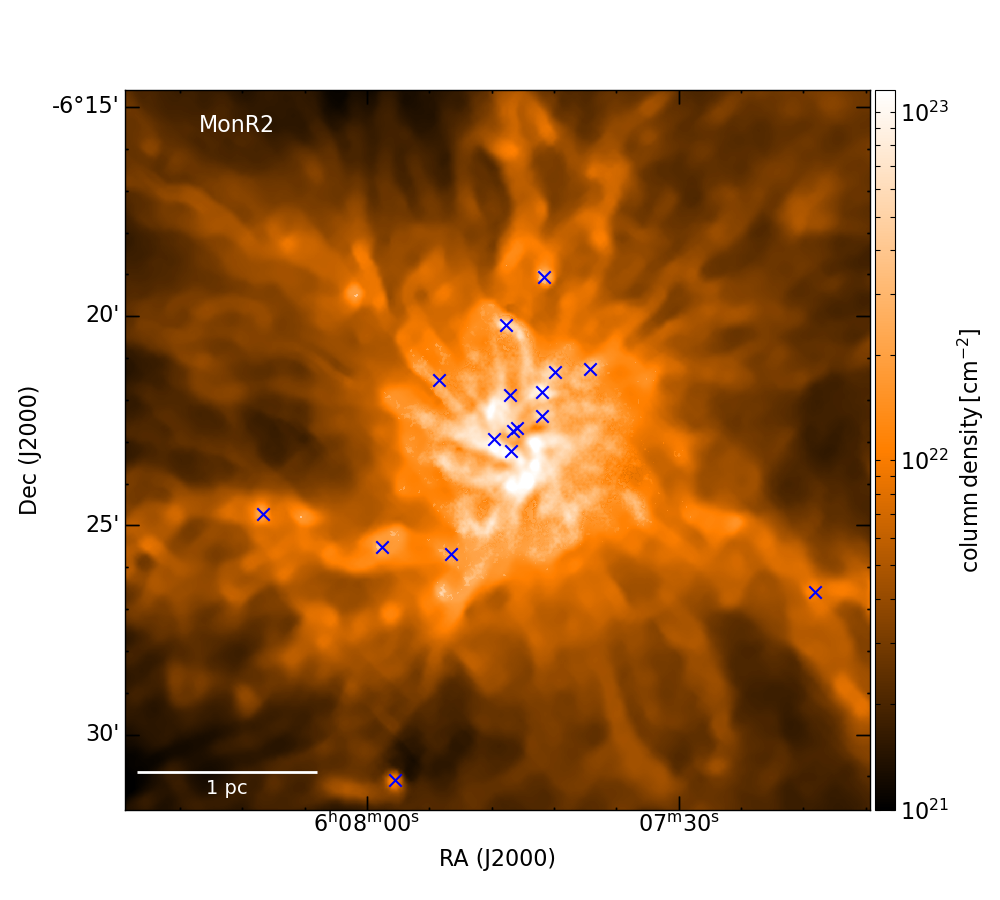}
    \caption{Multiresolution column density map of the MonR2 region  based on ArT\'eMiS $+$ {\it Herschel} data, with blue crosses marking the locations of identified protostellar clumps within the cloud area. 
    The resolution of the map ranges from $\sim$$8\arcsec$ at $N_{H_2}  > 4 \times 10^{22}\, {\rm cm}^{-2} $ to $18.2\arcsec$ resolution at lower column densities. 
    }
    \label{fig:map_clump_example}
\end{figure}

We utilized the reliable catalog of compact Hi-GAL sources frrom \cite{Elia2021} for a second, independent estimate of SFRs. 
For each clump, this catalog gives the mass, distance, velocity and evolutionary status (starless, prestellar, or protostellar). 
For MonR2 (not covered by Hi-GAL), we used the clumps catalog from \cite{Rayner2017} established from the {\it Herschel} HOBYS survey. 
We selected the protostellar clumps (detected by {\it Herschel}/PACS at 70\,$\mu$m) falling in the mask and with a velocity in the velocity range of the cloud, 
and then estimated an SFR using the SFR--protostellar clump mass relation given by \cite{Elia2022}.
As an example, we show the Hi-GAL protostellar clumps of the MonR2 region in Fig. \ref{fig:map_clump_example}.
The SFR of one protostellar clump was calculated from the \cite{Elia2022} relation, as follows:

\begin{equation}
    {\rm SFR}_\text{clump} = (5.6 \pm 1.4) \times 10^{-7} (M_\text{clump} / M_\odot)^{0.74 \pm 0.03} ~~ \rm M_\odot yr^{-1}  
    \label{equ:SFR-clump}
.\end{equation}

This relation 
is based on theoretical evolutionary tracks reported in the bolometric luminosity versus mass diagram (\citealt{Molinari2008}, \citealt{Baldeschi2017}, \citealt{Elia2021}).
Initially developed for the formation of a single massive protostellar object, the tracks have been updated by \cite{Veneziani2017} to account for the multiplicity of sources produced during the clump collapse.
Then \cite{Elia2022}, following \cite{Veneziani2017}, developed an algorithm which associates the final star mass to the input clump mass by interpolating the locus of final masses for known evolutionary tracks with a power law 
leading to Eq.~\ref{equ:SFR-clump} above.
An SFR$_\text{clump}$ value was then calculated from Eq. \ref{equ:SFR-clump} for each protostellar clump and then the SFR was estimated as the sum of the SFR$_\text{clump}$ over all clumps belonging to the region. 
Following Eq.~\ref{equ:SFR-clump} the uncertainty of the SFR per clump is given by

\begin{equation}
    \frac{\Delta{\rm SFR}_\text{clump}}{{\rm SFR}_\text{clump}}=\sqrt{\left(\frac{1.4}{5.6}\right)^2+\left(\ln(M_\text{clump})\cdot0.03\right)^2+\left(0.74\cdot \frac{\Delta M_\text{clump}}{M_\text{clump}}\right)^2}  \, ,
    \label{equ:dSFR-clump}
\end{equation}
and accordingly the uncertainty in the total SFR of the cloud was derived as $\Delta \text{SFR} = \sqrt{\sum_i^{N_\text{clump}} \Delta \text{SFR}_\text{clump}^2}$, where $N_\text{clump}$ is the number of protostellar clumps within the region.

\section{Results}
\label{sec:Results}

\subsection{Star formation efficiencies in CAFFEINE clouds }
\label{sec:CAFFEINE_obs}
Given the lists of Class II YSOs and protostellar clumps for each cloud in our sample, 
we constructed a set of column density contours and derive the SFR and gas mass within each contour. 
We followed the same approach as \cite{Pokhrel2021} but used different tracers for the SFR. 

To ensure a minimum level of statistical accuracy, we introduced limits for the SFR estimates. 
For a reliable SFR estimate derived from YSOs, we imposed a minimum number of YSOs of $N_\text{YSOs} > 5$ within the area of a contour as the fitting of the IMF to the mass distribution would otherwise be  too uncertain. 
We similarly imposed $N_\text{clumps} > 5$ as Eq.~\ref{equ:SFR-clump} \citep[from][]{Elia2022} refers to an average over several protostellar clumps 
and applying it to a region with a very small number of Hi-GAL protostellar clumps would become highly inaccurate. 
For a similar reason, we also excluded estimates for which the cloud mass is lower than twice the combined clump mass and imposed   
$M_\text{cloud} \ge 2 \times \sum_i^{N_\text{clumps}}  M_\text{clump}$.

For the two different SFR estimates, 
the cumulative SFE within each column density contour, $\rm SFE(>N_{H_2})$,  was then derived  as: 
\begin{equation}
    {\rm SFE}(>N_{\rm H_2}) = \frac{{\rm SFR}(>N_{\rm H_2})}{M_\text{cloud}{(>N_{\rm H_2})}} \,.
    \label{equ:SFE_def}
\end{equation}
This definition is however only valid for a well-defined cloud mass. 
We therefore considered only SFR and SFE estimates for which the corresponding column density contour is closed within the map of the cloud \citep[cf.][]{Alves2017}.
This limiting column density can be approximated by the peak of the column density probability density function (N-PDF) in the observed region. 

We derived the N-PDF ($\frac{\delta \tilde{N}}{\delta \log  N_{\rm H_2}}$) of each region from the background-corrected multiresolution column density map 
using a histogram of $\log N_{H_2}$ with a bin size of $\delta \log  N_{\rm H_2} =  0.05\, \rm dex$. 
We also translated column density to visual extinction, $A_\text{V}$, by adopting the standard \cite{Bohlin1978} relation, 
\hbox{$N_{\rm H_2}  = 0.94\times 10^{21}~{\rm cm^{-2}} \times A_\text{V} \rm[mag]$}. 
The number of pixels per bin $\delta {N}$
was then normalized to $\delta \tilde{N}$ so that $\sum \frac{\delta \tilde{N}}{\delta \log  N_{\rm H_2}} = 1$. 
The location of the N-PDF peak gives an indication of the lowest closed column density contour within the observed area. 
Furthermore, the power-law tail observed at higher column densities, described as $\frac{\delta \tilde{N}}{\delta \log  N_{\rm H_2}} \propto N_{\rm H_2}^s$  
where $s$ is the logarithmic slope of the power-law tail, is an indicator of the cloud's 
overall radial density structure 
\citep[e.g., see][]{Kainulainen2009,Schneider2013}. 
The median slope observed at high column densities in the whole CAFFEINE sample is $s \approx -2.3$, 
which is consistent with an average radial density profile $\rho(r) \propto r^{-1.9}$ at high densities (cf. Sect.~\ref{sec:models} below).

\begin{figure}
    \centering
    \includegraphics[width=0.49 \textwidth]{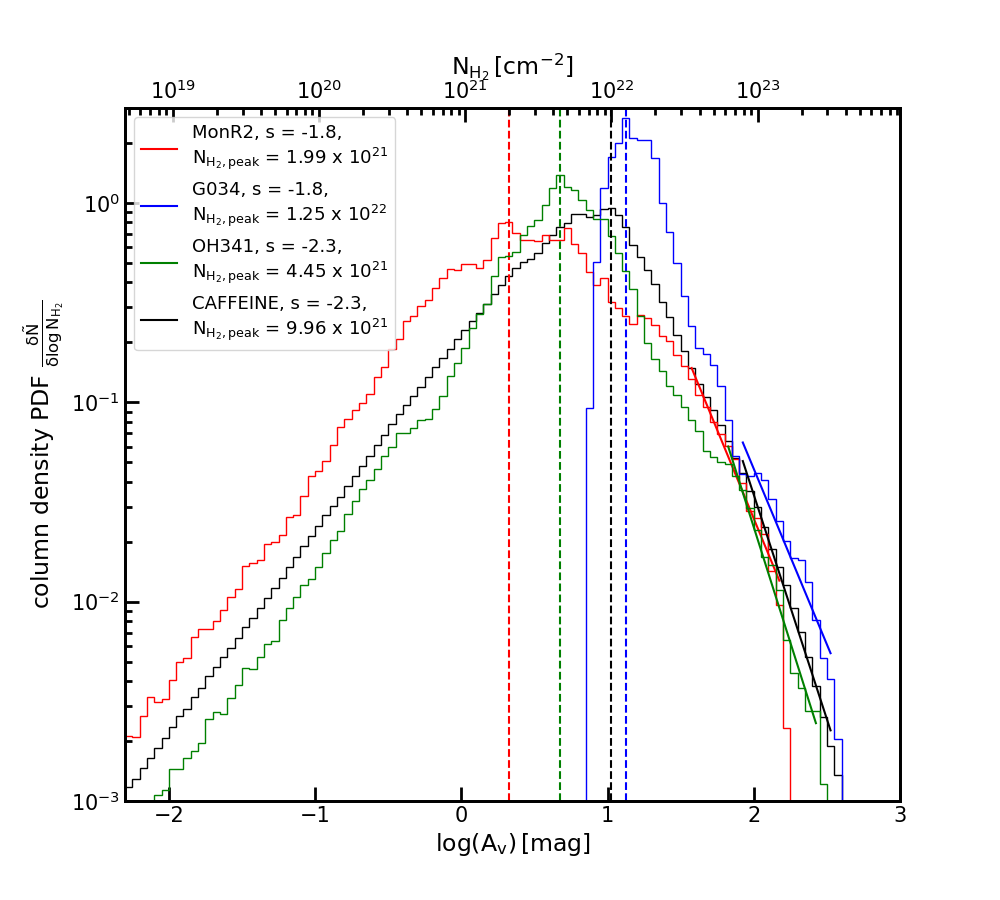}
    \caption{Column density probability density functions (N-PDFs) as a function of column density (top axis) and visual extinction (bottom axis) for the MonR2, G034, and OH341 regions 
 and for all CAFFEINE regions combined in red, blue, green and black, respectively. The peaks of the distributions are indicated by dashed lines with the column density $N_{\rm H_2, peak}$ presented in the legend. Additionally, solid lines indicate the fitted power-law tail at high column densities with the slope $s$ also mentioned in the legend.}
    \label{fig:PDF-example}
\end{figure}

In Fig.~\ref{fig:PDF-example}, we show the N-PDFs for three example regions and the overall CAFFEINE cloud sample. 
The peak of the N-PDF varies from cloud to cloud 
and is located at about $\rm 1 \times 10^{22} ~cm^{-2}$ for the combined sample. 
We therefore adopted this latter column density as the lower limit of reliable SFE estimates for the CAFFEINE clouds.  
Larger multiresolution maps would allow us to cover lower column density contours, 
but would increase the uncertainties due to line-of-sight confusion, in particular toward the Galactic Plane.

Over a broad range of column densities from $\sim \rm 10^{22} ~cm^{-2}$ to $\sim \rm 10^{23} ~cm^{-2}$, 
the standard deviation of the derived SFE values across the sample of CAFFEINE clouds is about 0.3 and 0.5 dex for the clump and YSO estimates,  respectively. 
These values reflect the complexity of SF on cloud scales where several parameters can have an influence. 
As we are mostly interested in the general trend between SFE and column density,  
we performed a weighted geometric average (in $\log$ space) over all the CAFFEINE clouds of the form:
\begin{equation}
    \log {\rm SFE}(>N_{\rm H_2})=\left(\frac{\sum_i^{N_{\rm reg}} \log_{10}({\rm SFE}_i(>N_{\rm H_2})) \cdot w_i}{\sum_i^{N_{\rm reg}} w_i}\right) ,
    \end{equation}
with the weight $w_i$ of a region $i$ defined as:
\begin{equation}
\begin{split}
    w_i&=1/\left(\frac{\Delta {\rm SFE}_i(>N_{\rm H_2})}{{\rm SFE}_i(>N_{\rm H_2})}\right)^2  \\
    \,&=1/\left(\left(\frac{\Delta {\rm SFR}_i(>N_{\rm H_2})}{{\rm SFR}_i(>N_{\rm H_2})}\right)^2+\left(\frac{1}{N_{i\rm,\,obj}(>N_{\rm H_2})}\right)\right), 
\end{split}
\end{equation}
where $\frac{\Delta N_\text{obj}}{N_\text{obj}}=\frac{1}{\sqrt{N_\text{obj}}}$ is the relative Poisson error given the number of detected objects $N_\text{obj}$ (clumps or YSOs, respectively), and $\frac{\Delta{\rm SFR}}{\rm SFR}$ is the relative uncertainty of the SFR as given by Eqs.~\ref{equ:dSFR-YSO} and~\ref{equ:dSFR-clump}.

\begin{figure}
    \centering
    \includegraphics[width=0.49 \textwidth]{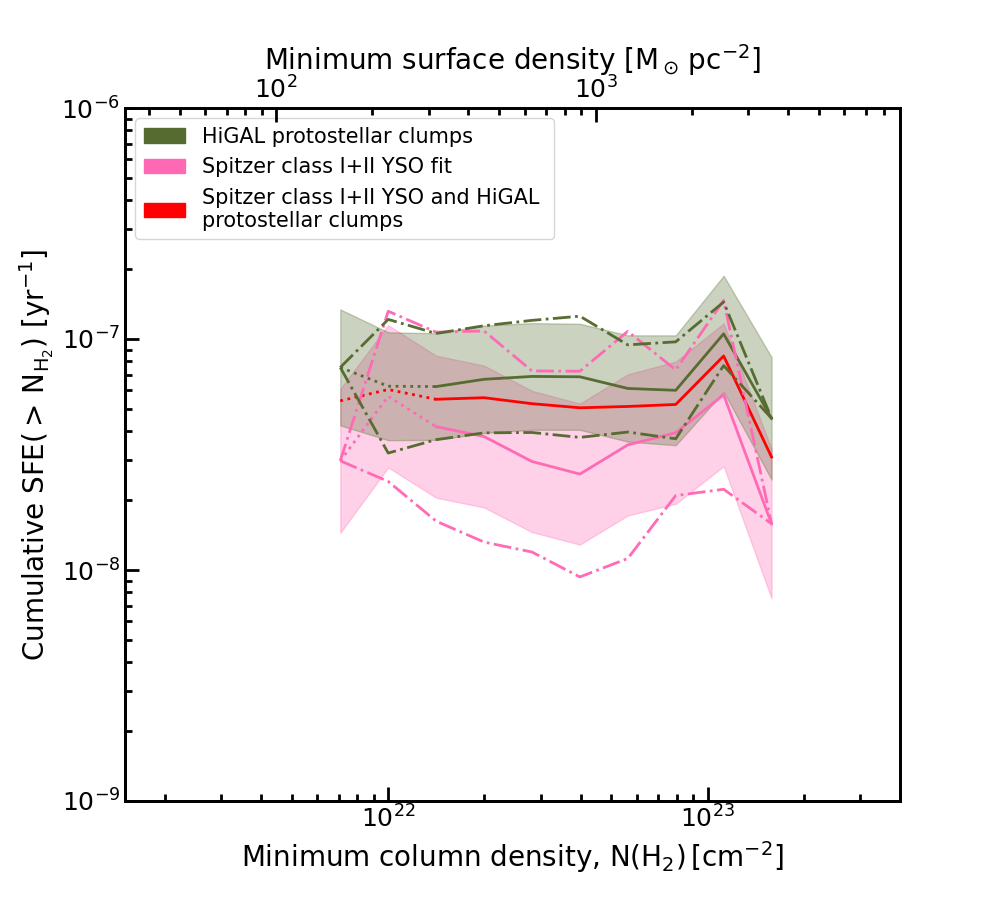}
    \caption{Cumulative SFE above a minimum column density ($N_{\rm H_2}$)
    contour  as a function of column density 
    for the CAFFEINE clouds. 
    The pink and dark green curves indicate weighted averages over all significant cloud estimates above each $N_{\rm H_2}$, 
    with SFRs derived from YSOs and Hi-GAL protostellar clumps, respectively. The shaded areas indicate the corresponding statistical uncertainties (cf. Eq. 8). 
    The dashed-dotted lines mark the standard deviations of the sample of cloud SFE estimates above each $N_{\rm H_2}$.
    The red curve marks the weighted geometric average of the two methods for estimating the SFRs and SFEs.}
    \label{fig:SFE-colden_CAF}
\end{figure}

We present the average SFE estimates for column density contours of $\rm 7.1\times 10^{21}~cm^{-2}$ to $\rm 2.2\times 10^{23}~cm^{-2}$ with steps of 0.15 dex in Fig. \ref{fig:SFE-colden_CAF}. 
We chose this column density step size to ensure that typically at least one protostellar clump is located in each bin. 
We give the numbers of protostellar clumps and YSOs and the respective estimates of the SFR and SFE for each cloud for the contours of $\rm 1.4\times 10^{22}~cm^{-2}$ and $\rm 7.9\times 10^{22}~cm^{-2}$ in Table~\ref{tab:CAFFEINE_sources_results} (Appendix~\ref{sect:CAFFEINE_SFR}).

The net uncertainties in the derived average SFE values were estimated from a combination of the Poisson statistical uncertainties 
and the systematic uncertainties inherent to each method. 
The relative uncertainty in the average SFE was derived as follows: 
\begin{equation}
    \frac{\Delta {\rm SFE}(>N_{\rm H_2})}{{\rm SFE}(>N_{\rm H_2})} = \sqrt{\frac{\sum_i^{N_\text{reg}} \left(\frac{w_i^2}{N_{i\rm,\,obj}(>N_{\rm H_2})}\right)}{\left(\sum_i^{N_\text{reg}} w_i\right)^2}+\frac{\sum_i^{N_\text{reg}}\left(\frac{\Delta{\rm SFR}_i(>N_{\rm H_2})}{{\rm SFR}_i(>N_{\rm H_2})}w_i\right)^2}{\left(\sum_i^{N_\text{reg}}w_i\right)^2}}, 
\end{equation}
where $\frac{\Delta N_\text{obj}}{N_\text{obj}}=\frac{1}{\sqrt{N_\text{obj}}}$ is the relative Poisson error, $\frac{\Delta{\rm SFR}}{\rm SFR}$ is the relative uncertainty of the SFR, and $N_\text{reg}$ is the number of regions with a valid 
SFE estimate at the given column density contour. 
We did not include the systematic uncertainty of a factor of $\sim 2$ in the cloud mass as it is identical for all SFE estimates.

Both methods for estimating the SFE use only tracer objects to infer the full population of YSOs. 
This inherent limitation causes the SFR and thus the SFE to be strongly dependent on the number of identified objects. 
By averaging over a large number of clouds, however, the statistical uncertainties decrease, but the systematic errors 
remain and quickly become the dominant source of uncertainty in the weighted averages. 

Despite a slight systematic disagreement between the two methods of deriving the SFR for the CAFFEINE clouds, 
both SFE curves presented in Fig.~\ref{fig:SFE-colden_CAF} show a consistent trend within the uncertainties. 
We find a relatively flat behavior around a mean of $\rm SFE_\text{clump} \approx 6.6\pm 1.2\times 10^{-8} \,yr^{-1}$ for the protostellar clump SFE. 
The YSO fitting method generally leads to lower SFE estimates with larger variations, but is in agreement with a flat trend 
around a mean of $\rm SFE_\text{YSO} \approx 3.5\pm 1.5\times 10^{-8} \,yr^{-1}$. 
At the highest column densities ($N_{\rm H_2} \ge \rm 8.0 \times 10^{22} \, cm^{-2}$), 
both methods yield almost identical estimates. 

\subsection{Comparing the two types of SFE estimates for the CAFFEINE sample}
\label{sec:Discussion_SFR_CAF}

As described above, both types of SFE estimates for the CAFFEINE clouds show a relatively flat behavior with increasing column density and larger variations at the highest column densities ($N_{\rm H_2} \ge \rm 10^{23} \, cm^{-2}$, Fig.~\ref{fig:SFE-colden_CAF}). These variations arise from larger statistical uncertainties, as fewer regions are able to probe these high column densities and the impact of single clouds increases. This situation is also reflected in the lower standard deviation of the cloud SFE distributions, as indicated by the dashed-dotted lines in Fig.~\ref{fig:SFE-colden_CAF}. Seeing an almost identical behavior for SFE estimates for $N_{\rm H_2} > \rm 10^{23}~cm^{-2}$ from clumps and YSOs separately indicates that the measurements are dominated by the same clouds, which shows that estimates at high column densities are limited by the number of clouds and the amount of dense gas that can be observed.

We also investigated the ability 
of the two methods to trace SF at high (column) densities. Here we looked at the number of clumps and YSOs that could be identified in each column density bin, that is, in the area between two 
consecutive column density contours (see Appendix~\ref{sect:Distribution_YSOs-clumps} and Fig.~\ref{fig:sources/area_CAFFEINE}).

The number of protostellar clumps and YSOs observed between column density contours  
decreases at higher densities, which is predominantly an effect of the smaller enclosed area at higher column densities. 
The SFE estimates are also affected by the distribution of clumps and YSOs within a cloud (i.e. see Figs. \ref{fig:map_YSO_example} and \ref{fig:map_clump_example}) and their detection. 
To remove the area effect, we also show the surface number densities of clumps and YSOs per unit area as a function of column density in Fig.~\ref{fig:sources/area_CAFFEINE}.

The surface density of YSOs per unit area increases with column density (Fig.~\ref{fig:sources/area_CAFFEINE}), indicating a concentration of YSOs toward denser areas,  
with the exception of the two highest column density bins. 
As mentioned in Sect.~\ref{sec:SFR-CAF-YSO} the WISE/{\it Spitzer} objects identified toward the CAFFEINE clouds
are not a complete sample of YSOs, but only the brightest, most massive and least obscured ones. 
We expect a higher degree of YSO incompleteness in denser areas
where extinction is higher. 
This effect was corrected for in our SFR estimates by fitting an IMF to the detected YSOs, but eventually leads to an underestimated SFR at very high column densities. 
The analysis in Appendix~A suggests that 
our method of estimating the SFR from massive YSOs is 
strongly impacted by incompleteness for column densities \hbox{$\ga $\ 7.5--10\ $\times 10^{22}\, \rm cm^{-2}$}, 
in agreement with the detection limit estimates made in Sect.~\ref{sec:SFR-CAF-YSO}.

Protostellar clumps, identified as bright peaks in far-infrared and 
submillimeter continuum emission, are not expected to suffer from a similar extinction limitation at high column densities. 
And, indeed, the surface density of Hi-GAL protostellar clumps continues to increase with column density up to 
the highest column density bins around $\sim $\ 1--2\ $\times 10^{23}\, \rm cm^{-2}$ (cf. Fig.~\ref{fig:sources/area_CAFFEINE}).
The method of \cite{Elia2022} was, however, designed to derive SFRs on large scales, including many protostellar clumps. 
Therefore, a minimum number of clumps 
is necessary to get accurate SFE$_{\rm clump}$ estimates.  
This was ensured by imposing $N_{\text{clumps}\, > \, N_{\rm H_2}} > 5$ for a region to be considered 
in the calculations of the SFR in Sect.~\ref{sec:Meth_clump_relation} and the average SFE value above a given column density $N_{\rm H_2}$.

Toward lower column densities, both types of SFE estimates are limited by the spatial extent of the ArT\'eMiS observations 
and the ability to correct the data for line-of-sight confusion.
We took this limitation into account by considering only measurements within closed column density contours in our analysis.
As a result, the number of clouds contributing to the average SFE value in Eq.~(6) 
is also strongly reduced for column densities $\la 10^{22}\, \rm cm^{-2}$. 
This column density limit is also reflected in the peak of the global column density PDF obtained 
when combining all CAFFEINE regions together (Fig.~\ref{fig:PDF-example}). 

In summary, we conclude that the YSO SFE estimates are reliable in the range of column densities between $\sim$\,$10^{22}~\rm cm^{-2}$ and $\ga$\,$7.5 \times 10^{22}~\rm cm^{-2}$
and the clump SFE estimates are reliable from $\sim$\,$10^{22}~\rm cm^{-2}$ to $\sim$\,$10^{23}~\rm cm^{-2}$.
Both methods indicate a similar trend between SFE and column density in this column density range.
The weighted mean of the two types of SFE estimates 
is presented as the red line in Fig.~\ref{fig:SFE-colden_CAF}.
For reliable SFE estimates at column densities lower than $10^{22}~\rm cm^{-2}$, we 
analyze the results of {\it Spitzer} YSO surveys in 
nearby molecular clouds in the following section.

\subsection{Star formation rates and efficiencies in nearby clouds}
\label{sec:SFR_nearby}
For a more complete picture of the SFE over a wide range of column densities, we also derived estimates 
for nearby ($d <500~\rm pc$) 
clouds from the Gould Belt (see Table \ref{tab:GB_clouds}). 
Extensive {\it Spitzer}-based catalogs of Class~I and Class~II YSOs are available for these clouds, which are complete down to the brown dwarf mass limit of 0.08\,$M_\odot$ 
out to  $d \la 500~\rm pc$ (for Class~Is) and $d \la 300~\rm pc$  (for Class~IIs) \citep[cf.][]{Evans2009}. 
Furthermore, nearby clouds can be traced down to lower column densities, but have only small amounts of gas at high column densities \citep[i.e. see][]{Pokhrel2021}.  
For the analysis of these clouds, 
we followed a very similar approach to the method employed with CAFFEINE data (Sect.~5.1) 
but we concentrated on lower column densities down to 
$\rm 2.5\times 10^{21}~cm^{-2}$,  
again with steps of 0.15\,dex. 
We estimated the cloud masses in various column density bins using the $18.2\arcsec$-resolution column density maps 
from the {\it Herschel} Gould Belt survey \citep[HGBS --][]{Andre2010}. 
The SFRs were derived from the YSO catalogs of \cite{Megeath2012, Dunham2015, Lada2017}. 
The references for the adopted maps and catalogs for each molecular cloud can be found in Table~\ref{tab:GB_clouds}.

\begin{table}
\caption{Nearby clouds considered for comparison with the CAFFEINE sample}
\label{tab:GB_clouds}
\begin{tabular}{llrrc}
\hline \hline
Cloud & dist & RA & DEC & References\tablefootmark{a} \\
 & kpc & $\degree$ & $\degree$ & YSOs, $N_{\rm H_2}$ map\\
\hline
Aquila           & 0.44  & 277.42 &  -2.80 & 1, 4, 5, 6  \\
Ophiuchus        & 0.13 & 246.88 & -24.21 & 1, 6, 7  \\
Orion B          & 0.42  &  86.90 &   0.10 & 2, 6, 8  \\
Lupus I          & 0.15  & 235.26 & -34.09 & 1, 6, 9  \\
Lupus III        & 0.20  & 242.51 & -39.08 & 1, 6, 9  \\
Lupus IV         & 0.15  & 241.15 & -42.07 & 1, 6, 9  \\
Corona Australis & 0.13  & 287.15 & -37.12 & 1, 6, 10 \\
Serpens          & 0.48 & 278.75 &   0.00 & 1, 4, 6, 11 \\
Perseus          & 0.25  &  53.92 &  31.53 & 1, 6, 12 \\
Chamaeleon I     & 0.15  & 163.92 & -77.12 & 1, 6 \\
California       & 0.47  &  65.00 &  37.69 & 3, 13 \\
\hline
\end{tabular}
\tablefoot{
\tablefoottext{a}{References for the adopted distances, lists of YSOs and column density maps: (1) \cite{Dunham2015}; (2) \cite{Megeath2012}; (3) \cite{Lada2017}; (4) \cite{Ortiz-Leon2023}; (5) \cite{Konyves2015}; 
(6) \url{http://gouldbelt-herschel.cea.fr/archives} (HGBS); 
(7) \cite{Ladjelate2020}; (8) \cite{Konyves2020}; (9) \cite{Benedettini2018}; (10) \cite{Bresnahan2018}; (11) \cite{Fiorellino2021}; (12) \cite{Pezzuto2021}; 
(13) \cite[][]{ZhanG2024, Harvey2013}.}
}
\end{table}

Here we also derived two independent estimates of the SFR for each region, one based on Class~II YSOs  and the other based on Class~I YSOs. 
In both cases, we counted the number of YSOs belonging to the relevant class and located within a given column density contour. 
Especially at low $A_V$ (low $N_{\rm H_2}$), a fraction of the YSOs classified as Class~I objects 
on the basis of their SEDs are known to lack a significant envelope of dense gas and are not bona fide protostars \citep{Motte2001,Heiderman2015}. 
As we want to focus on bona fide Class~I protostars, we corrected the observed numbers of Class~I sources in each column density bin using 
the contamination curve derived by \citet{Heiderman2015}. (This contamination is very large $> 90\%$ at $A_V < 4$ but negligible $\la 10\%$ at $A_V > 10$.)
We then multiplied 
the number of YSOs in each class by the same mean stellar mass of $M_\star =\rm 0.5\pm0.1~M_\odot$ as 
adopted by \citet{Evans2009}\footnote{For comparison, the mean stellar mass is $\sim 0.4\, \rm M_\odot $ for the \citet{Kroupa2002} IMF 
in the mass range 0.01--120$\, \rm M_\odot $ and $\sim 0.5\, \rm M_\odot $ for the \citet{Chabrier2005} IMF in the same mass range. 
A recent {\it Gaia} determination of the IMF in the solar neighborhood \citep{Kirkpatrick2024} finds a mean stellar mass of $0.41\, \rm M_\odot $ but is insensitive to massive stars $> 8\, \rm M_\odot$.}
and divided the result by the typical evolutionary timescale of the corresponding YSO phase, i.e., $t_{\star\text{Class II}} = \rm 2.0\pm1.0~Myr$ for Class~II
and $t_{\star\text{,Class I}} = \rm 0.5\pm 0.2~Myr$  for Class~I objects \citep{Evans2009, Dunham2014, Dunham2015}.

As all of the selected Gould Belt clouds are nearby and located off the Galactic Plane, we did not subtract any background. 
In the cases of Lupus~III, Corona Australis and Chameleon~I, 
we used multiresolution column density maps where the HGBS maps were embedded in wider-scale column density maps from {\it Planck} data. 
The use of such multiresolution maps was necessary to ensure closed column density contours  at $2.5\times 10^{21} \rm~cm^{-2}$ and $3.5\times 10^{21} \rm~cm^{-2}$. 
For Aquila and Ophiuchus, these two column density contours were ignored as YSO catalogs are only available over locations of higher column densities \citep{Evans2003, Dunham2015}.

For each cloud, the SFE within each column density contour was derived separately using either the Class~II or Class~I YSO SFRs following Eq.~\ref{equ:SFE_def}. 
For the Class~II-based estimates, the cloud-to-cloud standard deviation of the SFEs is in the range of 0.15 -- 0.25 dex below column densities of $\le 1.4\times 10^{22}\rm ~cm^{-2}$ 
but increases to $\sim$\,0.3 -- 0.6 dex at higher column densities. 
This latter column density is also roughly the limit above which Class~II YSOs are no longer observed in all of the 
clouds. 
On the other hand, Class~I-based estimates show a cloud to cloud standard deviation increasing from $\sim0.25$ dex to 0.6 dex up 
to a column density of $N_{\rm H_2} = 4.0\times 10^{22}\rm ~cm^{-2}$ and a standard deviation of $\sim 0.4$ dex above that.

For the two different methods, we again averaged the results over the whole sample of clouds to identify a more general trend 
of the SFE with column density (see Fig.~\ref{fig:SFE-colden_GB} below). 

For each cloud,  the uncertainty in the SFE estimate was derived by combining 
the Poisson-error associated with the number of identified YSOs within each closed contour, $N_\text{YSO}$, 
the uncertainty in the average stellar mass, $\Delta M_\star$, and the uncertainty in the evolutionary timescale, $\Delta t_\star$. 
The uncertainty in the global SFE averaged over the clouds of Table~\ref{tab:GB_clouds} is thus given by

\begin{equation}
    \frac{{\rm \Delta SFE}(>N_{\rm H_2})}{{\rm SFE}(>N_{\rm H_2})} = \sqrt{\frac{\sum_i^{N_\text{reg}} \left(\left(\frac{w_i^2}{N_{i,\rm\,YSO}(>N_{H_2})}\right)\right)}{\left(\sum_i^{N_\text{reg}} w_i\right)^2}+\left(\frac{\Delta M_\star}{M_\star} \right)^2+\left(\frac{\Delta t_\star}{t_\star} \right)^2}
,\end{equation}

\noindent
where $N_\text{reg}$ is the number of clouds with valid SFE estimates at the given $N_{\rm H_2}$ contour. 
As the evolutionary timescale of Class~I YSOs is based on the lifetime of Class~II YSOs, this dependency needs to be reflected in the uncertainties. 
We adopted relative uncertainties of $(\frac{\Delta t_{\star}}{t_\star})_{\rm Class\,II}=50\%$ and $(\frac{\Delta t_{\star}}{t_\star})_{\rm Class\,I}=70\%$.

For both types of estimates, the derived SFE$(>N_{\rm H_2})$ tends to increase with column density up 
to $N_{\rm H_2} = 1.4\times 10^{22}\rm ~cm^{-2}$ (cf. Fig.~\ref{fig:SFE-colden_GB}).
This increase is slightly more pronounced in the Class~I-based estimates.
Above $N_{\rm H_2} = 1.4\times 10^{22}\rm ~cm^{-2}$, however, 
the SFE estimates based on Class~II YSOs start to decline, while the Class I SFEs exhibit a near power-law increase for column densities 
between $5.0 \times 10^{21}\rm ~cm^{-2}$ and $5.6 \times 10^{22}\rm ~cm^{-2}$ with a slope of about $0.91\pm 0.06$, reaching a plateau at  higher column densities. 
This significant difference is discussed in more detail in Sect.~\ref{sec:Discussion_SFE_GB} below.

\section{Discussion}
\label{sec:Discussion}
In Sect.~\ref{sec:Results}, 
we used two independent methods to estimate SFRs in a wide range of column densities 
for a sample of 49 clouds observed as part of the CAFFEINE survey with ArT\'eMiS. 
These methods are YSO distribution fitting (Section \ref{sec:SFR-CAF-YSO}) and Hi-GAL protostellar clump modeling \citep[Section \ref{sec:Meth_clump_relation}][]{Elia2022}. 
Our SFR estimates 
enable us to probe the SFE at higher column densities than previous works \citep[e.g.,][]{Evans2014, Pokhrel2021}, but are limited at low column densities. 
Therefore, we also derived SFE estimates for a sample of nearby clouds using Class I and II YSO counting based on  published  {\it Spitzer} YSO catalogs and {\it Herschel} column density maps (see Sect.~\ref{sec:SFR_nearby}).
Combining the results obtained for CAFFEINE clouds and nearby clouds provides a more comprehensive picture of how the SFE varies as function of gas surface density. 
Here, we first put our observational results  in the context of two simple theoretical models for the SFE-density relation. 
We then compare the two models with the observations and discuss the limitations of the observational methods.

\subsection{Simple models for the star formation efficiency}
\label{sec:models}

As briefly mentioned in Sect.~1, there are two leading simple models for the star formation efficiency ${\rm SFE} \equiv {\rm SFR}/M_{\rm gas}$ in molecular gas. 
In the first one, the SFE per free-fall time, $\epsilon_{\rm ff} \equiv ({\rm SFR}/M_{\rm gas}) \times t_{\rm ff}$, 
is constant $\sim$\,1--2\%, independent of cloud density \citep[][]{Krumholz2007,Krumholz2014}, and therefore SFE increases with gas density, since $t_{\rm ff} \propto \rho^{-0.5}$. 
We refer to this first model as the $\epsilon_{\rm ff}$ scenario. 
In the second model, SFE quickly becomes negligible below some threshold or transition
density $n_{\rm H_2}^{\rm crit} \sim 2 \times 10^4\, {\rm cm}^{-3} $  
and is roughly constant $\sim 5 \times 10^{-8}\, {\rm yr^{-1}} $ above the 
threshold 
density \citep[][]{Lada2010,Evans2014}. 
There is a natural interpretation of the above empirical threshold density in the context of the filament paradigm of SF 
supported by {\it Herschel} results in nearby molecular clouds \citep[][]{Andre2014}. 
In this observationally-driven paradigm, the threshold density corresponds to the typical volume density of $\sim$0.1-pc-wide transcritical filaments 
with masses per unit length within a factor of 2 of the thermally critical mass per unit length $M_{\rm line, crit} = 2\, c_{\rm s}^2/G \sim 16\, M_\odot/{\rm pc}$ of 
nearly isothermal molecular gas cylinders at a temperature $T \sim 10\,$K (i.e., a sound speed $c_{\rm s} \sim 0.2\,$km/s). 
In the present paper, we therefore refer to the second model as the filament or transition scenario.
In the following, we derive the predictions of these two models for the dependence of the 
cumulative star formation efficiency ${\rm SFE}(>N_{\rm H_2})$ [cf. Eq.~(3)] on column density $N_{\rm H_2}$ or 
equivalently molecular gas surface density $\Sigma_{\rm H_2} = \mu_{\rm H_2}\,m_{\rm p}\, N_{\rm H_2}$. 

We can relate ${\rm SFE}(>N_{\rm H_2})$ to the differential star formation efficiency, ${\rm SFE}(\Sigma_{\rm H_2}) \equiv \Sigma_{\rm SFR}/\Sigma_{\rm H_2} $, 
and the column density PDF in the cloud~$\propto dA/d{\rm ln}\,\Sigma_{\rm H_2} $, where $\Sigma_{\rm SFR}$ is the SFR surface density 
and $dA$ is the surface area of the cloud in a logarithmic range of gas surface densities $d{\rm ln}\,\Sigma_{\rm H_2} $ around $\Sigma_{\rm H_2} $, as follows. 
${\rm SFE}(>\Sigma_{\rm H_2})$ is the ratio of the cumulative star formation rate ${\rm SFR}(>\Sigma_{\rm H_2})$ 
to the cumulative cloud mass  $M(>\Sigma_{\rm H_2})$ which can be expressed as 
\begin{equation}
\begin{split}
{\rm SFR}(>\Sigma_{\rm H_2}) \,& \,= \int_{\Sigma_{\rm H_2}}^{+\infty}\, \Sigma_{\rm SFR} \, \frac{dA}{d\Sigma_{\rm H_2}^\prime}\, d\Sigma_{\rm H_2}^\prime\\  
&\,= \int_{\Sigma_{\rm H_2}}^{+\infty}\, {\rm SFE}(\Sigma_{\rm H_2}^\prime) \, \frac{dA}{d{\rm ln}\,\Sigma_{\rm H_2}^\prime}\, d\Sigma_{\rm H_2}^\prime   \, {\rm , and}
\end{split}
\end{equation}
\begin{equation}
{\rm M}(>\Sigma_{\rm H_2}) = \int_{\Sigma_{\rm H_2}}^{+\infty}\, \Sigma_{\rm H_2}^\prime \, dA = \int_{\Sigma_{\rm H_2}}^{+\infty}\, \frac{dA}{d{\rm ln}\,\Sigma_{\rm H_2}^\prime}\, d\Sigma_{\rm H_2}^\prime  \, ,
\end{equation}
\noindent
respectively. Therefore, SFE$(>\Sigma_{\rm H_2})$ appears to be a weighted-average version of ${\rm SFE}(\Sigma_{\rm H_2}) $, 
with a weighting corresponding the column density PDF of the cloud.

In the $\epsilon_{\rm ff}$ scenario, the differential SFE depends on the gas volume density $\rho$ as 
${\rm SFE}(\rho) = \epsilon_{\rm ff}/t_{\rm ff}(\rho) $,  
with $t_{\rm ff}(\rho) \propto \rho^{-1/2} $. 
Assuming a centrally condensed, spheroidal molecular cloud with an average radial density profile $\rho(r) \propto r^{-\alpha}$ (typically $1.5 < \alpha < 2$), 
the gas surface density scales as $\Sigma_{\rm H_2} \propto r^{1-{\alpha}}$ with radius and the volume density can thus be expressed in terms of surface density
as $\rho \propto \Sigma_{\rm H_2}^{\frac{\alpha}{\alpha-1}}$.
Moreover, the index of the cloud density profile $\alpha$ can be conveniently related to the slope $s$ 
of the column density PDF (see Sect.~5.1) as $\alpha = (s-2)/s$ (see, e.g., Federrath \& Klessen 2013). 
In the regime of column densities where the PDF is a power law, 
the differential SFE therefore scales as:
\begin{equation}
 {\rm SFE}_{\epsilon_{\rm ff}}(\Sigma_{\rm H_2}) \propto  \epsilon_{\rm ff} \times \Sigma_{\rm H_2}^{\frac{2-s}{4}} \, .
\end{equation}
Provided that $s < -2$, it can also be easily shown using Eqs.~(5), (10), (11), (12)  
that the cumulative star formation efficiency ${\rm SFE}_{\epsilon_{\rm ff}}(>\Sigma_{\rm H_2})$ 
is expected to scale in the same way as 
the differential form  ${\rm SFE}_{\epsilon_{\rm ff}}(\Sigma_{\rm H_2})$ in this picture.

In contrast, in the filament scenario, ${\rm SFE}_{\rm fil}(\Sigma_{\rm H_2}) \approx 0$ for \hbox{$\Sigma_{\rm H_2} < \Sigma_{\rm th}$}
and ${\rm SFE}_{\rm fil}(\Sigma_{\rm H_2}) \approx {\rm SFE_{max}}$ 
for $\Sigma_{\rm H_2} > \Sigma_{\rm th}$, 
where $\Sigma_{\rm th} \approx M_{\rm line, crit}/W_{\rm fil} \sim 160\, M_\odot\, {\rm pc}^{-2} $ is 
the typical surface density of thermally transcritical filaments with width $W_{\rm fil} \sim 0.1\,$pc \citep[][]{Andre2014,Arzoumanian2011,Arzoumanian2019}, 
\hbox{${\rm SFE_{max}} = {\rm CFE_{max}} \times \epsilon_{\rm core} / t_{\rm pre\star} \sim  5 \times 10^{-8}\, {\rm yr^{-1}} $} 
the typical SFE expected in thermally supercritical filaments \citep[e.g.,][]{Shimajiri2017},  
${\rm CFE_{max}} \sim 20\% $ the typical core formation efficacy in thermally supercritical filaments \citep[e.g.,][]{Konyves2015},
$\epsilon_{\rm core} \sim 30\% $ the typical conversion efficacy from core mass to stellar mass \citep[e.g.,][]{Alves2007}, 
and $t_{\rm pre\star} \sim 1.2\,$Myr the typical lifetime of prestellar cores \citep{Ward-Thompson2007,Konyves2015}.\\
Using again Eqs.~(5), (10), (11), we see that 
\begin{equation}
{\rm SFE_{fil}}(>\Sigma_{\rm H_2})  \approx {\rm SFE_{max}} \ \, {\rm for}\ \Sigma_{\rm H_2} > \Sigma_{\rm th} \,, 
\end{equation}
while
\begin{equation}
\begin{split}
{\rm SFE}_{\rm fil}(>\Sigma_{\rm H_2}) \,& \,\approx {\rm SFE_{max}} \times 
\int_{\Sigma_{\rm th}}^{+\infty}\, \frac{dA}{d{\rm ln}\,\Sigma}\, d\Sigma \ / \int_{\Sigma_{\rm H_2}}^{+\infty} \, \frac{dA}{d{\rm ln}\,\Sigma} \, d\Sigma\\
&\,\approx {\rm SFE_{max}} \times (\Sigma_{\rm H_2}/\Sigma_{\rm th})^{-s-1} 
\end{split}
\end{equation}
\noindent
for $\Sigma_{\rm H_2} < \Sigma_{\rm th}$.\\ 
In reality, the differential star formation efficiency ${\rm SFE}_{\rm fil}(\Sigma_{\rm H_2}) $
is not a true step function but a ``smooth step function'', reflecting the form of the differential prestellar core formation efficacy 
in molecular filaments \citep[cf.][]{Konyves2015,Andre2019}, which we take to be 
\hbox{${\rm CFE}(M_{\rm line})  = {\rm CFE}_{\rm max} \times \left[1\, -\, {\rm exp}\, \left(\frac{1}{2} - \frac{2\,M_{\rm line}}{3\,M_{\rm line, crit}}\right) \right]$} here.
This leads to a smooth transition between the two regimes expressed by Eqs.~(13) and (14)
for ${\rm SFE_{fil}}(>\Sigma_{\rm H_2})$.

The predictions of the above two models for the SFE are compared with observations in Sect.~\ref{sec:global_picture}
and Figs.~\ref{fig:SFE-colden} \& \ref{fig:diff_SFE_GB} below. 
It should be noted here that, for typical N-PDF slopes such as the median slope $s \approx -2.3$ observed for the CAFFEINE clouds (see Sect.~\ref{sec:CAFFEINE_obs}), 
the two models behave very similarly at low column densities (below $ \Sigma_{\rm th}$) in terms of their cumulative SFE$(>\Sigma_{\rm H_2})$, 
scaling as power laws $\propto \Sigma_{\rm H_2}^{1.1}$ and $\propto \Sigma_{\rm H_2}^{1.3}$ according to Eqs.~(12) and (14), respectively.
In the regime of column densities around $ \Sigma_{\rm th}$, the two models are thus easier to discriminate through their predictions for the differential ${\rm SFE}(\Sigma_{\rm H_2})$.

\subsection{Global picture of the star formation efficiency}
\label{sec:global_picture}

\begin{figure}
    \centering
    \includegraphics[width=0.49 \textwidth]{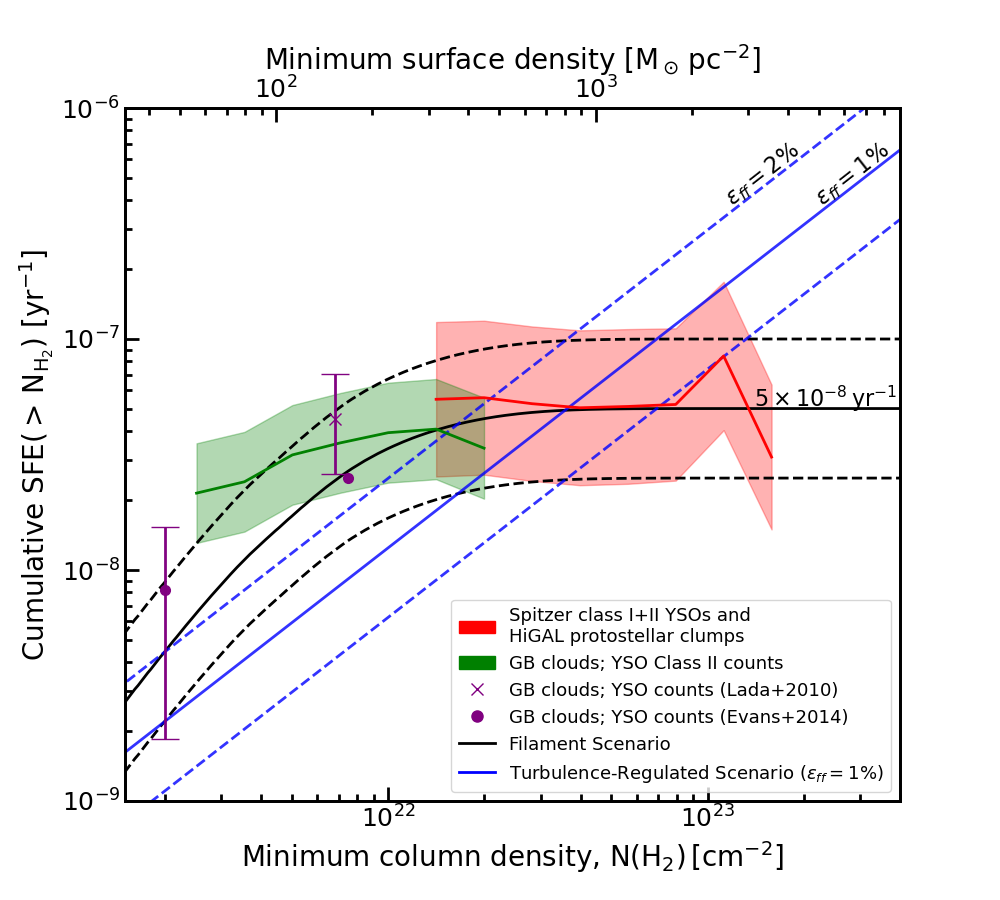}
    \caption{Average SFE above a minimum column density contour as a function of column density 
    for the selected CAFFEINE (red) and nearby clouds (green). The shaded areas mark the corresponding  uncertainties (cf. Eq.~(8) in Sect.~\ref{sec:CAFFEINE_obs}).
    For the CAFFEINE sample, the red line shows the weighted average of the two SFE estimates. 
    The purple dots indicate averages of former works, where the bars show the interquartile range of each cloud sample. For comparison, the blue and black curves 
    represent the two simple models for the SFE ($\epsilon_{\rm ff}$ and filament scenarios -- see Sect.~\ref{sec:models}) and a factor 2 uncertainty. 
    See Figs.~\ref{fig:SFE-colden_CAF} and \ref{fig:SFE-colden_GB} for more details on the observed SFEs.}
    \label{fig:SFE-colden}
\end{figure}

In Sect.~\ref{sec:Results}, 
we derived cumulative SFE estimates as a function of column density  
for both the CAFFEINE clouds and nearby clouds, to cover the widest possible range of column densities. 
As cumulative SFE estimates for nearby clouds based on Class\,I YSOs may be misleading  
below $\sim$$10^{22} \rm ~cm^{-2}$ 
(see Sect.~\ref{sec:Discussion_SFE_GB} below),  
we concentrate on Class\,II-based SFE estimates in this section, but discuss the differences with Class\,I-based SFE estimates 
in the next subsection. 
A combined plot of SFE against column density for the CAFFEINE clouds and the sample of nearby clouds in Table \ref{tab:GB_clouds}  
is shown in Fig.~\ref{fig:SFE-colden}, along 
with additional published estimates for nearby clouds at two column density levels ($A_V \sim 2$~mag and $A_V \sim 8$~mag) from  \citet{Lada2010} and \citet{Evans2014}.
The predictions of the $\epsilon_{\rm ff}$ \citep{Krumholz2014} and filament \citep{Andre2014} scenarios
described in Sect.~\ref{sec:models} are also overlaid as solid and dashed lines. 

The SFE estimates by \cite{Lada2010} and \cite{Evans2014} were based on SFRs derived from Class II YSO counting similar to this paper, but used cloud masses obtained 
from extended dust extinction maps as opposed to {\it Herschel} column density maps. 
Additionally, in each cloud, \cite{Lada2010} counted all YSOs within the whole cloud area, i.e., the observed area corresponding to $A_V \ge 2 \rm ~mag$ ($N_{\rm H_2} \ge 2.0 \times 10^{21}\rm ~cm^{-2}$).
This difference leads to slightly higher SFEs at high $A_V$ than our present estimates (as we count only YSOs observed within a given $A_V$ contour to estimate the SFE above that $A_V$ value).
The cloud-to-cloud standard deviation is, however, comparable to our findings. \cite{Evans2014} provide average SFE estimates at column densities 
of $N_{\rm H_2} = 2.0\times 10^{21}\rm ~cm^{-2}$ and $N_{\rm H_2} = 7.5\times 10^{21}\rm ~cm^{-2}$, 
but quote SFR estimates for each cloud only at the lower column density. 
At low column densities, the cloud-to-cloud dispersion of SFE values is significantly higher, as pointed out by \cite{Lada2010}. 
This higher dispersion implies that the mean SFE value averaged over all clouds depends more strongly on the cloud sample at low column densities than at high column densities. 
Therefore, the slight disagreement between our SFE estimates and those of \cite{Evans2014} at $A_V \sim 2$~mag may be partly due to differing cloud samples.

Comparing the CAFFEINE and nearby-cloud measurements with the two theoretical models described in Sect.~\ref{sec:models}, 
we find significantly better agreement with the filament scenario. 
In particular, 
the SFE traced by Class~II YSOs in nearby clouds shows a significant increase with column density from $N_{\rm H_2} \sim 2\times 10^{21}\rm ~cm^{-2}$ 
to $N_{\rm H_2} \sim 10^{22}\rm ~cm^{-2}$ and subsequently levels off, following the predictions of the filament scenario rather closely. 
Class~II YSOs, however, may be in the process of migrating away from their parent molecular cloud 
(see Sect.~\ref{sec:Discussion_SFE_GB} below for discussion), 
and Class~II YSO counting may thus not directly reflect 
the instantaneous SFR within a cloud. 
As it is not possible to correct for this migration effect without a detailed analysis of YSO dynamics (which is out of the scope of the present paper), 
Class\,II-based SFEs may be underestimated at the high end of the column density range presented in Fig.~\ref{fig:SFE-colden}.
Therefore, the turbulence-regulated $\epsilon_{\rm ff}$ scenario may also account for the measurements obtained in nearby clouds if a relatively high value 
of the SFE per free-fall time is adopted, $\epsilon_{\rm ff} \approx 2\%$ -- $4\%$.

The difference between the two models is more pronounced at higher column densities. 
This shows the importance of the present CAFFEINE results, which probe the behavior of the SFE in a regime of column densities that is 
virtually unconstrained by
nearby cloud data. 
The relatively flat behavior of the observed cumulative SFE curve at a value of about 
$5.4\pm 1.5 \times 10^{-8}~\rm yr^{-1}$ for $N_{\rm H_2} \ga 10^{22}\rm ~cm^{-2}$ 
is in close agreement with 
the predictions of the filament scenario of SF (Fig.~\ref{fig:SFE-colden}).

Thus, while SFE estimates in nearby clouds are not able to disprove either of the two models, 
the CAFFEINE data clearly favor the filament scenario. 
Even if observational uncertainties in SFR estimates are larger for the CAFFEINE clouds due to their larger distances, 
the similarity of the SFE estimates 
we obtained in Sect.~\ref{sec:Results} using two independent methods (Fig.~\ref{fig:SFE-colden_CAF})  
strengthens the result. 
Here, the high spatial resolution and large dynamic range of the combined \artemis \,and \textit{Herschel} data have been instrumental in 
allowing us to probe the SFE in molecular clouds at higher column densities than is possible in nearby clouds.

In this study, we concentrated on the SFE averaged over several molecular clouds. 
But we also showed in Figs.~\ref{fig:SFE-colden_CAF} and \ref{fig:SFE-colden_GB} that there is a significant spread of SFE values between the clouds. 
An obvious caveat of observational studies is the limitation to surface densities compared to volume densities that are more closely related to SF. 
The relation between column density and volume density is strongly dependent on the 3D geometrical structure 
of the clouds \citep{Hu2022}, e.g., spherical or filamentary. 
In the filament scenario of SF, the dense gas is assumed to be primarily structured in the form 
of filaments with typical half-power widths $\sim$\,0.1\,pc, while the low-density gas in molecular clouds may 
be more accurately described as spheroidal on average.

\subsection{Caveat: Difference between Class I and Class II YSO SFE estimates in nearby clouds}
\label{sec:Discussion_SFE_GB}

\begin{figure}
    \centering
    \includegraphics[width=0.49\textwidth]{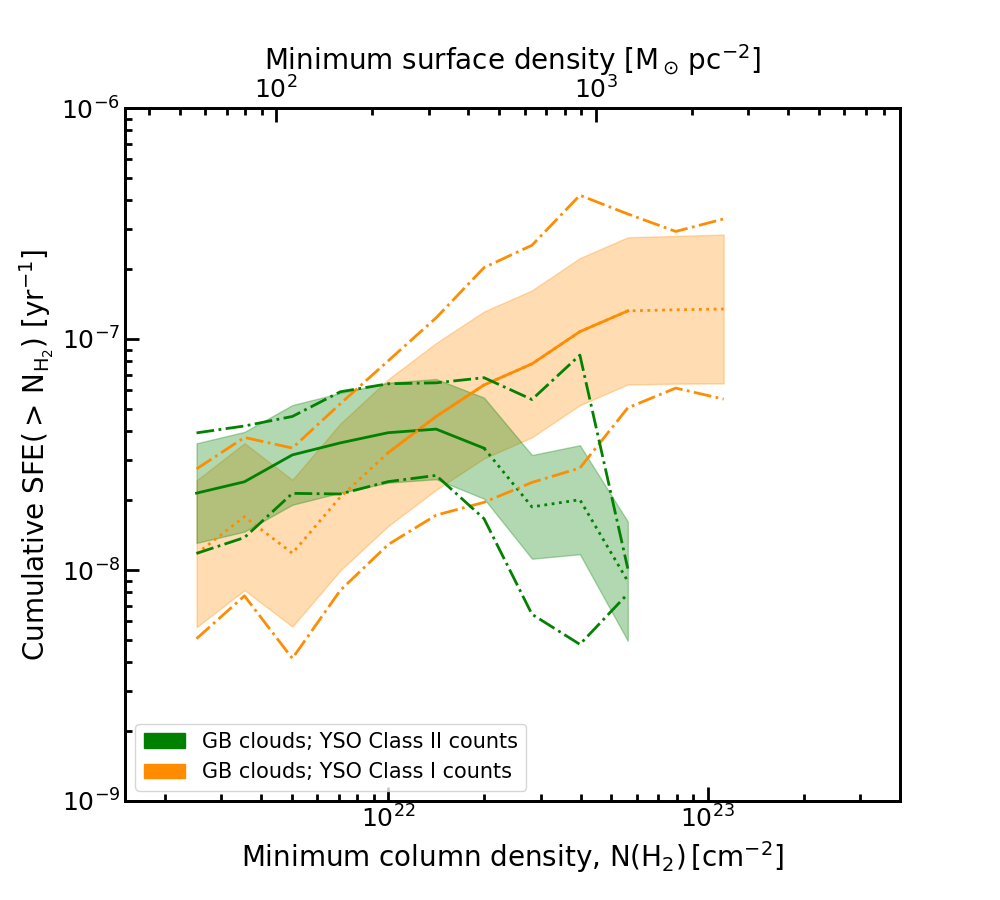}
    \caption{Cumulative SFE above a minimum column density contour as a function of column density for nearby clouds. The orange and green curves indicate the weighted averages over all significant cloud estimates with SFRs derived 
    from Class~I and Class~II YSO counting, respectively. The shaded areas indicate the corresponding statistical uncertainties (cf. Eq.~(9) in Sect.~\ref{sec:SFR_nearby}). The dashed-dotted lines mark the standard deviations of the sample of cloud SFE estimates above each $N_{\rm H_2}$.}
    \label{fig:SFE-colden_GB}
\end{figure}

The two sets of SFE estimates we derived in Sect.~\ref{sec:SFR_nearby} for the distance-limited sample of nearby clouds in Table~\ref{tab:GB_clouds},  
using Class\,II- and Class\,I-YSO counting, respectively, deviate at high column densities (Fig.~\ref{fig:SFE-colden_GB}). 
Comparing both estimates with previous studies, we find that, taken at face value, our Class\,II-based estimates are consistent with the results of \cite{Lada2010} and \cite{Evans2014}, 
while the near power-law increase of the Class I SFE estimates is reminiscent of 
the results presented by \cite{Pokhrel2021}.

\begin{figure*}
    \centering
    \begin{minipage}{0.49\textwidth}
    \includegraphics[width=\textwidth]{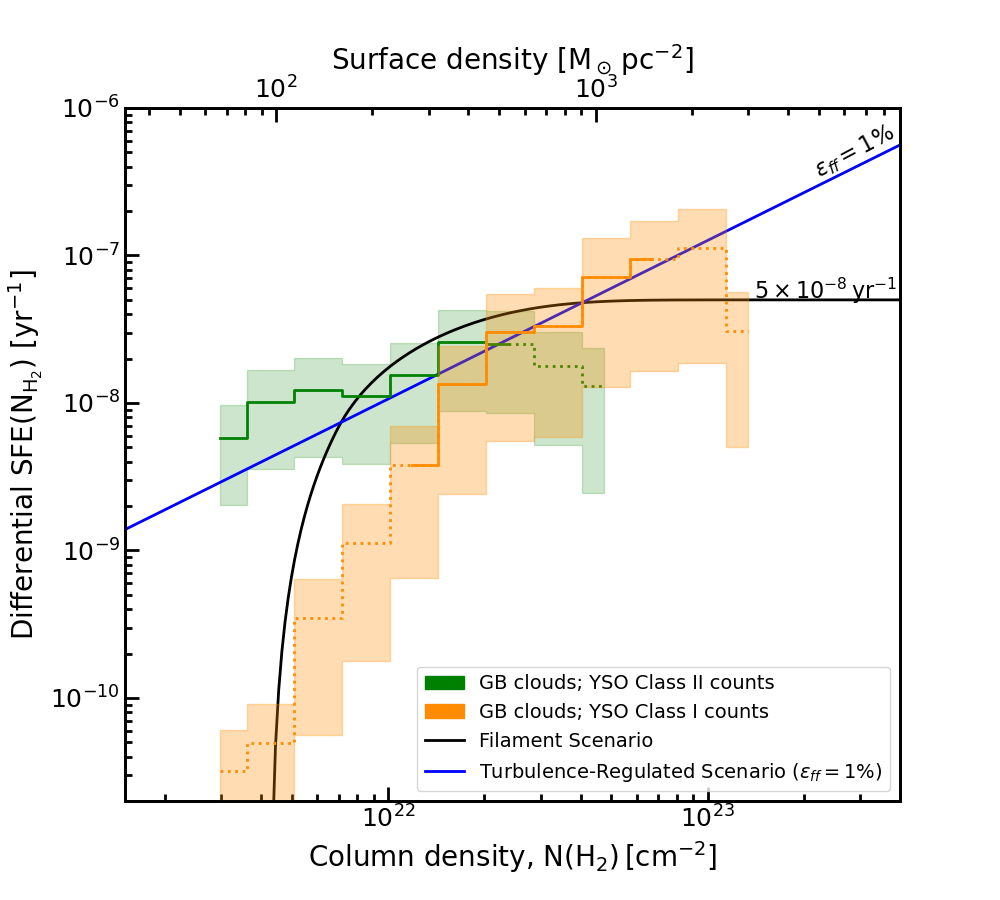}
    \end{minipage}
    \begin{minipage}{0.49\textwidth}
    \includegraphics[width=\textwidth]{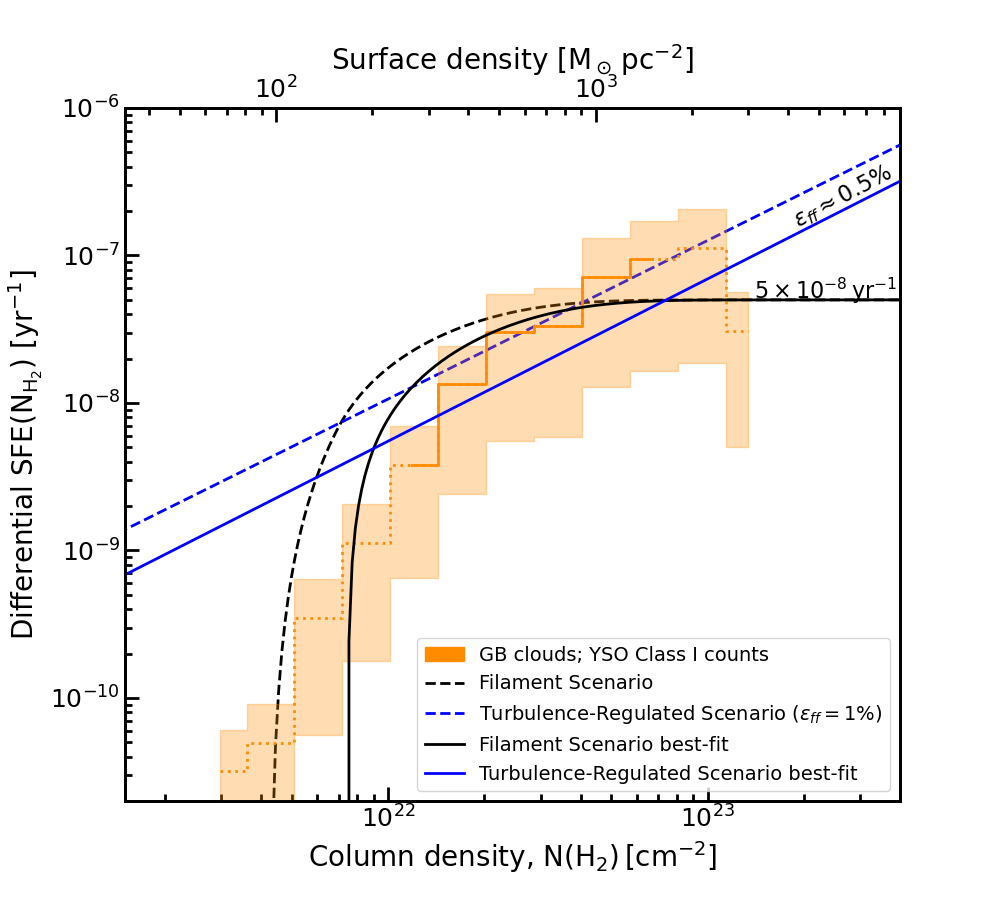}
    \end{minipage}
    \caption{Differential SFE per column density bin from 
    YSO-counting in nearby clouds vs. column density compared to both the filament or transition model (black curve) 
    and the $\epsilon_\text{ff}$ model (blue line).
    \textbf{Left:} Comparison of both Class\,I-based (orange shading) and  Class\,II-based (green shading) SFE estimates with the nominal models discussed in Sect.~\ref{sec:models}.
    \textbf{Right:} Comparison of the Class\,I-based SFE estimates (orange shading) with the ``best-fit'' transition and $\epsilon_\text{ff}$ models (black and blue solid lines, see text). 
    The nominal models shown in the left panel are displayed as dashed lines. 
    }
    \label{fig:diff_SFE_GB}
\end{figure*}

To investigate the origin of this difference between Class\,II- and Class\,I-based SFE estimates, 
we first look at the distributions of the numbers of Class~II and Class~I YSOs with increasing column density. 
We find that Class~I YSOs are 
more strongly concentrated toward the dense areas than Class~II YSOs (Fig.~\ref{fig:sources/area_HGBS}).
Despite the slightly different distance limits used in Sect.~\ref{sec:SFR_nearby} for the cloud samples considered for the Class\,II- and Class\,I-based SFE estimates, 
the observed difference in the spatial distributions of these two classes of 
YSOs seems to be a robust, general trend \citep[see, e.g.,][]{Salji2015a, Gupta2022}. 
This is likely due to the dynamics of young stellar populations. 
YSOs have been observed to decouple progressively from the dense gas environment in which they were born and to migrate away from their parent molecular cloud. 
As a result, we expect the younger Class~I YSOs to be located closer to their birth locations than the older Class~II objects.  
In turn, the estimated SFEs per column density bin may also be expected to exhibit a significant difference depending 
on which class of YSOs is used to trace SF.  

In Fig.~\ref{fig:diff_SFE_GB}, we show how the differential SFE varies with column density in nearby clouds based on Class\,II and Class\,I SFR estimates. 
Given the low number statistics per column density bin, we here treat the whole sample of nearby clouds 
as a single cloud, unlike the geometrical averaging we performed in Sect.~\ref{sec:SFR_nearby} 
for the cumulative SFE vs. column density plots shown in Figs.~\ref{fig:SFE-colden} and \ref{fig:SFE-colden_GB}.
At low column densities, a steep rise of the differential SFE derived from Class\,I YSOs is observed 
from a value of only $3 \pm 3 \times 10^{-11}\rm ~yr^{-1}$ at $N_{\rm H_2} \sim 3 \times 10^{21}\rm ~cm^{-2}$, 
jumping by more than two orders of magnitude to 0.5--1.5\ $ \times 10^{-8}\rm ~yr^{-1}$  at $N_{\rm H_2} \sim 1.4 \times 10^{22}\rm ~cm^{-2}$. 
This jump reflects the very low number of Class~I YSOs found at  $N_{\rm H_2} < 10^{22}\rm ~cm^{-2}$ 
and the fact that most protostars are formed in the densest parts of the clouds. 
At higher column densities, a much shallower increase of the differential SFE based on Class\,I YSOs
is seen, from  $3 \pm 2 \times 10^{-8}\rm ~yr^{-1}$ at $N_{\rm H_2} \sim 2 \times 10^{22}\rm ~cm^{-2}$ 
to $9 \pm 7 \times 10^{-8}\rm ~yr^{-1}$ at $N_{\rm H_2} \sim 7 \times 10^{22}\rm ~cm^{-2}$.
The very low number of Class~I YSOs observed below $10^{22}\rm ~cm^{-2}$ (see Fig.~\ref{fig:sources/area_HGBS}) implies 
that the near power-law increase seen in the cumulative SFE versus column density curve SFE$(>\Sigma_{\rm H_2})$ derived from Class~I YSOs 
(Fig.~\ref{fig:SFE-colden_GB}) is primarily driven by the marked decline in cloud mass 
with gas density imposed by the column density PDF, e.g., $M(>N_{\rm H_2}) \propto N_{\rm H_2}^{s+1}$ 
with $s+1 \approx -1.3$ (see Eq.~[11] and Fig.~\ref{fig:PDF-example}), 
at least in the low column density regime. 
This conclusion is further supported by Fig.~1c of \cite{Pokhrel2021}. 
Accordingly, as mentioned earlier, the cumulative SFE$(>\Sigma_{\rm H_2})$ curve is not a good discriminator of the two models of Sect.~\ref{sec:models} 
at low column densities because both models predict very similar power laws in this regime (cf. Eqs.~[12] and [14]).
Figure~\ref{fig:diff_SFE_GB} also suggests that the Class\,I-based differential SFE levels off
at column densities of $\ga 5\times 10^{22}\rm ~cm^{-2}$, albeit with lower statistical significance. 

The average differential SFE derived from Class\,II-based SFR estimates 
is significantly higher ($\sim 1\pm0.5\times 10^{-8}\rm ~yr^{-1}$) than its Class\,I counterpart 
in the low column density regime ($3\times 10^{21} \la N_{\rm H_2} \la 10^{22}\rm ~cm^{-2}$).
Above $N_{\rm H_2} \ga 10^{22}\rm ~cm^{-2}$, 
the Class\,II-based differential SFE increases up to $2.3 \times 10^{-8}\rm ~yr^{-1}$ at a column density of $\sim 2 \times 10^{22}\rm ~cm^{-2}$, 
but becomes less reliable at higher column densities as the total number of Class~II YSOs observed in the clouds of Table~\ref{tab:GB_clouds} 
drops to low values ($<~50$, see Fig.~\ref{fig:sources/area_HGBS}).

Above $N_{\rm H_2} \ga 2 \times 10^{22}\rm ~cm^{-2}$, both the differential and the cumulative SFE 
derived from Class~I YSOs  exceed their Class~II counterparts by a factor $\sim$\,2--3. 
This may be due to two effects. First, as already pointed out, Class~II objects may have significantly drifted away
from their dense birthplaces. Second, there is some evidence that SF has 
been accelerating on a typical $e$-folding timescale of $\sim$\,1--3\,Myr in nearby star-forming regions, 
possibly due to overall cloud contraction \citep[e.g.,][]{Palla2000,Watkins2019}. 
Since Class~II YSOs measure SF averaged over a timescale $\sim$\,2\,Myr, 
while Class~I YSOs measure SF integrated over a timescale of only $\sim$\,0.5\,Myr,
one may expect the Class~I SFRs to exceed  the Class~II SFRs  
by a significant factor 
in dense regions undergoing accelerated SF.

Another possible effect that can contribute to the observed difference between Class\,II-based and Class\,I-based estimates of the SFE 
at high column densities is extinction. The circumstellar envelopes of Class\,I YSOs causes bright far-infrared emission 
which is less affected by extinction. 
This can lead to a detection bias in favor of Class~I YSOs 
compared to Class\,II YSOs 
in high column density regions. 
In summary,  
we consider that 
Class\,II-based SFE estimates are 
reliable between $\sim$$3 \times 10^{21}\rm ~cm^{-2}$ 
and $\sim$$2 \times 10^{22}\rm ~cm^{-2}$, 
while our Class\,I-based SFE estimates seem reliable 
up to $\sim$\,3--$6 \times 10^{22}\rm ~cm^{-2}$.

\begin{figure}
    \centering
    \includegraphics[width=0.49\textwidth]{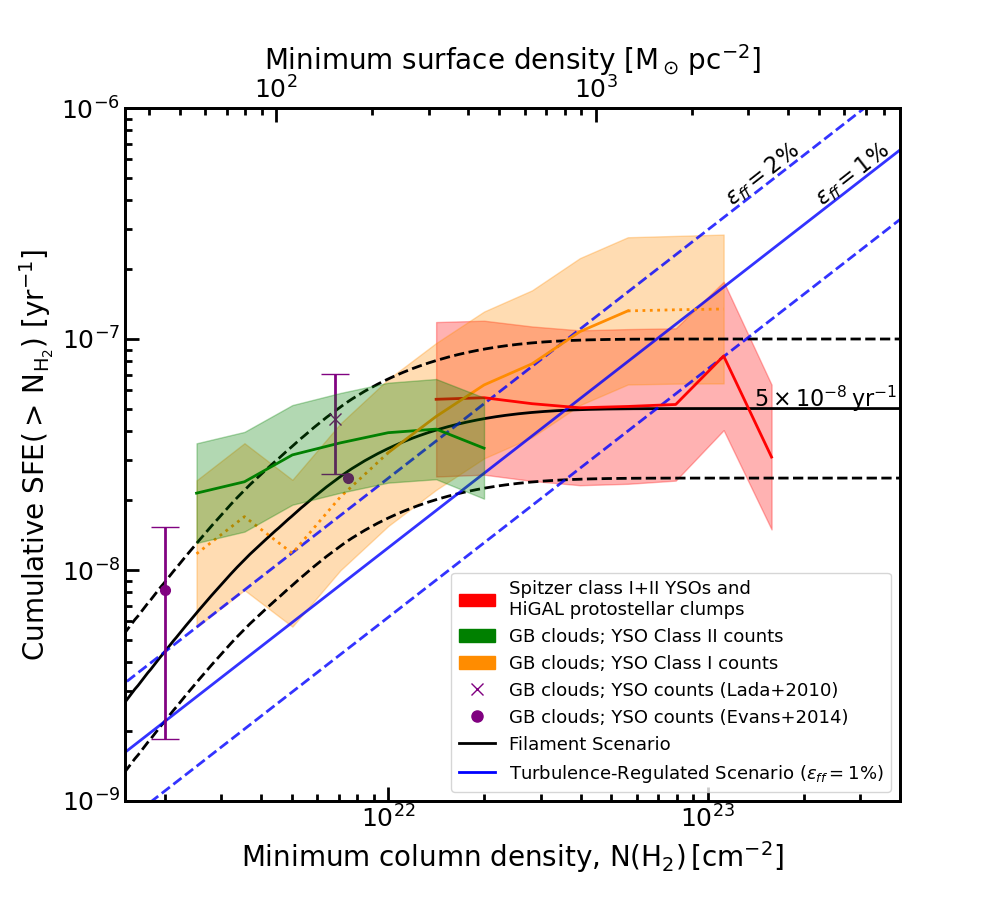}
    \caption{Cumulative average SFE above a minimum column density contour as a function of column density 
    for a selection of CAFFEINE (red) and nearby clouds (green and orange). The shaded areas mark the corresponding  uncertainties. For the CAFFEINE sample, the weighted average of the two SFE estimates are shown. 
    The purple markers indicate averages of former works, where the bars show the interquartile range of the cloud sample. For comparison, the blue and black curves 
    show the two simple models of Sect.~\ref{sec:models} for the SFE ($\epsilon_{\rm ff}$ and filament scenarios). 
    See Figs.~\ref{fig:SFE-colden_CAF} and \ref{fig:SFE-colden_GB} for more details on the observed SFEs.}
    \label{fig:SFE-colden_wCl2}
\end{figure}

Figure~\ref{fig:SFE-colden_wCl2} compares the cumulative SFE traced by Class\,I YSOs in nearby clouds with 
both the CAFFEINE results and the two models introduced in Sect.~\ref{sec:models}. 
Between $N_{\rm H_2} \sim 10^{22}\rm ~cm^{-2}$ and $N_{\rm H_2} \sim 5 \times 10^{22}\rm ~cm^{-2}$, 
our Class\,I-based SFE estimates for nearby clouds remain within the error bars of the filament-scenario predictions but
are suggestive of a near power-law increase in SFE, which is consistent with 
the results of \cite{Pokhrel2021} and the $\epsilon_{\rm ff}$ SF model.
Above $\sim 5 \times 10^{22}\rm ~cm^{-2}$, however, the relation seems to flatten out. 
Again a similar behavior is found by \cite{Pokhrel2021}, but in both studies the number of Class~I YSOs 
at $N_{\rm H_2} > 5 \times 10^{22}\rm ~cm^{-2}$ 
and the total gas mass available at these high column densities 
are insufficient for a robust conclusion. 

Overall, if we confront the two models of Sect.~\ref{sec:models} with the differential form of our Class\,I-based SFE 
estimates for nearby clouds (Fig.~\ref{fig:diff_SFE_GB}), which is a better discriminator at low column densities, 
we find 
better agreement with the transition or filament scenario. 
The normalized chi-squared statistic $\chi^2 = \sum\, \rm \left(\frac{SFE_{model}-SFE_{obs}}{\sigma_{SFE}}\right)^2$ is indeed a factor of $> 3$ lower 
for the nominal filament or transition model 
than for the $\epsilon_{\rm ff}$=1\% model.
Moreover, if we relax both models somewhat and allow the threshold column density to be a free parameter of the transition scenario 
and the characteristic $\epsilon_{\rm ff}$ value to be a free parameter of the other model (see right panel of Fig.~\ref{fig:diff_SFE_GB})), 
we find that the best-fit transition model has $\chi^2_{\rm best, fil} < 10$
while the best-fit $\epsilon_{\rm ff}$ model 
has $\chi^2_{\rm best, \epsilon_{\rm ff}} >  50$ 
(for 10 degrees of freedom in both cases). 
The best-fit transition model, which has a threshold column density of $\sim 7.5 \times 10^{21}\rm ~cm^{-2}$, is thus statistically consistent 
with the data to better than 1$\sigma$ ($p$-value = 0.49). 
In contrast, the best-fit $\epsilon_{\rm ff}$ model, for which $\epsilon_{\rm ff} \approx 0.5\%$, 
can be rejected at a $> 5\sigma$ 
confidence level ($p$-value $<$$< 6\times 10^{-6}$).

Counting Class~I YSOs is probably one of the most direct methods available to estimate the SFR in nearby molecular clouds. 
Class~I YSOs are younger and brighter than Class~II sources in the mid- to far-infrared range, 
which render them easier to detect closer to their birthplaces.
Therefore, we interpret the abrupt increase in the differential SFE (Fig. \ref{fig:diff_SFE_GB}) and in the surface density of Class~I objects (Fig. \ref{fig:sources/area_HGBS}) 
above $N_{\rm H_2} \sim 10^{22}\rm ~cm^{-2}$ as a column density threshold for SF. 
This interpretation is in agreement with the findings of several earlier studies
which suggested threshold values between $0.65 \times 10^{22}\rm ~cm^{-2}$ and $2.2 \times 10^{22}\rm ~cm^{-2}$ 
\citep[e.g.,][]{Onishi2001, Johnstone2004, Andre2010, Heiderman2010, Lada2010, Heiderman2015, Konyves2015}. 
As already mentioned in Sect.~\ref{sec:models}, 
such a SF threshold is quite naturally accounted for in the context of the filament scenario, in which stars form primarily in dense, $\sim$0.1-pc-wide filaments 
with line masses exceeding the thermally critical mass per unit length of isothermal gas cylinders ($\sim 16\, M_\odot/$pc for gas at $\sim$10\,K) \citep{Andre2014}.

Considering all of these comparisons  
together, we conclude 
that SF traced by nearby Class\,I YSOs does not provide decisive evidence 
for one or the other of the two presented scenarios, but tends to support the transition model.
The combination of our CAFFEINE findings with the results obtained 
for nearby clouds 
clearly favors the filament scenario, as can be seen in Figs.~\ref{fig:SFE-colden} and \ref{fig:SFE-colden_wCl2}. 
In the range of column densities between $\sim$$10^{22}\rm ~cm^{-2}$ and $\sim 6 \times 10^{22}\rm ~cm^{-2}$ where Class\,I-based SFE estimates for nearby clouds
overlap with CAFFEINE SFE estimates and suggest slightly different trends, we give more weight to the CAFFEINE results because the CAFFEINE cloud sample 
contains a factor of $\sim$100--200 more dense gas mass in each column density bin than the nearby-cloud sample (cf. Fig.~\ref{fig:mass_cd}). 
The CAFFEINE results are thus statistically more significant.
Our YSO-based SFE estimates for the CAFFEINE sample use both Class\,II and Class\,I YSOs 
and therefore should remain reliable at high column densities, at least up to $N_{\rm H_2} \sim 7.5 \times 10^{22}~\rm cm^{-2}$. 
Furthermore, we also estimate the SFEs of CAFFEINE clouds using Hi-GAL protostellar clumps, which are not limited by extinction 
and remain reliable up to $N_{\rm H_2} \ga 10^{23}~\rm cm^{-2}$ (see Sect.~\ref{sec:Discussion_SFR_CAF}).

Finally, we note that the reliability of observational SFR and SFE estimates and the effect of potential observational biases 
has been investigated in several recent works using numerical simulations or Monte-Carlo models \citep{Grudic2022,Suin2023,Dib2024}. 
The most significant potential bias is that any particular SFE estimate derived from observations is an average value over a finite time 
span $\sim$\,0.5--2\,Myr\footnote{More specifically, our YSO-based and clump-based SFE estimates for CAFFEINE clouds 
correspond to average values over $\sim$2.5\,Myr and $\sim$0.3--1\,Myr, respectively. Class\,II-based and Class\,I-based SFE estimates 
for nearby clouds are average values over $\sim$2\,Myr and $\sim$0.5\,Myr, respectively.}
around a specific cloud evolutionary stage and cannot take into account the time dependence of SF in a cloud. 
For example, over the full 9\,Myr duration of the simulation discussed by \citet{Grudic2022} for a $2\times 10^4\, \rm M_\odot$ GMC, 
the true SFE per free-fall time $\epsilon_{\rm ff}$ exhibits very significant variations, rising from an initial value $\epsilon_{\rm ff}$\,$<$<$\,$0.1\% 
to a peak value of $\sim$10\% at 5--6 Myr, before declining to $\sim$1\% by the end of the simulation. 
While the observational proxy $\epsilon_{\rm ff, YSO}$ calculated by \citet[][see the orange curve in their Fig.~7]{Grudic2022} to mimick an observer's Class\,I-based 
estimate fails to track the detailed variations of $\epsilon_{\rm ff}$ accurately and underestimates the peak $\sim$10\% value by a factor of $\ga 3$, 
it nevertheless follows the general evolution quite well and has an average value over the full duration of the simulation which is close to the true average value. 
It should also be pointed out that \citet{Grudic2022} define the ``true'' SFE and true $\epsilon_{\rm ff}$ 
with respect to the total initial cloud mass and initial free-fall time in the simulation, while we are more concerned with the SFE in dense gas above a limiting column density 
and with the $\epsilon_{\rm ff}$ value with respect to the free-fall time of dense gas in the present work. 
In the context of the filament or transition scenario advocated here, 
the rather steep rise of both SFE and $\epsilon_{\rm ff}$ seen during the first $\sim$5~Myr of the \citet{Grudic2022} simulation is easily interpreted 
as the direct result of a corresponding increase in the mass of dense gas (initially strictly 0) 
due to large-scale turbulent shock compression and gravitational collapse.  
In this paper, we have averaged our measurements over 49 CAFFEINE clouds, irrespective of their evolutionary state, 
as we are not trying to study the time evolution of the SFE. 
The analysis presented in Sect.~3.5 of  \citet{Grudic2022} suggests that our results should describe the general trend between SFE and column density, 
averaged over the whole period of active SF in GMCs, to within a factor of $\sim$2 accuracy, 
even if the typical values of $\sim$1--4\% we infer for $\epsilon_{\rm ff}$ in dense molecular gas may underestimate 
the actual peak value achieved during the entire SF history of a GMC.

\subsection{Role of stellar feedback}

Given the large and diverse sample of molecular clouds, it is also possible to investigate how environmental effects and feedback from high-mass stars  
impact the SFR and SFE of molecular clouds. On one hand, feedback from high-mass stars can promote further SF 
by compressing the surrounding gas, enhancing the dense gas fraction, and weighting the associated $N_{\rm H_2}$--PDF 
toward higher column densities. 
On the other hand, radiative feedback drives H\,{\sc{ii}} regions which disperse molecular gas and therefore tend to inhibit SF in their immediate vicinity. 
Using high-resolution MHD simulations, \citet{Suin2023} and \citet{Chevance2023} have shown that the presence of stellar feedback can indeed strongly influence 
the Schmidt-Kennicutt relation and the typical SFE per free-fall time, mainly through its influence on the cloud's density structure. 
The role of the cloud's column density PDF is easy to understand in the context of the two models described 
in Sect.~\ref{sec:models} since in both models the logarithmic slope $s$ of the PDF partly controls the expected variations 
of the SFE with column density (see Eqs.~[12] and [14]).

\begin{figure}
    \centering
    \includegraphics[width=0.49\textwidth]{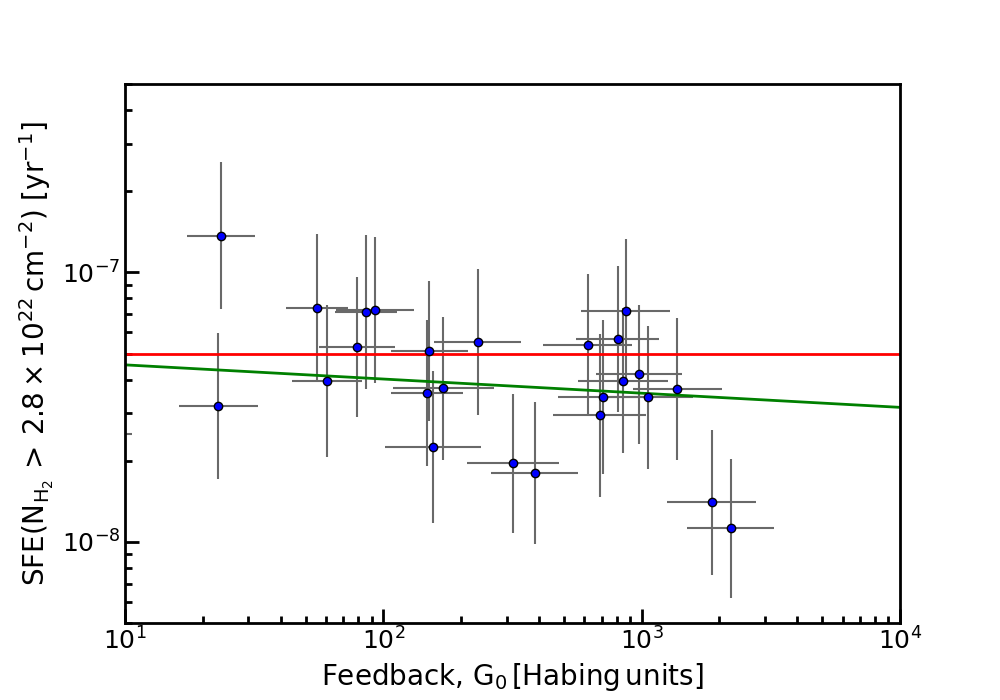}    
    \caption{Cumulative SFE above $N_{\rm H_2} >2.8\times 10^{22} \rm cm^{-2}$ against the 90th percentile of FUV flux over the region observed by CAFFEINE. The red line indicates the SFE predicted by the filament scenario and the green line shows the result of a fitted power-law with a slope of $-0.05\pm0.05$ and a $p$-value of 0.34.}

    \label{fig:feedback}
\end{figure}

Radiative feedback in the CAFFEINE clouds is best traced by the emission of far-UV (FUV) photons, which then heat the surrounding gas. 
We can estimate the flux of the FUV radiation $F_{\rm FUV}$ by
\begin{equation}
    F_{\rm FUV}[G_0] = 4000\, \pi\, I_{\rm FIR}/1.6 \, ,
\end{equation}
where $I_{\rm FIR}$ is the far infrared emission observed by \textit{Herschel}/PACS at $\rm 70~\mu m$ and $\rm 160~ \mu m$ \citep{Habing1968,Kramer2008}. As the regions imaged by CAFFEINE typically cover fields significantly larger 
than the area directly impacted by early feedback from high-mass stars, and over which large variations of $G_0$ are observed, 
we use $G_0^{90\%}$, the 90th percentile of the observed $G_0$ values in each field, as a quantitative estimate of the strength of radiative feedback on dense gas. 
The uncertainty in $G_0^{90\%}$ is derived from the dispersion of $G_0$ estimates over the cloud. 
We plot the SFE estimated above $N_{\rm H_2} >2.8\times 10^{22} \rm cm^{-2}$ 
against $G_0^{90\%}$ in Fig.~\ref{fig:feedback}. 

We find no significant correlation between our SFE and $G_0$ estimates, which is indicated by a Pearson correlation coefficient of only $r=-0.19$. 
We also checked that this finding does not depend on the adopted minimum column density 
and exact percentile in the $G_0$ distribution per cloud: Upon varying these two parameters, 
the derived slope of the best-fit linear correlation (in log space) remained consistent with zero within errors, 
but also slightly negative. 
Therefore, to first order, it appears that radiative feedback has 
little, if any, influence on the SFE 
in dense gas. Its main effect may be in regulating the amount of dense gas in a given region.  
A more dedicated study of the role of feedback in the CAFFEINE clouds 
would nevertheless be needed to draw a firm conclusion on this topic.
Such a study, including a comparison with numerical simulations, is currently planned and will be published in a future paper.


\section{Summary and conclusions}
\label{sec:Conclusion}

We imaged the submillimeter dust continuum emission of 49 dense molecular clouds with APEX-\artemis \,as part of the CAFFEINE survey. 
We then combined the \artemis \,maps with \textit{Herschel} data to create 
high-dynamic-range column density maps of the clouds, which are free of any saturation effect and have a factor of  $\sim$\,4 higher resolution ($8\arcsec$) 
than standard \textit{Herschel} column density maps. This improved resolution is essential to resolve the regions of high column density targeted in our study.
 
After deriving two independent sets of SFR estimates for 
the CAFFEINE clouds, based on {\it Spitzer} YSO distribution fitting and Hi-GAL protostellar clump modeling, respectively, we used these high-quality maps to analyze the relation between SFE and gas (column) density.
For a more complete view of the SFE at low column densities, we also used Class~II- and Class~I-YSO counting in combination 
with column density maps from the {\it Herschel} Gould Belt survey for 11 nearby clouds. 
Comparing the observational data with the 
predictions of the filament (or transition) and $\epsilon_{\rm ff}$ SF scenarios, we draw the following conclusions: 
   \begin{enumerate}
    \item There is broad agreement between 
       the two sets of SFRs we derive for the CAFFEINE molecular clouds (at distances $0.6 \la d \la 3$\,kpc), that is, between those derived
      from fitting a standard stellar IMF to identified {\it Spitzer} YSOs
      and from identified Hi-GAL protostellar clumps with the relation of \cite{Elia2022}, respectively.
    \item Star formation efficiency is found to be essentially independent of column density in the regime of column densities probed by the CAFFEINE cloud sample, 
      namely $10^{22} \la N_{\rm H_2} \la 10^{23}\rm ~cm^{-2}$. 
      At these column densities, the CAFFEINE clouds comprise two orders of magnitude more dense gas mass than the nearby-cloud sample (cf. Fig.~\ref{fig:mass_cd}).  
    \item In nearby molecular clouds, Class~I YSOs 
       are concentrated in areas of high column density ($ N_{\rm H_2} \ga 10^{22}\rm ~cm^{-2}$), suggesting the presence of a column density transition 
       for SF
       at about $N_{\rm H_2} \sim 10^{22}\rm ~cm^{-2}$ (Figs.~\ref{fig:diff_SFE_GB} and \ref{fig:sources/area_HGBS}), as also pointed out by earlier studies. 
       Above this transition column density,  
       the SFE estimates based on nearby Class~I YSOs tend to increase with column density, but become uncertain for $N_{\rm H_2} \ga $\,3--$6 \times 10^{22}\rm ~cm^{-2}$
       due to the small gas masses available in each cloud at these densities (cf. Fig.~\ref{fig:mass_cd}).
    \item Comparing these observational results with the two simple scenarios proposed for the SFE,  
      we find better agreement with the transition picture and the filament scenario  
      (Figs.~\ref{fig:SFE-colden} and ~\ref{fig:diff_SFE_GB}). 
      This conclusion is mainly based on two findings: the presence of an apparent SF threshold at $ N_{\rm H_2} \approx 10^{22}\rm ~cm^{-2}$,
      below which the observed SFE quickly declines (cf. point~3),  and the observation of a nearly constant SFE at higher column densities in the CAFFEINE sample (cf. point~2). 
      A caveat is that available Class\,I-based estimates for nearby clouds are suggestive of a near power-law increase in SFE between 
      $ N_{\rm H_2} \approx 10^{22}\rm ~cm^{-2}$ and $ N_{\rm H_2} \approx  5 \times 10^{22}\rm ~cm^{-2}$, which is reminiscent of the $\epsilon_{\rm ff}$ scenario,  
      while remaining consistent with the transition model within error bars (see Figs.~\ref{fig:diff_SFE_GB} and \ref{fig:SFE-colden_wCl2}).
    \item We do not find any significant correlation between our measured SFEs 
      at high column densities $ N_{\rm H_2} \ga 10^{22}\rm ~cm^{-2}$ and the strength of local radiative feedback quantified by
      the level of FUV-radiation $G_0$ (cf. Fig.~\ref{fig:feedback}). This suggests that feedback has little 
      influence on the SFE in dense gas, even if it may control the amount of dense gas available for SF. 
      A more dedicated study of the role of feedback in the CAFFEINE clouds 
      is needed to confirm this preliminary conclusion.
\end{enumerate}

\begin{acknowledgements}
We thank the referee
for useful comments which helped us strengthen the paper.
We are very thankful for the continuous support provided by the APEX and ESO staff during ArTéMiS operations. M.M. and Ph.~A. acknowledge financial support by CNES and “Ile de France” regional funding (DIM-ACAV+ Program). AZ thanks the support of the Institut Universitaire de France. We also acknowledge support from the French national programs of CNRS/INSU on stellar and ISM physics (PNPS and PCMI). 
\\

\textit{Software:} APLpy \citep{aplpy2019}, astropy \citep{astropy2013}, Matplotlib \citep{matplotlib}, NumPy \citep{numpy}, SciPy \citep{SciPy2020}. This document was prepared using the Overleaf web application, which can be
found at www.overleaf.com.
\end{acknowledgements}

%
%
\bibliographystyle{aa} 
\bibliography{SFE_references} 

\begin{thebibliography}{108}
\expandafter\ifx\csname natexlab\endcsname\relax\def\natexlab#1{#1}\fi

\bibitem[{Alves {et~al.}(2007)Alves, Lombardi, \& Lada}]{Alves2007}
Alves, J., Lombardi, M., \& Lada, C.~J. 2007, \aap, 462, L17

\bibitem[{Alves {et~al.}(2017)Alves, Lombardi, \& Lada}]{Alves2017}
Alves, J., Lombardi, M., \& Lada, C.~J. 2017, \aap, 606, L2

\bibitem[{Andr{\'{e}} {et~al.}(2019)Andr{\'{e}}, Arzoumanian, K{\"{o}}nyves,
  Shimajiri, \& Palmeirim}]{Andre2019}
Andr{\'{e}}, P., Arzoumanian, D., K{\"{o}}nyves, V., Shimajiri, Y., \&
  Palmeirim, P. 2019, \aap, 629, L4

\bibitem[{{Andr{\'e}} {et~al.}(2014){Andr{\'e}}, {Di Francesco},
  {Ward-Thompson}, {Inutsuka}, {Pudritz}, \& {Pineda}}]{Andre2014}
{Andr{\'e}}, P., {Di Francesco}, J., {Ward-Thompson}, D., {et~al.} 2014, in
  Protostars and Planets VI, ed. H.~{Beuther}, R.~S. {Klessen}, C.~P.
  {Dullemond}, \& T.~{Henning}, 27--51

\bibitem[{Andr{\'{e}} {et~al.}(2010)Andr{\'{e}}, Men'shchikov, Bontemps,
  K{\"{o}}nyves, Motte, Schneider, Didelon, Minier, Henning, Royer,
  Mer{\'{i}}n, Vavrek, Attard, Arzoumanian, Wilson, Ade, Aussel, Men, Bontemps,
  K{\"{o}}nyves, Motte, Schneider, Didelon, Minier, Henning, Royer,
  Mer{\'{i}}n, Vavrek, Attard, Arzoumanian, Wilson, Ade, \& Aussel}]{Andre2010}
Andr{\'{e}}, P., Men'shchikov, A., Bontemps, S., {et~al.} 2010, \aap, 518, 1

\bibitem[{Andr{\'{e}} {et~al.}(2016)Andr{\'{e}}, Rev{\'{e}}ret, K{\"{o}}nyves,
  Arzoumanian, Tig{\'{e}}, Gallais, Roussel, {Le Pennec}, Rodriguez, Doumayrou,
  Dubreuil, Lortholary, Martignac, Talvard, Delisle, Visticot, Dumaye, {De
  Breuck}, Shimajiri, Motte, Bontemps, Hennemann, Zavagno, Russeil, Schneider,
  Palmeirim, Peretto, Hill, Minier, Roy, \& Rygl}]{Andre2016}
Andr{\'{e}}, P., Rev{\'{e}}ret, V., K{\"{o}}nyves, V., {et~al.} 2016, \aap,
  592, A54

\bibitem[{Arzoumanian {et~al.}(2011)Arzoumanian, Andr{\'{e}}, Didelon,
  K{\"{o}}nyves, Schneider, Men'shchikov, Sousbie, Zavagno, Bontemps, {Di
  Francesco}, Griffin, Hennemann, Hill, Kirk, Martin, Minier, Molinari, Motte,
  Peretto, Pezzuto, Spinoglio, Ward-Thompson, White, Wilson, Men, Sousbie,
  Zavagno, Motte, Peretto, Pezzuto, Spinoglio, White, Wilson, Men'shchikov,
  Sousbie, Zavagno, Bontemps, {Di Francesco}, Griffin, Hennemann, Hill, Kirk,
  Martin, Minier, Molinari, Motte, Peretto, Pezzuto, Spinoglio, Ward-Thompson,
  White, \& Wilson}]{Arzoumanian2011}
Arzoumanian, D., Andr{\'{e}}, P., Didelon, P., {et~al.} 2011, \aap, 529, L6

\bibitem[{Arzoumanian {et~al.}(2019)Arzoumanian, Andr{\'{e}}, K{\"{o}}nyves,
  Palmeirim, Roy, Schneider, Benedettini, Didelon, {Di Francesco}, Kirk, \&
  Ladjelate}]{Arzoumanian2019}
Arzoumanian, D., Andr{\'{e}}, P., K{\"{o}}nyves, V., {et~al.} 2019, \aap, 621,
  A42

\bibitem[{{Astropy Collaboration} {et~al.}(2013){Astropy Collaboration},
  Robitaille, Tollerud, Greenfield, Droettboom, Bray, Aldcroft, Davis,
  Ginsburg, Price-Whelan, Kerzendorf, Conley, Crighton, Barbary, Muna,
  Ferguson, Grollier, Parikh, Nair, Unther, Deil, Woillez, Conseil, Kramer,
  Turner, Singer, Fox, Weaver, Zabalza, Edwards, {Azalee Bostroem}, Burke,
  Casey, Crawford, Dencheva, Ely, Jenness, Labrie, Lim, Pierfederici, Pontzen,
  Ptak, Refsdal, Servillat, \& Streicher}]{astropy2013}
{Astropy Collaboration}, Robitaille, T., Tollerud, E., {et~al.} 2013, \aap,
  558, A33

\bibitem[{Baldeschi {et~al.}(2017)Baldeschi, Molinari, Elia, Pezzuto, \&
  Schisano}]{Baldeschi2017}
Baldeschi, A., Molinari, S., Elia, D., Pezzuto, S., \& Schisano, E. 2017,
  \mnras, 472, 1778

\bibitem[{{Baraffe} {et~al.}(2009){Baraffe}, {Chabrier}, \&
  {Gallardo}}]{Baraffe2009}
{Baraffe}, I., {Chabrier}, G., \& {Gallardo}, J. 2009, \apjl, 702, L27

\bibitem[{Barnes {et~al.}(2015)Barnes, Muller, Indermuehle, O'Dougherty, Lowe,
  Cunningham, Hernandez, \& Fuller}]{Barnes2015}
Barnes, P.~J., Muller, E., Indermuehle, B., {et~al.} 2015, \apj, 812, 6

\bibitem[{{Benedettini} {et~al.}(2018){Benedettini}, {Pezzuto}, {Schisano},
  {Andr{\'e}}, {K{\"o}nyves}, {Men'shchikov}, {Ladjelate}, {Di Francesco},
  {Elia}, {Arzoumanian}, {Louvet}, {Palmeirim}, {Rygl}, {Schneider},
  {Spinoglio}, \& {Ward-Thompson}}]{Benedettini2018}
{Benedettini}, M., {Pezzuto}, S., {Schisano}, E., {et~al.} 2018, \aap, 619, A52

\bibitem[{Bigiel {et~al.}(2011)Bigiel, Leroy, Walter, Brinks, de~Blok, Kramer,
  Rix, Schruba, Schuster, Usero, \& Wiesemeyer}]{Bigiel2011}
Bigiel, F., Leroy, A., Walter, F., {et~al.} 2011, \apjl, 730, L13

\bibitem[{Bohlin {et~al.}(1978)Bohlin, Savage, \& Drake}]{Bohlin1978}
Bohlin, R.~C., Savage, B.~D., \& Drake, J.~F. 1978, \apj, 224, 132

\bibitem[{{Bracco} {et~al.}(2020){Bracco}, {Bresnahan}, {Palmeirim},
  {Arzoumanian}, {Andr{\'e}}, {Ward-Thompson}, \& {Marchal}}]{Bracco2020}
{Bracco}, A., {Bresnahan}, D., {Palmeirim}, P., {et~al.} 2020, \aap, 644, A5

\bibitem[{{Bresnahan} {et~al.}(2018){Bresnahan}, {Ward-Thompson}, {Kirk},
  {Pattle}, {Eyres}, {White}, {K{\"o}nyves}, {Men'shchikov}, {Andr{\'e}},
  {Schneider}, {Di Francesco}, {Arzoumanian}, {Benedettini}, {Ladjelate},
  {Palmeirim}, {Bracco}, {Molinari}, {Pezzuto}, \& {Spinoglio}}]{Bresnahan2018}
{Bresnahan}, D., {Ward-Thompson}, D., {Kirk}, J.~M., {et~al.} 2018, \aap, 615,
  A125

\bibitem[{Bressan {et~al.}(2012)Bressan, Marigo, Girardi, Salasnich, {Dal
  Cero}, Rubele, \& Nanni}]{Bressan2012}
Bressan, A., Marigo, P., Girardi, L., {et~al.} 2012, \mnras, 427, 127

\bibitem[{{Chabrier}(2005)}]{Chabrier2005}
{Chabrier}, G. 2005, in Astrophysics and Space Science Library, Vol. 327, The
  Initial Mass Function 50 Years Later, ed. E.~{Corbelli}, F.~{Palla}, \&
  H.~{Zinnecker}, 41

\bibitem[{Chevance {et~al.}(2023)Chevance, Krumholz, McLeod, Ostriker,
  Rosolowsky, \& Sternberg}]{Chevance2023}
Chevance, M., Krumholz, M.~R., McLeod, A.~F., {et~al.} 2023, in Astronomical
  Society of the Pacific Conference Series, Vol. 534, Protostars Planets VII,
  ed. S.~Inutsuka, Y.~Aikawa, T.~Muto, K.~Tomida, \& M.~Tamura, 1

\bibitem[{{Dib} {et~al.}(2024){Dib}, {Zhou}, {Comer{\'o}n}, {Gardu{\~n}o},
  {Kravtsov}, {Clark}, {Li}, {Lara-L{\'o}pez}, {Liu}, {Shadmehri}, \&
  {Doughty}}]{Dib2024}
{Dib}, S., {Zhou}, J.~W., {Comer{\'o}n}, S., {et~al.} 2024, \aap , submitted,
  arXiv:2405.00095

\bibitem[{{Duarte-Cabral} {et~al.}(2021){Duarte-Cabral}, {Colombo}, {Urquhart},
  {Ginsburg}, {Russeil}, {Schuller}, {Anderson}, {Barnes}, {Beltr{\'a}n},
  {Beuther}, {Bontemps}, {Bronfman}, {Csengeri}, {Dobbs}, {Eden}, {Giannetti},
  {Kauffmann}, {Mattern}, {Medina}, {Menten}, {Lee}, {Pettitt}, {Riener},
  {Rigby}, {Traficante}, {Veena}, {Wienen}, {Wyrowski}, {Agurto}, {Azagra},
  {Cesaroni}, {Finger}, {Gonzalez}, {Henning}, {Hernandez}, {Kainulainen},
  {Leurini}, {Lopez}, {Mac-Auliffe}, {Mazumdar}, {Molinari}, {Motte}, {Muller},
  {Nguyen-Luong}, {Parra}, {Perez-Beaupuits}, {Montenegro-Montes}, {Moore},
  {Ragan}, {S{\'a}nchez-Monge}, {Sanna}, {Schilke}, {Schisano}, {Schneider},
  {Suri}, {Testi}, {Torstensson}, {Venegas}, {Wang}, \&
  {Zavagno}}]{Duarte-Cabral2021}
{Duarte-Cabral}, A., {Colombo}, D., {Urquhart}, J.~S., {et~al.} 2021, \mnras,
  500, 3027

\bibitem[{Dunham {et~al.}(2014)Dunham, Stutz, Allen, {Evans N.$\sim$J.},
  Fischer, Megeath, Myers, Offner, Poteet, Tobin, \& Vorobyov}]{Dunham2014}
Dunham, M., Stutz, A., Allen, L., {et~al.} 2014, in Protostars Planets VI, ed.
  H.~Beuther, R.~S. Klessen, C.~P. Dullemond, \& T.~Henning, 195--218

\bibitem[{Dunham {et~al.}(2015)Dunham, Allen, Evans, Broekhoven-Fiene, Cieza,
  {Di Francesco}, Gutermuth, Harvey, Hatchell, Heiderman, Huard, Johnstone,
  Kirk, Matthews, Miller, Peterson, \& Young}]{Dunham2015}
Dunham, M.~M., Allen, L.~E., Evans, N.~J., {et~al.} 2015, \apjs, 220, 26 pp

\bibitem[{Elia {et~al.}(2021)Elia, Merello, Molinari, Schisano, Zavagno,
  Russeil, M{\`{e}}ge, Martin, Olmi, Pestalozzi, Plume, Ragan, Benedettini,
  Eden, Moore, Noriega-Crespo, Paladini, Palmeirim, Pezzuto, Pilbratt, Rygl,
  Schilke, Strafella, Tan, Traficante, Baldeschi, Bally, di~Giorgio,
  Fiorellino, Liu, Piazzo, \& Polychroni}]{Elia2021}
Elia, D., Merello, M., Molinari, S., {et~al.} 2021, \mnras, 504, 2742

\bibitem[{Elia {et~al.}(2022)Elia, Molinari, Schisano, Soler, Merello, Russeil,
  Veneziani, Zavagno, Noriega-Crespo, Olmi, Benedettini, Hennebelle, Klessen,
  Leurini, Paladini, Pezzuto, Traficante, Eden, Martin, Sormani, Coletta,
  Colman, Plume, Maruccia, Mininni, \& Liu}]{Elia2022}
Elia, D., Molinari, S., Schisano, E., {et~al.} 2022, \apj, 941, 162

\bibitem[{Evans {et~al.}(2003)Evans, Allen, Blake, Boogert, Bourke, Harvey,
  Kessler, Koerner, Lee, Mundy, Myers, Padgett, Pontoppidan, Sargent,
  Stapelfeldt, van Dishoeck, Young, \& Young}]{Evans2003}
Evans, N.~J., Allen, L.~E., Blake, G.~A., {et~al.} 2003, \pasp, 115, 965

\bibitem[{Evans {et~al.}(2009)Evans, Dunham, J{\o}rgensen, Enoch, Mer{\'{i}}n,
  van Dishoeck, Alcal{\'{a}}, Myers, Stapelfeldt, Huard, Allen, Harvey, van
  Kempen, Blake, Koerner, Mundy, Padgett, \& Sargent}]{Evans2009}
Evans, N.~J., Dunham, M.~M., J{\o}rgensen, J.~K., {et~al.} 2009, \apj, 181, 321

\bibitem[{Evans {et~al.}(2014)Evans, Heiderman, \& Vutisalchavakul}]{Evans2014}
Evans, N.~J., Heiderman, A., \& Vutisalchavakul, N. 2014, \apj, 782, 114

\bibitem[{{Fiorellino} {et~al.}(2021){Fiorellino}, {Elia}, {Andr{\'e}},
  {Men'shchikov}, {Pezzuto}, {Schisano}, {K{\"o}nyves}, {Arzoumanian},
  {Benedettini}, {Ward-Thompson}, {Bracco}, {Di Francesco}, {Bontemps}, {Kirk},
  {Motte}, \& {Molinari}}]{Fiorellino2021}
{Fiorellino}, E., {Elia}, D., {Andr{\'e}}, P., {et~al.} 2021, \mnras, 500, 4257

\bibitem[{{Greene} {et~al.}(1994){Greene}, {Wilking}, {Andre}, {Young}, \&
  {Lada}}]{Greene1994}
{Greene}, T.~P., {Wilking}, B.~A., {Andre}, P., {Young}, E.~T., \& {Lada},
  C.~J. 1994, \apj, 434, 614

\bibitem[{{Grudi{\'c}} {et~al.}(2022){Grudi{\'c}}, {Guszejnov}, {Offner},
  {Rosen}, {Raju}, {Faucher-Gigu{\`e}re}, \& {Hopkins}}]{Grudic2022}
{Grudi{\'c}}, M.~Y., {Guszejnov}, D., {Offner}, S. S.~R., {et~al.} 2022,
  \mnras, 512, 216

\bibitem[{{Gupta} \& {Chen}(2022)}]{Gupta2022}
{Gupta}, A. \& {Chen}, W.-P. 2022, \aj, 163, 233

\bibitem[{G{\"{u}}sten {et~al.}(2006)G{\"{u}}sten, Nyman, Schilke, Menten,
  Cesarsky, \& Booth}]{Guesten2006}
G{\"{u}}sten, R., Nyman, L.~A., Schilke, P., {et~al.} 2006, \aap, 454, L13

\bibitem[{{Habing}(1968)}]{Habing1968}
{Habing}, H.~J. 1968, \bain, 19, 421

\bibitem[{Harris {et~al.}(2020)Harris, Millman, van~der Walt, Gommers,
  Virtanen, Cournapeau, Wieser, Taylor, Berg, Smith, Kern, Picus, Hoyer, van
  Kerkwijk, Brett, Haldane, del R{\'{i}}o, Wiebe, Peterson,
  G{\'{e}}rard-Marchant, Sheppard, Reddy, Weckesser, Abbasi, Gohlke, \&
  Oliphant}]{numpy}
Harris, C.~R., Millman, K.~J., van~der Walt, S.~J., {et~al.} 2020, Nature, 585,
  357

\bibitem[{{Harvey} {et~al.}(2013){Harvey}, {Fallscheer}, {Ginsburg}, {Terebey},
  {Andr{\'e}}, {Bourke}, {Di Francesco}, {K{\"o}nyves}, {Matthews}, \&
  {Peterson}}]{Harvey2013}
{Harvey}, P.~M., {Fallscheer}, C., {Ginsburg}, A., {et~al.} 2013, \apj, 764,
  133

\bibitem[{{Heiderman} \& {Evans}(2015)}]{Heiderman2015}
{Heiderman}, A. \& {Evans}, Neal~J., I. 2015, \apj, 806, 231

\bibitem[{Heiderman {et~al.}(2010)Heiderman, Evans, Allen, Huard, \&
  Heyer}]{Heiderman2010}
Heiderman, A., Evans, N.~J., Allen, L.~E., Huard, T., \& Heyer, M. 2010, \apj,
  723, 1019

\bibitem[{{Hildebrand}(1983)}]{Hildebrand1983}
{Hildebrand}, R.~H. 1983, \qjras, 24, 267

\bibitem[{Hu {et~al.}(2022)Hu, Krumholz, Pokhrel, \& Gutermuth}]{Hu2022}
Hu, Z., Krumholz, M.~R., Pokhrel, R., \& Gutermuth, R.~A. 2022, \mnras, 511,
  1431

\bibitem[{Hunter(2007)}]{matplotlib}
Hunter, J.~D. 2007, Comput. Sci. \& Eng., 9, 90

\bibitem[{Immer {et~al.}(2012)Immer, Menten, Schuller, \& Lis}]{Immer2012b}
Immer, K., Menten, K.~M., Schuller, F., \& Lis, D.~C. 2012, \aap, 548, A120

\bibitem[{Jackson {et~al.}(2006)Jackson, Rathborne, Shah, Simon, Bania,
  Clemens, Chambers, Johnson, Dormody, Lavoie, \& Heyer}]{Jackson2006}
Jackson, J.~M., Rathborne, J.~M., Shah, R.~Y., {et~al.} 2006, \apjs, 163, 145

\bibitem[{Johnstone {et~al.}(2004)Johnstone, {Di Francesco}, \&
  Kirk}]{Johnstone2004}
Johnstone, D., {Di Francesco}, J., \& Kirk, H. 2004, \apjl, 611, L45

\bibitem[{Kainulainen {et~al.}(2009)Kainulainen, Beuther, Henning, \&
  Plume}]{Kainulainen2009}
Kainulainen, J., Beuther, H., Henning, T., \& Plume, R. 2009, \aap, 508, L35

\bibitem[{{Kang} {et~al.}(2009){Kang}, {Bieging}, {Povich}, \&
  {Lee}}]{Kang2009}
{Kang}, M., {Bieging}, J.~H., {Povich}, M.~S., \& {Lee}, Y. 2009, \apj, 706, 83

\bibitem[{{Kang} {et~al.}(2017){Kang}, {Kerton}, {Choi}, \& {Kang}}]{Kang2017}
{Kang}, S.-J., {Kerton}, C.~R., {Choi}, M., \& {Kang}, M. 2017, \apj, 845, 21

\bibitem[{Kauffmann {et~al.}(2008)Kauffmann, Bertoldi, Bourke, Evans, \&
  Lee}]{Kauffmann2008}
Kauffmann, J., Bertoldi, F., Bourke, T.~L., Evans, N.~J., \& Lee, C.~W. 2008,
  \aap, 487, 993

\bibitem[{Kennicutt(1998)}]{Kennicutt1998}
Kennicutt, R. C.~J. 1998, \apj, 498, 541

\bibitem[{{Kirkpatrick} {et~al.}(2024){Kirkpatrick}, {Marocco}, {Gelino},
  {Raghu}, {Faherty}, {Bardalez Gagliuffi}, {Schurr}, {Apps}, {Schneider},
  {Meisner}, {Kuchner}, {Caselden}, {Smart}, {Casewell}, {Raddi}, {Kesseli},
  {Stevnbak Andersen}, {Antonini}, {Beaulieu}, {Bickle}, {Bilsing}, {Chieng},
  {Colin}, {Deen}, {Dereveanco}, {Doll}, {Durantini Luca}, {Frazer}, {Gantier},
  {Gramaize}, {Grant}, {Hamlet}, {Higashimura}, {Hyogo}, {Ja{\l}owiczor},
  {Jonkeren}, {Kabatnik}, {Kiwy}, {Martin}, {Michaels}, {Pendrill}, {Pessanha
  Machado}, {Pumphrey}, {Rothermich}, {Russwurm}, {Sainio}, {Sanchez},
  {Sapelkin-Tambling}, {Sch{\"u}mann}, {Selg-Mann}, {Singh}, {Stenner}, {Sun},
  {Tanner}, {Th{\'e}venot}, {Ventura}, {Voloshin}, {Walla}, {W{\k{e}}dracki},
  {Adorno}, {Aganze}, {Allers}, {Brooks}, {Burgasser}, {Calamari}, {Connor},
  {Costa}, {Eisenhardt}, {Gagn{\'e}}, {Gerasimov}, {Gonzales}, {Hsu}, {Kiman},
  {Li}, {Low}, {Mamajek}, {Pantoja}, {Popinchalk}, {Rees}, {Stern},
  {Su{\'a}rez}, {Theissen}, {Tsai}, {Vos}, {Zurek}, \& {The Backyard Worlds:
  Planet 9 Collaboration}}]{Kirkpatrick2024}
{Kirkpatrick}, J.~D., {Marocco}, F., {Gelino}, C.~R., {et~al.} 2024, \apjs,
  271, 55

\bibitem[{Koenig \& Leisawitz(2014)}]{Koenig2014}
Koenig, X. \& Leisawitz, D. 2014, \apj, 791, 27 pp

\bibitem[{{K{\"o}nyves} {et~al.}(2020){K{\"o}nyves}, {Andr{\'e}},
  {Arzoumanian}, {Schneider}, {Men'shchikov}, {Bontemps}, {Ladjelate},
  {Didelon}, {Pezzuto}, {Benedettini}, {Bracco}, {Di Francesco}, {Goodwin},
  {Rygl}, {Shimajiri}, {Spinoglio}, {Ward-Thompson}, \& {White}}]{Konyves2020}
{K{\"o}nyves}, V., {Andr{\'e}}, P., {Arzoumanian}, D., {et~al.} 2020, \aap,
  635, A34

\bibitem[{{K{\"o}nyves} {et~al.}(2015){K{\"o}nyves}, {Andr{\'e}},
  {Men'shchikov}, {Palmeirim}, {Arzoumanian}, {Schneider}, {Roy}, {Didelon},
  {Maury}, {Shimajiri}, {Di Francesco}, {Bontemps}, {Peretto}, {Benedettini},
  {Bernard}, {Elia}, {Griffin}, {Hill}, {Kirk}, {Ladjelate}, {Marsh}, {Martin},
  {Motte}, {Nguy{\^e}n Luong}, {Pezzuto}, {Roussel}, {Rygl}, {Sadavoy},
  {Schisano}, {Spinoglio}, {Ward-Thompson}, \& {White}}]{Konyves2015}
{K{\"o}nyves}, V., {Andr{\'e}}, P., {Men'shchikov}, A., {et~al.} 2015, \aap,
  584, A91

\bibitem[{{Kramer} {et~al.}(2008){Kramer}, {Cubick}, {R{\"o}llig}, {Sun},
  {Yonekura}, {Aravena}, {Bensch}, {Bertoldi}, {Bronfman}, {Fujishita},
  {Fukui}, {Graf}, {Hitschfeld}, {Honingh}, {Ito}, {Jakob}, {Jacobs}, {Klein},
  {Koo}, {May}, {Miller}, {Miyamoto}, {Mizuno}, {Onishi}, {Park}, {Pineda},
  {Rabanus}, {Sasago}, {Schieder}, {Simon}, {Stutzki}, {Volgenau}, \&
  {Yamamoto}}]{Kramer2008}
{Kramer}, C., {Cubick}, M., {R{\"o}llig}, M., {et~al.} 2008, \aap, 477, 547

\bibitem[{Kroupa(2002)}]{Kroupa2002}
Kroupa, P. 2002, Science (80-. )., 295, 82

\bibitem[{{Krumholz}(2014)}]{Krumholz2014}
{Krumholz}, M.~R. 2014, \physrep, 539, 49

\bibitem[{Krumholz \& McKee(2005)}]{Krumholz2005}
Krumholz, M.~R. \& McKee, C.~F. 2005, \apj, 630, 250

\bibitem[{{Krumholz} \& {Tan}(2007)}]{Krumholz2007}
{Krumholz}, M.~R. \& {Tan}, J.~C. 2007, \apj, 654, 304

\bibitem[{{Kuhn} {et~al.}(2021){Kuhn}, {de Souza}, {Krone-Martins},
  {Castro-Ginard}, {Ishida}, {Povich}, {Hillenbrand}, \& {COIN
  Collaboration}}]{Kuhn2021}
{Kuhn}, M.~A., {de Souza}, R.~S., {Krone-Martins}, A., {et~al.} 2021, \apjs,
  254, 33

\bibitem[{Lada {et~al.}(2012)Lada, Forbrich, Lombardi, \& Alves}]{Lada2012}
Lada, C.~J., Forbrich, J., Lombardi, M., \& Alves, J.~F. 2012, \apj, 745, 190

\bibitem[{Lada {et~al.}(2017)Lada, Lewis, Lombardi, \& Alves}]{Lada2017}
Lada, C.~J., Lewis, J.~A., Lombardi, M., \& Alves, J. 2017, \aap, 606, A100

\bibitem[{Lada {et~al.}(2010)Lada, Lombardi, \& Alves}]{Lada2010}
Lada, C.~J., Lombardi, M., \& Alves, J.~F. 2010, \apj, 724, 687

\bibitem[{{Ladjelate} {et~al.}(2020){Ladjelate}, {Andr{\'e}}, {K{\"o}nyves},
  {Ward-Thompson}, {Men'shchikov}, {Bracco}, {Palmeirim}, {Roy}, {Shimajiri},
  {Kirk}, {Arzoumanian}, {Benedettini}, {Di Francesco}, {Fiorellino},
  {Schneider}, {Pezzuto}, {Motte}, \& {Herschel Gould Belt Survey
  Team}}]{Ladjelate2020}
{Ladjelate}, B., {Andr{\'e}}, P., {K{\"o}nyves}, V., {et~al.} 2020, \aap, 638,
  A74

\bibitem[{Marton {et~al.}(2019)Marton, {\'{A}}brah{\'{a}}m, Szegedi-Elek,
  Varga, Kun, K{\'{o}}sp{\'{a}}l, Varga-Vereb{\'{e}}lyi, Hodgkin, Szabados,
  Beck, \& Kiss}]{Marton2019}
Marton, G., {\'{A}}brah{\'{a}}m, P., Szegedi-Elek, E., {et~al.} 2019, \mnras,
  487, 2522

\bibitem[{Marton {et~al.}(2016)Marton, T{\'{o}}th, Paladini, Kun, Zahorecz,
  McGehee, \& Kiss}]{Marton2016}
Marton, G., T{\'{o}}th, L., Paladini, R., {et~al.} 2016, \mnras, 458, 3479

\bibitem[{{Mattern} {et~al.}(2018){Mattern}, {Kauffmann}, {Csengeri},
  {Urquhart}, {Leurini}, {Wyrowski}, {Giannetti}, {Barnes}, {Beuther},
  {Bronfman}, {Duarte-Cabral}, {Henning}, {Kainulainen}, {Menten}, {Schisano},
  \& {Schuller}}]{Mattern2018b}
{Mattern}, M., {Kauffmann}, J., {Csengeri}, T., {et~al.} 2018, \aap, 619, A166

\bibitem[{Megeath {et~al.}(2016)Megeath, Gutermuth, Muzerolle, Kryukova, Hora,
  Allen, Flaherty, Hartmann, Myers, Pipher, Stauffer, Young, \&
  Fazio}]{Megeath2016}
Megeath, S., Gutermuth, R., Muzerolle, J., {et~al.} 2016, \aj, 151, 5

\bibitem[{Megeath {et~al.}(2012)Megeath, Gutermuth, Muzerolle, Kryukova,
  Flaherty, Hora, Allen, Hartmann, Myers, Pipher, Stauffer, Young, \&
  Fazio}]{Megeath2012}
Megeath, S.~T., Gutermuth, R., Muzerolle, J., {et~al.} 2012, Astron. J., 144,
  27 pp

\bibitem[{Molinari {et~al.}(2010)Molinari, Consortium, Bally, Barlow, Bernard,
  Martin, Moore, Noriega-Crespo, Plume, Testi, Zavagno, Abergel, Ali,
  Andr{\'{e}}, Baluteau, Benedettini, Bern{\'{e}}, Billot, Blommaert, Bontemps,
  Boulanger, Brand, Brunt, Burton, Campeggio, Carey, Caselli, Cesaroni,
  Cernicharo, Chakrabarti, Chrysostomou, Codella, Cohen, Compiegne, Davis,
  de~Bernardis, de~Gasperis, {Di Francesco}, di~Giorgio, Elia, Faustini,
  Fischera, Fukui, Fuller, Ganga, Garcia-Lario, Giard, Giardino, Glenn,
  Goldsmith, Griffin, Hoare, Huang, Jiang, Joblin, Joncas, Juvela, Kirk,
  Lagache, Li, Lim, Lord, Lucas, Maiolo, Marengo, Marshall, Masi, Massi,
  Matsuura, Meny, Minier, Miville-Desch{\^{e}}nes, Montier, Motte,
  M{\"{u}}ller, Natoli, Neves, Olmi, Paladini, Paradis, Pestalozzi, Pezzuto,
  Piacentini, Pomar{\`{e}}s, Popescu, Reach, Richer, Ristorcelli, Roy, Royer,
  Russeil, Saraceno, Sauvage, Schilke, Schneider-Bontemps, Schuller, Schultz,
  Shepherd, Sibthorpe, Smith, Smith, Spinoglio, Stamatellos, Strafella,
  Stringfellow, Sturm, Taylor, Thompson, Tuffs, Umana, Valenziano, Vavrek,
  Viti, Waelkens, Ward-Thompson, White, Wyrowski, Yorke, Zhang, Swinyard,
  Bally, Barlow, Bernard, Martin, Moore, Noriega-Crespo, Plume, Testi, Zavagno,
  Abergel, Ali, Andr{\'{e}}, Baluteau, Benedettini, Bern{\'{e}}, Billot,
  Blommaert, Bontemps, Boulanger, Brand, Brunt, Burton, Campeggio, Carey,
  Caselli, Cesaroni, Cernicharo, Chakrabarti, Chrysostomou, Codella, Cohen,
  Compiegne, Davis, de~Bernardis, de~Gasperis, {Di Francesco}, di~Giorgio,
  Elia, Faustini, Fischera, Fukui, Fuller, Ganga, Garcia-Lario, Giard,
  Giardino, Glenn, Goldsmith, Griffin, Hoare, Huang, Jiang, Joblin, Joncas,
  Juvela, Kirk, Lagache, Li, Lim, Lord, Lucas, Maiolo, Marengo, Marshall, Masi,
  Massi, Matsuura, Meny, Minier, Miville-Desch{\^{e}}nes, Montier, Motte,
  M{\"{u}}ller, Natoli, Neves, Olmi, Paladini, Paradis, Pestalozzi, Pezzuto,
  Piacentini, Pomar{\`{e}}s, Popescu, Reach, Richer, Ristorcelli, Roy, Royer,
  Russeil, Saraceno, Sauvage, Schilke, Schneider-Bontemps, Schuller, Schultz,
  Shepherd, Sibthorpe, Smith, Smith, Spinoglio, Stamatellos, Strafella,
  Stringfellow, Sturm, Taylor, Thompson, Tuffs, Umana, Valenziano, Vavrek,
  Viti, Waelkens, Ward-Thompson, White, Wyrowski, Yorke, \&
  Zhang}]{Molinari2010}
Molinari, S., Consortium, t. H.-G., Bally, J., {et~al.} 2010, \pasp, 122, 314

\bibitem[{Molinari {et~al.}(2008)Molinari, Pezzuto, Cesaroni, Brand, Faustini,
  \& Testi}]{Molinari2008}
Molinari, S., Pezzuto, S., Cesaroni, R., {et~al.} 2008, \aap, 481, 345

\bibitem[{{Motte} \& {Andr{\'e}}(2001)}]{Motte2001}
{Motte}, F. \& {Andr{\'e}}, P. 2001, \aap, 365, 440

\bibitem[{Motte {et~al.}(2010)Motte, Zavagno, Bontemps, Schneider, Hennemann,
  di~Francesco, Andr{\'{e}}, Saraceno, Griffin, Marston, Ward-Thompson, White,
  Minier, Men'shchikov, Hill, Abergel, Anderson, Aussel, Balog, Baluteau,
  Bernard, Cox, Csengeri, Deharveng, Didelon, di~Giorgio, Hargrave, Huang,
  Kirk, Leeks, Li, Martin, Molinari, Nguyen-Luong, Olofsson, Persi, Peretto,
  Pezzuto, Roussel, Russeil, Sadavoy, Sauvage, Sibthorpe, Spinoglio, Testi,
  Teyssier, Vavrek, Wilson, \& Woodcraft}]{Motte2010}
Motte, F., Zavagno, A., Bontemps, S., {et~al.} 2010, \aap, 518, L77

\bibitem[{Onishi {et~al.}(2001)Onishi, Yoshikawa, Yamamoto, Kawamura, Mizuno,
  \& Fukui}]{Onishi2001}
Onishi, T., Yoshikawa, N., Yamamoto, H., {et~al.} 2001, \pasj, 53, 1017

\bibitem[{Ortiz-Le{\'{o}}n {et~al.}(2023)Ortiz-Le{\'{o}}n, Dzib, Loinard, Gong,
  Pillai, \& Plunkett}]{Ortiz-Leon2023}
Ortiz-Le{\'{o}}n, G.~N., Dzib, S.~A., Loinard, L., {et~al.} 2023, \aap, 673, L1

\bibitem[{{Ossenkopf} \& {Henning}(1994)}]{Ossenkopf1994}
{Ossenkopf}, V. \& {Henning}, T. 1994, \aap, 291, 943

\bibitem[{{Palla} \& {Stahler}(2000)}]{Palla2000}
{Palla}, F. \& {Stahler}, S.~W. 2000, \apj, 540, 255

\bibitem[{Palmeirim {et~al.}(2013)Palmeirim, Andr{\'{e}}, Kirk, Ward-Thompson,
  Arzoumanian, K{\"{o}}nyves, Didelon, Schneider, Benedettini, Bontemps, {Di
  Francesco}, Elia, Griffin, Hennemann, Hill, Martin, Men'shchikov, Molinari,
  Motte, {Nguyen Luong}, Nutter, Peretto, Pezzuto, Roy, Rygl, Spinoglio, White,
  Men'shchikov, Molinari, Motte, {Nguyen Luong}, Nutter, Peretto, Pezzuto, Roy,
  Rygl, Spinoglio, \& White}]{Palmeirim2013}
Palmeirim, P., Andr{\'{e}}, P., Kirk, J., {et~al.} 2013, \aap, 550, A38

\bibitem[{{Peretto} {et~al.}(2023){Peretto}, {Rigby}, {Louvet}, {Fuller},
  {Traficante}, \& {Gaudel}}]{Peretto2023}
{Peretto}, N., {Rigby}, A.~J., {Louvet}, F., {et~al.} 2023, \mnras, 525, 2935

\bibitem[{{Pezzuto} {et~al.}(2021){Pezzuto}, {Benedettini}, {Di Francesco},
  {Palmeirim}, {Sadavoy}, {Schisano}, {Li Causi}, {Andr{\'e}}, {Arzoumanian},
  {Bernard}, {Bontemps}, {Elia}, {Fiorellino}, {Kirk}, {K{\"o}nyves},
  {Ladjelate}, {Men'shchikov}, {Motte}, {Piccotti}, {Schneider}, {Spinoglio},
  {Ward-Thompson}, \& {Wilson}}]{Pezzuto2021}
{Pezzuto}, S., {Benedettini}, M., {Di Francesco}, J., {et~al.} 2021, \aap, 645,
  A55

\bibitem[{{Planck early results I.}(2011)}]{Planck2011}
{Planck early results I.} 2011, \aap, 536, A1

\bibitem[{Pokhrel {et~al.}(2021)Pokhrel, Gutermuth, Krumholz, Federrath, Heyer,
  Khullar, Megeath, Myers, Offner, Pipher, Fischer, Henning, \&
  Hora}]{Pokhrel2021}
Pokhrel, R., Gutermuth, R.~A., Krumholz, M.~R., {et~al.} 2021, \apjl, 912, L19

\bibitem[{{Preibisch} {et~al.}(1993){Preibisch}, {Ossenkopf}, {Yorke}, \&
  {Henning}}]{Preibisch1993}
{Preibisch}, T., {Ossenkopf}, V., {Yorke}, H.~W., \& {Henning}, T. 1993, \aap,
  279, 577

\bibitem[{Rayner {et~al.}(2017)Rayner, Griffin, Schneider, Motte,
  K{\"{o}}nyves, Andr{\'{e}}, {Di Francesco}, Didelon, Pattle, Ward-Thompson,
  Anderson, Benedettini, Bernard, Bontemps, Elia, Fuente, Hennemann, Hill,
  Kirk, Marsh, Men'shchikov, {Nguyen Luong}, Peretto, Pezzuto, Rivera-Ingraham,
  Roy, Rygl, S{\'{a}}nchez-Monge, Spinoglio, Tig{\'{e}}, Trevi{\~{n}}o-Morales,
  \& White}]{Rayner2017}
Rayner, T. S.~M., Griffin, M.~J., Schneider, N., {et~al.} 2017, \aap, 607, A22

\bibitem[{Rev{\'{e}}ret {et~al.}(2014)Rev{\'{e}}ret, Andr{\'{e}}, {Le Pennec},
  Talvard, Agn{\`{e}}se, Arnaud, Clerc, de~Breuck, Cigna, Delisle, Doumayrou,
  Duband, Dubreuil, Dumaye, Ercolani, Gallais, Groult, Jourdan, Leriche,
  Maffei, Lortholary, Martignac, Rabaud, Relland, Rodriguez, Vandeneynde, \&
  Visticot}]{Reveret2014}
Rev{\'{e}}ret, V., Andr{\'{e}}, P., {Le Pennec}, J., {et~al.} 2014, in Society
  of Photo-Optical Instrumentation Engineers (SPIE) Conference Series, Vol.
  9153, Millimeter, Submillimeter, Far-Infrared Detect. Instrum. Astron. VII,
  ed. W.~S. Holland \& J.~Zmuidzinas, 915305

\bibitem[{Robitaille(2017)}]{Robitaille2017}
Robitaille, T. 2017, \aap, 600, A11

\bibitem[{Robitaille(2019)}]{aplpy2019}
Robitaille, T. 2019, {APLpy v2.0: The Astronomical Plotting Library in Python}

\bibitem[{Roussel(2013)}]{Roussel2013}
Roussel, H. 2013, \pasp, 125, 1126

\bibitem[{Roussel(2018)}]{Roussel2018}
Roussel, H. 2018, arXiv e-prints, arXiv:1803.04264

\bibitem[{Roy {et~al.}(2014)Roy, Andr{\'{e}}, Palmeirim, Attard, K{\"{o}}nyves,
  Schneider, Peretto, Men'shchikov, Ward-Thompson, Kirk, Griffin, Marsh,
  Abergel, Arzoumanian, Benedettini, Hill, Motte, {Nguyen Luong}, Pezzuto,
  Rivera-Ingraham, Roussel, Rygl, Spinoglio, Stamatellos, \& White}]{Roy2014}
Roy, A., Andr{\'{e}}, P., Palmeirim, P., {et~al.} 2014, \aap, 562, A138

\bibitem[{{Russeil} {et~al.}(2011){Russeil}, {Pestalozzi}, {Mottram},
  {Bontemps}, {Anderson}, {Zavagno}, {Beltr{\'a}n}, {Bally}, {Brand}, {Brunt},
  {Cesaroni}, {Joncas}, {Marshall}, {Martin}, {Massi}, {Molinari}, {Moore},
  {Noriega-Crespo}, {Olmi}, {Thompson}, {Wienen}, \& {Wyrowski}}]{Russeil2011}
{Russeil}, D., {Pestalozzi}, M., {Mottram}, J.~C., {et~al.} 2011, \aap, 526,
  A151

\bibitem[{{Salji} {et~al.}(2015){Salji}, {Richer}, {Buckle}, {Hatchell},
  {Kirk}, {Beaulieu}, {Berry}, {Broekhoven-Fiene}, {Currie}, {Fich}, {Jenness},
  {Johnstone}, {Mottram}, {Nutter}, {Pattle}, {Pineda}, {Quinn}, {Tisi},
  {Walker-Smith}, {di Francesco}, {Hogerheijde}, {Ward-Thompson}, {Bastien},
  {Butner}, {Chen}, {Chrysostomou}, {Coude}, {Davis}, {Drabek-Maunder},
  {Duarte-Cabral}, {Fiege}, {Friberg}, {Friesen}, {Fuller}, {Graves},
  {Greaves}, {Gregson}, {Holland}, {Joncas}, {Kirk}, {Knee}, {Mairs}, {Marsh},
  {Matthews}, {Moriarty-Schieven}, {Rawlings}, {Robertson}, {Rosolowsky},
  {Rumble}, {Sadavoy}, {Thomas}, {Tothill}, {Viti}, {White}, {Wilson},
  {Wouterloot}, {Yates}, \& {Zhu}}]{Salji2015a}
{Salji}, C.~J., {Richer}, J.~S., {Buckle}, J.~V., {et~al.} 2015, \mnras, 449,
  1769

\bibitem[{Sault {et~al.}(1995)Sault, Teuben, \& Wright}]{Sault1995}
Sault, R.~J., Teuben, P.~J., \& Wright, M. C.~H. 1995, in Astronomical Society
  of the Pacific Conference Series, Vol.~77, Astron. Data Anal. Softw. Syst.
  IV, ed. R.~A. Shaw, H.~E. Payne, \& J.~J.~E. Hayes, 433

\bibitem[{Schneider {et~al.}(2013)Schneider, Andr{\'{e}}, K{\"{o}}nyves,
  Bontemps, Motte, Federrath, Ward-Thompson, Arzoumanian, Benedettini,
  Bressert, Didelon, {Di Francesco}, Griffin, Hennemann, Hill, Palmeirim,
  Pezzuto, Peretto, Roy, Rygl, Spinoglio, \& White}]{Schneider2013}
Schneider, N., Andr{\'{e}}, P., K{\"{o}}nyves, V., {et~al.} 2013, \apjl, 766,
  L17

\bibitem[{{Schruba} {et~al.}(2011){Schruba}, {Leroy}, {Walter}, {Bigiel},
  {Brinks}, {de Blok}, {Dumas}, {Kramer}, {Rosolowsky}, {Sandstrom},
  {Schuster}, {Usero}, {Weiss}, \& {Wiesemeyer}}]{Schruba2011}
{Schruba}, A., {Leroy}, A.~K., {Walter}, F., {et~al.} 2011, \aj, 142, 37

\bibitem[{Schuller {et~al.}(2021{\natexlab{a}})Schuller, Andr{\'{e}},
  Shimajiri, Zavagno, Peretto, Arzoumanian, Csengeri, K{\"{o}}nyves, Palmeirim,
  Pezzuto, Rigby, Roussel, Ajeddig, Dumaye, Gallais, {Le Pennec}, Martignac,
  Mattern, Rev{\'{e}}ret, Rodriguez, \& Talvard}]{Schuller2021b}
Schuller, F., Andr{\'{e}}, P., Shimajiri, Y., {et~al.} 2021{\natexlab{a}},
  \aap, 651, A36

\bibitem[{{Schuller} {et~al.}(2009){Schuller}, {Menten}, {Contreras},
  {Wyrowski}, {Schilke}, {Bronfman}, {Henning}, {Walmsley}, {Beuther},
  {Bontemps}, {Cesaroni}, {Deharveng}, {Garay}, {Herpin}, {Lefloch}, {Linz},
  {Mardones}, {Minier}, {Molinari}, {Motte}, {Nyman}, {Reveret}, {Risacher},
  {Russeil}, {Schneider}, {Testi}, {Troost}, {Vasyunina}, {Wienen}, {Zavagno},
  {Kovacs}, {Kreysa}, {Siringo}, \& {Wei{\ss}}}]{Schuller2009}
{Schuller}, F., {Menten}, K.~M., {Contreras}, Y., {et~al.} 2009, \aap, 504, 415

\bibitem[{Schuller {et~al.}(2021{\natexlab{b}})Schuller, Urquhart, Csengeri,
  Colombo, Duarte-Cabral, Mattern, Ginsburg, Pettitt, Wyrowski, Anderson,
  Azagra, Barnes, Beltran, Beuther, Billington, Bronfman, Cesaroni, Dobbs,
  Eden, Lee, Medina, Menten, Moore, Montenegro-Montes, Ragan, Rigby, Riener,
  Russeil, Schisano, Sanchez-Monge, Traficante, Zavagno, Agurto, Bontemps,
  Finger, Giannetti, Gonzalez, Hernandez, Henning, Kainulainen, Kauffmann,
  Leurini, Lopez, Mac-Auliffe, Mazumdar, Molinari, Motte, Muller, Nguyen-Luong,
  Parra, Perez-Beaupuits, Schilke, Schneider, Suri, Testi, Torstensson, Veena,
  Venegas, Wang, \& Wienen}]{Schuller2021a}
Schuller, F., Urquhart, J., Csengeri, T., {et~al.} 2021{\natexlab{b}}, \mnras,
  500, 3064

\bibitem[{Sewi{\l}o {et~al.}(2019)Sewi{\l}o, Whitney, Yung, Robitaille, Elia,
  Indebetouw, Schisano, Szczerba, Karska, Wiseman, Babler, Boyer, Fischer,
  Meade, Olmi, Padgett, \& Si{\'{o}}dmiak}]{Sewilo2019}
Sewi{\l}o, M., Whitney, B.~A., Yung, B.~H., {et~al.} 2019, \apjs, 240, 26

\bibitem[{Shimajiri {et~al.}(2017)Shimajiri, Andr{\'{e}}, Braine,
  K{\"{o}}nyves, Schneider, Bontemps, Ladjelate, Roy, Gao, \&
  Chen}]{Shimajiri2017}
Shimajiri, Y., Andr{\'{e}}, P., Braine, J., {et~al.} 2017, \aap, 604, A74

\bibitem[{{Stutz}(2018)}]{Stutz2018}
{Stutz}, A.~M. 2018, \mnras, 473, 4890

\bibitem[{{Suin} {et~al.}(2024){Suin}, {Zavagno}, {Colman}, {Hennebelle},
  {Verliat}, \& {Russeil}}]{Suin2023}
{Suin}, P., {Zavagno}, A., {Colman}, T., {et~al.} 2024, \aap, 682, A76

\bibitem[{Talvard {et~al.}(2018)Talvard, Rev{\'{e}}ret, Le-Pennec, Andr{\'{e}},
  Arnaud, Clerc, de~Breuck, Delisle, Doumayrou, Duband, Dubreuil, Dumaye,
  Ercolani, Gallais, Lortholary, Martignac, Relland, Rodriguez, Roussel,
  Schuller, \& Visticot}]{Talvard2018}
Talvard, M., Rev{\'{e}}ret, V., Le-Pennec, Y., {et~al.} 2018, in Society of
  Photo-Optical Instrumentation Engineers (SPIE) Conference Series, Vol. 10708,
  Millimeter, Submillimeter, Far-Infrared Detect. Instrum. Astron. IX, ed.
  J.~Zmuidzinas \& J.-R. Gao, 1070838

\bibitem[{Veneziani {et~al.}(2017)Veneziani, Schisano, Elia, Noriega-Crespo,
  Carey, {Di Giorgio}, Fukui, Maiolo, Maruccia, Mizuno, Mizuno, Molinari,
  Mottram, Moore, Onishi, Paladini, Paradis, Pestalozzi, Pezzuto, Piacentini,
  Plume, Russeil, \& Strafella}]{Veneziani2017}
Veneziani, M., Schisano, E., Elia, D., {et~al.} 2017, \aap, 599, A7

\bibitem[{Virtanen {et~al.}(2020)Virtanen, Gommers, Oliphant, Haberland, Reddy,
  Cournapeau, Burovski, Peterson, Weckesser, Bright, {van der Walt}, Brett,
  Wilson, Millman, Mayorov, Nelson, Jones, Kern, Larson, Carey, Polat, Feng,
  Moore, {VanderPlas}, Laxalde, Perktold, Cimrman, Henriksen, Quintero, Harris,
  Archibald, Ribeiro, Pedregosa, {van Mulbregt}, \& {SciPy 1.0
  Contributors}}]{SciPy2020}
Virtanen, P., Gommers, R., Oliphant, T.~E., {et~al.} 2020, Nature Methods, 17,
  261

\bibitem[{Ward-Thompson {et~al.}(2007)Ward-Thompson, {Di Francesco}, Hatchell,
  Hogerheijde, Nutter, Bastien, Basu, Bonnell, Bowey, Brunt, Buckle, Butner,
  Cavanagh, Chrysostomou, Curtis, Davis, Dent, van Dishoeck, Edmunds, Fich,
  Fiege, Fissel, Friberg, Friesen, Frieswijk, Fuller, Gosling, Graves, Greaves,
  Helmich, Hills, Holland, Houde, Jayawardhana, Johnstone, Joncas, Kirk, Kirk,
  Knee, Matthews, Matthews, Matzner, Moriarty-Schieven, Naylor, Padman, Plume,
  Rawlings, Redman, Reid, Richer, Shipman, Simpson, Spaans, Stamatellos,
  Tsamis, Viti, Weferling, White, Whitworth, Wouterloot, Yates, \&
  Zhu}]{Ward-Thompson2007}
Ward-Thompson, D., {Di Francesco}, J., Hatchell, J., {et~al.} 2007, \pasp, 119,
  855

\bibitem[{{Watkins} {et~al.}(2019){Watkins}, {Peretto}, {Marsh}, \&
  {Fuller}}]{Watkins2019}
{Watkins}, E.~J., {Peretto}, N., {Marsh}, K., \& {Fuller}, G.~A. 2019, \aap,
  628, A21

\bibitem[{{Zhang} {et~al.}(2024){Zhang}, {Andr{\'e}}, {Men'shchikov}, {Li},
  {Kravtsov}, \& {Clark}}]{ZhanG2024}
{Zhang}, G.-Y., {Andr{\'e}}, P., {Men'shchikov}, A., {et~al.} 2024, \aap ,
  submitted

\end{thebibliography}

\begin{appendix} 

\section{Statistics of YSOs, protostellar clumps, and gas mass against column density}
\label{sect:Distribution_YSOs-clumps}
In Sect.~\ref{sec:SFR-CAF-YSO}, 
we derived the SFRs of CAFFEINE molecular clouds from YSOs and 
Hi-GAL protostellar clumps identified within the cloud boundaries. 
To understand the limitations of these two methods it is essential to look at the distribution of objects as a function of column density.

\begin{figure}
    \centering
    \includegraphics[width=0.49 \textwidth]{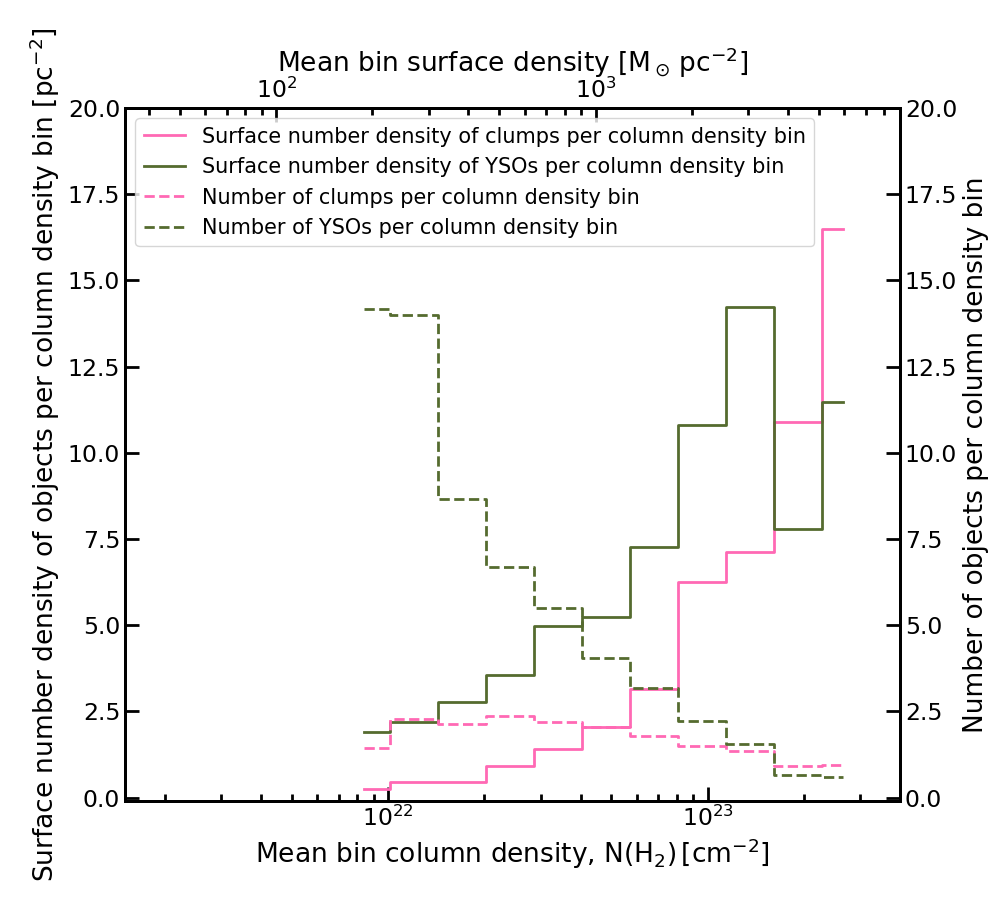}
    \caption{Average number statistics of YSOs (olive) and Hi-GAL protostellar clumps (pink) in the CAFFEINE clouds of this study. 
    The solid histograms indicate the surface number densities 
    in each column density bin (left axis), while the dashed histograms show the absolute numbers of objects in each column density bin (right axis).}
    \label{fig:sources/area_CAFFEINE}
\end{figure}

As most CAFFEINE clouds are quite distant (up to $d \la 3$\,kpc), we expect the identified YSOs to be biased toward relatively luminous and massive objects
given the resolution and sensitivity limits of the underlying infrared surveys. 
This limitation is accounted for by fitting an IMF to the detected YSOs and 
extrapolating the result to the whole stellar mass spectrum to estimate the total YSO mass and corresponding SFR, 
as described in Sect.~\ref{sec:SFR-CAF-YSO}. 
Additionally, the dust in the dense inner parts of the clouds will absorb the near-/mid-infrared emission from embedded YSOs and 
render their detection more difficult at higher column densities. 
The limiting column density above which YSOs can no longer be used to trace the SFR 
may be identified by plotting the average number and average surface density 
of detected YSOs per cloud and column density bin against column density, which is shown in Fig.~\ref{fig:sources/area_CAFFEINE}. 
It can be seen that the surface density of detected YSOs per unit area increases with column density up to $ N_{\rm H_2} \la 10^{23}\rm ~cm^{-2}$
and starts dropping and fluctuating in the two highest column density bins. 
We thus find that our SFR estimates from massive YSOs are strongly impacted by incompleteness for column densities $\ga 10^{23}\, \rm cm^{-2}$, 
which is consistent 
with the detection limit estimates made in Sect.~\ref{sec:SFR-CAF-YSO}. 
We do not see such an incompleteness effect  
for the protostellar clumps, 
as they are identified as peaks in submillimeter dust emission and not affected by dust extinction. 
The lower number of protostellar clumps observed at low column densities 
reflects the strong connection between SF and dense gas.

\begin{figure}
    \centering
   \includegraphics[width=0.49 \textwidth]{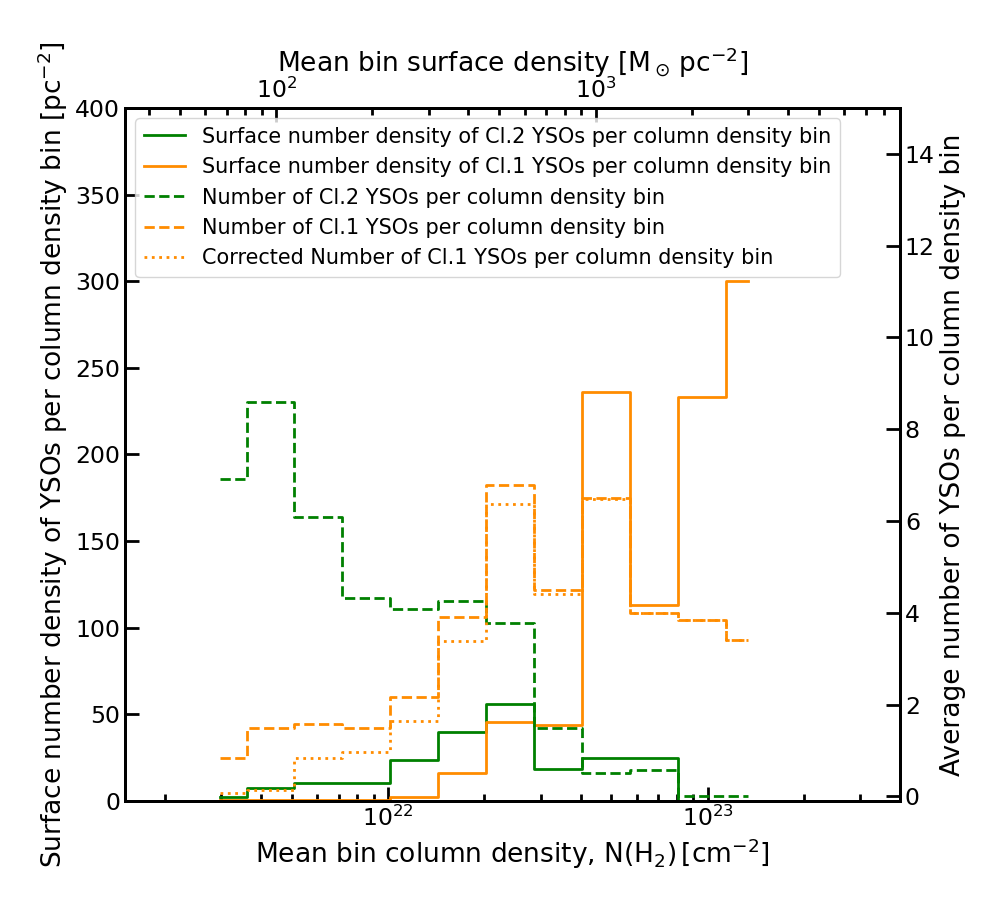}
    \caption{Average number statistics of Class~I (orange) and Class~II (green) YSOs for the nearby clouds considered in this study. 
    The solid histograms show the surface number densities 
    in each column density bin (left axis), while the dashed and dotted histograms give the numbers of objects in each column density bin (right axis). 
    The dashed and dotted orange histograms show the number sof Class~I objects per bin before and after correction for contamination 
    by the fraction of Class~I sources with no significant protostellar envelope \citep[cf. Sect.~\ref{sec:SFR_nearby} and][]{Heiderman2015}, respectively.
    }
    \label{fig:sources/area_HGBS}
\end{figure}

Similarly, 
we also investigate the distribution of Class~I and Class~II YSOs with respect to column density in nearby clouds (Fig.~\ref{fig:sources/area_HGBS}). 
We find significant differences in these distributions. While the surface density of Class~I YSOs peaks at the highest density bin ($ N_{\rm H_2} \approx 1.1\times10^{23}\rm ~cm^{-2}$) the peak of Class~II YSOs is at $N_{\rm H_2} \approx 2.5\times10^{22}\rm ~cm^{-2}$. We interpret this difference as a sign of migration of YSOs after decoupling from their parent gas structures
\cite[cf.][]{Salji2015a, Stutz2018}.

\begin{figure}
    \centering
   \includegraphics[width=0.49 \textwidth]{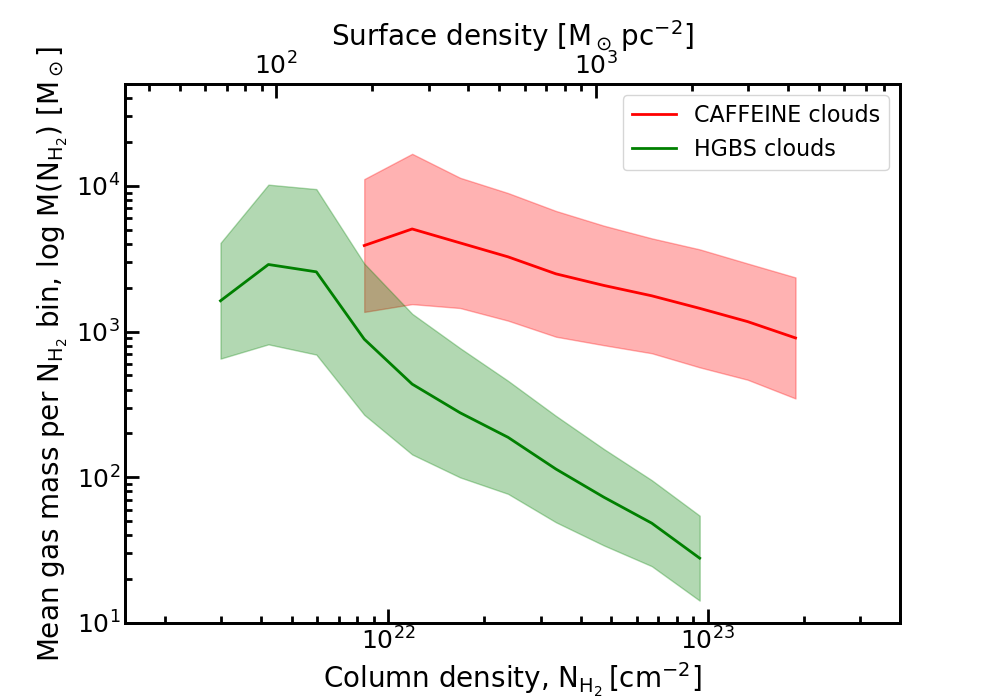}
    \caption{Mean gas mass per cloud in each column density bin against column density for the CAFFEINE sample (red) and the nearby clouds considered here (green). 
    The shaded areas indicate the standard deviations of the sample of differential gas mass values.
    }
    \label{fig:mass_cd}
\end{figure}

Figure~\ref{fig:mass_cd} shows the mean gas mass per cloud per column density bin as a function of column density, for both the CAFFEINE clouds of Table \ref{tab:CAFFEINE_sources} 
and the nearby clouds of Table~\ref{tab:GB_clouds}.
It illustrates that the CAFFEINE clouds include one to two orders of magnitude more dense gas mass than the nearby clouds 
in the range of column densities of $10^{22}~{\rm cm^{-2}}\la N_{\rm H_2} \la10^{23}~{\rm cm^{-2}}$.

\section{Alternative method for combining \artemis\, and \textit{Herschel} data}
\label{sect:alt_combination}
In cases where the baseline method did not produce satisfactory results, the data processing of ArTéMiS data and the hybridization of SPIRE and ArTéMiS maps were performed with an extension 
of the \textit{Scanamorphos} software for ArTéMiS \citep{Roussel2013, Roussel2018}.
The first step is to compute astrometric offsets for each scan, stopping the data processing after the subtraction of the average drift and the uncorrelated drifts on the longest timescale. After projecting separately each scan, the brightest source within the corresponding field of view is fitted with a Gaussian in both the SPIRE 350\,$\mu$m map and the ArTéMiS map (convolved to the angular resolution of the SPIRE 350\,$\mu$m map). The data processing is then restarted with these input astrometric corrections.
The main difference introduced with respect to the normal algorithm is that the SPIRE map (interpolated into the ArTéMiS passband at 450\,$\mu$m, as described earlier) is used to ease the baseline subtraction: time series simulated from the SPIRE map are subtracted from the ArTéMiS time series (after appropriate scaling to the unit of Jy per ArTéMiS beam), baselines are subtracted from the resulting time series with the usual method, and then the simulated SPIRE time series are added back to the data. The drift subtraction on smaller spatial and temporal scales is unchanged.

The hybridization with SPIRE and beam area recalibration are done directly in the spatial domain as described below. The hybrid map $H$ is designed to preserve the total flux of any type of source (on any characteristic spatial scale), as measured with SPIRE. The final ArTéMiS map and the SPIRE map are first convolved to bring them to a common FWHM angular resolution (in the case of SPIRE, the FWHM increase is very modest, of 4 -- 5\%). Then, a filtered version of the ArTéMiS map is created, containing only pedestal emission under bright sources (to which ArTéMiS is not as sensitive as SPIRE, and that cannot be reliably recovered), and subtracted from the ArTéMiS map to create a map containing only compact sources, which we hereafter call the "compact map", $C$ ($C_{\rm conv}$ in its lower-resolution version). A flux calibration corrective factor is then computed as the ratio of the total emission of the SPIRE map $S_{\rm conv}$ (after subtraction of a large-scale zero-level) within a certain mask to the total emission of $C_{\rm conv}$ (with the same FWHM as $S_{\rm conv}$) within the same mask. The brightness threshold used to define the mask selects only the brightest sources, well above the sensitivity limits of both instruments. The fiducial (Gaussian) beam area is then divided by this corrective factor (on the order of 0.5 at both wavelengths) to obtain the effective beam area, before converting the ArTéMiS map unit from Jy/beam to MJy/sr. A map containing only extended emission (to which SPIRE is uniquely sensitive) is then computed as $E = S_{\rm conv} - C_{\rm conv~recal}$, and the hybrid map is constructed as $H = C_{\rm recal} + W~ S_{\rm conv}$ where the weight map $W$ multiplying the SPIRE map is $W = E / S_{\rm conv}$. In that way, SPIRE fluxes are preserved at any location in the hybrid map.

To deal with saturation in some SPIRE fields, we modified the above process, to identify pixels affected by saturation or nonlinear response, and to set the SPIRE weight map to zero in these pixels. Care is taken to avoid discontinuities in the $W$ map, by working temporarily with a version of $S_{\rm conv}$ where the pixels to be discarded are filled with $C_{\rm conv~recal}$, with a scalar offset.

\section{Detailed SFR and SFE estimates for each CAFFEINE cloud}
\label{sect:CAFFEINE_SFR}

Table~\ref{tab:CAFFEINE_sources_results} provides the numbers of protostellar clumps and YSOs as well as our respective estimates of the SFR and SFE for each CAFFEINE cloud 
within two specific column density contours: $\rm 1.4\times 10^{22}~cm^{-2}$ and $\rm 7.9\times 10^{22}~cm^{-2}$.

\clearpage
\onecolumn

\begin{landscape}
\begin{longtable}{l|cccccc|cccccc}
\caption{\label{tab:CAFFEINE_sources_results} Derived properties for the CAFFEINE clouds}\\
\hline \hline
 & \multicolumn{6}{c|}{Clumps} & \multicolumn{6}{c}{YSOs}\\
Cloud & $N_{22.15}$\tablefootmark{a} & $N_{22.90}$\tablefootmark{a} & SFR$_{22.15}$\tablefootmark{b} & SFR$_{22.90}$\tablefootmark{b} & SFE$_{22.15}$ & SFE$_{22.90}$ & $N_{22.15}$\tablefootmark{c} & $N_{22.90}$\tablefootmark{c} & SFR$_{22.15}$\tablefootmark{d} & SFR$_{22.90}$\tablefootmark{d} & SFE$_{22.15}$ & SFE$_{22.90}$ \\
 &  &  & M$_\odot$\,Myr$^{-1}$ & M$_\odot$\,Myr$^{-1}$ & $\times 10^{-8}$Myr$^{-1}$ & $\times 10^{-8}$Myr$^{-1}$ &  &  & M$_\odot$\,Myr$^{-1}$ & M$_\odot$\,Myr$^{-1}$ & $\times 10^{-8}$Myr$^{-1}$ & $\times 10^{-8}$Myr$^{-1}$\\
\hline
\endfirsthead
\caption{continued.}\\
\hline\hline
 & \multicolumn{6}{c|}{Clumps} & \multicolumn{6}{c}{YSOs}\\
Cloud & $N_{22.15}$\tablefootmark{a} & $N_{22.90}$\tablefootmark{a} & SFR$_{22.15}$\tablefootmark{b} & SFR$_{22.90}$\tablefootmark{b} & SFE$_{22.15}$ & SFE$_{22.90}$ & $N_{22.15}$\tablefootmark{c} & $N_{22.90}$\tablefootmark{c} & SFR$_{22.15}$\tablefootmark{d} & SFR$_{22.90}$\tablefootmark{d} & SFE$_{22.15}$ & SFE$_{22.90}$ \\
 &  &  & M$_\odot$\,Myr$^{-1}$ & M$_\odot$\,Myr$^{-1}$ & $\times 10^{-8}$Myr$^{-1}$ & $\times 10^{-8}$Myr$^{-1}$ &  &  & M$_\odot$\,Myr$^{-1}$ & M$_\odot$\,Myr$^{-1}$ & $\times 10^{-8}$Myr$^{-1}$ & $\times 10^{-8}$yr$^{-1}$\\
\hline
\endhead
\hline
\multicolumn{13}{l}{
\tablefoot{
\tablefoottext{a}{Numbers of protostellar clumps identified within contours of $N_{\rm H_2}=10^{22.15}~\rm cm^{-2}$ and $N_{\rm H_2}=10^{22.90}~\rm cm^{-2}$, respectively.}
\tablefoottext{b}{SFRs within the same contours 
derived from protostellar clumps using the method described in Sect.~\ref{sec:Meth_clump_relation}.}
\tablefoottext{c}{Numbers of Class~I and II YSOs identified within contours of $N_{\rm H_2}=10^{22.15}~\rm cm^{-2}$ and $N_{\rm H_2}=10^{22.90}~\rm cm^{-2}$, respectively.}
\tablefoottext{b}{SFRs within the same contours 
derived from YSOs using the method described in Sect.~\ref{sec:SFR-CAF-YSO}.}
}}
\endfoot
MonR2 & 15 & 4 & 20 & 9 & 2.06 & 6.19 & 98 & 7 & 105 & 13 & 10.86 & 8.71 \\
R5180 & 6 & 3 & 159 & 98 & 2.45 & 2.3 & 5 & 0 & 106 & - & 1.64 & - \\
NGC2264 & 6 & 4 & 34 & 28 & 1.75 & 6.87 & 50 & 6 & 62 & 32 & 3.16 & 7.59 \\
GAL316 & 19 & 7 & 1152 & - & 10.71 & - & 30 & 5 & 84 & 89 & 0.78 & 3.06 \\
I14453 & 6 & 1 & 266 & - & 11.25 & - & 4 & 0 & - & - & - & - \\
SDC317p7 & 6 & 1 & 323 & - & 7.39 & - & 8 & 1 & 276 & - & 6.32 & - \\
SDC317p9 & 10 & 3 & 304 & - & 11.54 & - & 23 & 10 & 102 & 57 & 3.87 & 8.53 \\
RCW87 & 16 & 1 & 595 & - & 9.16 & - & 22 & 0 & 50 & - & 0.77 & - \\
G322 & 3 & 2 & - & - & - & - & 3 & 0 & - & - & - & - \\
SDC326 & 71 & 19 & 3700 & - & 6.5 & - & 91 & 18 & 574 & 190 & 1.01 & 1.16 \\
G327 & 11 & 3 & 446 & 179 & 2.23 & 1.82 & 14 & 2 & 187 & - & 0.94 & - \\
I15541 & 15 & 6 & 760 & 344 & 3.43 & 4.24 & 40 & 5 & 249 & 141 & 1.12 & 1.74 \\
GAL332 & 14 & 3 & 577 & 207 & 3.56 & 5.55 & 24 & 3 & 186 & - & 1.15 & - \\
G332 & 22 & 2 & 865 & - & 8.14 & - & 32 & 1 & 178 & - & 1.67 & - \\
RCW106 & 38 & 4 & 2136 & - & 3.35 & - & 64 & 2 & 456 & - & 0.72 & - \\
G333 & 5 & 1 & 236 & - & 7.38 & - & 6 & 0 & 266 & - & 8.33 & - \\
G333.2 & 36 & 10 & 3032 & 726 & 4.23 & 2.75 & 28 & 4 & 271 & - & 0.38 & - \\
I16175 & 4 & 1 & 148 & - & 2.51 & - & 6 & 2 & 273 & - & 4.6 & - \\
G337.1 & 6 & 0 & 54 & - & 5.09 & - & 18 & 0 & 338 & - & 31.91 & - \\
I16351 & 4 & 1 & 83 & 49 & 1.8 & 4.42 & 17 & 0 & 246 & - & 5.35 & - \\
I16367 & 6 & 1 & - & - & - & - & 21 & 5 & 233 & 72 & 7.14 & 17.36 \\
SDC338 & 30 & 9 & 930 & 414 & 4.42 & 9.76 & 62 & 9 & 922 & 142 & 4.38 & 3.34 \\
G337.9 & 11 & 1 & 683 & - & 5.88 & - & 22 & 2 & 356 & - & 3.07 & - \\
G339 & 8 & 2 & 286 & - & 9.95 & - & 24 & 4 & 251 & - & 8.71 & - \\
G340.8 & 6 & 2 & 284 & - & 3.2 & - & 13 & 2 & 103 & - & 1.17 & - \\
OH341 & 16 & 5 & 1144 & 398 & 7.04 & 9.68 & 33 & 7 & 256 & 178 & 1.57 & 4.34 \\
SDC340 & 15 & 7 & 370 & - & 7.1 & - & 40 & 9 & 76 & 105 & 1.46 & 7.74 \\
G341.9 & 2 & 2 & 370 & - & 4.24 & - & 4 & 0 & - & - & - & - \\
RCW116B & 14 & 5 & 550 & - & 9.07 & - & 26 & 5 & 291 & 290 & 4.81 & 17.59 \\
I16572 & 12 & 3 & 266 & - & 7.95 & - & 31 & 6 & 142 & 101 & 4.24 & 21.54 \\
SDC344 & 13 & 6 & 409 & 224 & 1.43 & 4.46 & 80 & 15 & 293 & 36 & 1.03 & 0.72 \\
SDC345.4 & 5 & 2 & 195 & - & 3.57 & - & 14 & 2 & 48 & - & 0.88 & - \\
SDC345.0 & 10 & 2 & 424 & - & 3.28 & - & 66 & 7 & 226 & 26 & 1.75 & 2.29 \\
SDC348 & 34 & 16 & 1891 & - & 9.75 & - & 54 & 20 & 141 & 179 & 0.73 & 2.66 \\
NGC6334N & 8 & 3 & 162 & - & 2.04 & - & 24 & 1 & 94 & - & 1.18 & - \\
I17233a & 4 & 3 & 61 & - & 8.93 & - & 12 & 7 & 99 & 161 & 14.48 & 61.92 \\
I17233b & 8 & 0 & 246 & - & 9.71 & - & 1 & 0 & - & - & - & - \\
NGC6357SE & 14 & 3 & 376 & - & 6.86 & - & 13 & 1 & 95 & - & 1.74 & - \\
OH353 & 10 & 5 & 1671 & - & 6.57 & - & 33 & 4 & 114 & - & 0.45 & - \\
G358.5 & 17 & 6 & 760 & - & 8.84 & - & 31 & 2 & 396 & - & 4.6 & - \\
G005 & 15 & 2 & - & - & - & - & 14 & 0 & 172 & - & 3.58 & - \\
G008 & 11 & 3 & 470 & - & 7.05 & - & 5 & 0 & 108 & - & 1.62 & - \\
SDC010 & 40 & 12 & 768 & 367 & 3.58 & 8.51 & 45 & 9 & 190 & 71 & 0.89 & 1.65 \\
G010 & 8 & 0 & 357 & - & 1.79 & - & 16 & 1 & 225 & - & 1.13 & - \\
W33 & 35 & 17 & 1671 & 905 & 1.07 & 2.84 & 142 & 18 & 513 & 258 & 0.33 & 0.81 \\
SDC014 & 22 & 4 & 708 & - & 4.82 & - & 82 & 13 & 237 & 83 & 1.61 & 3.97 \\
W42 & 5 & 0 & 267 & - & 2.58 & - & 5 & 1 & 1827 & - & 17.68 & - \\
G034 & 19 & 4 & 1193 & 474 & 1.24 & 3.34 & 65 & 5 & 645 & 476 & 0.67 & 3.35 \\
W48B & 11 & 4 & 323 & 240 & 2.8 & 6.21 & 40 & 5 & 314 & 262 & 2.72 & 6.79 \\
\hline
\end{longtable}
\end{landscape}

\end{appendix}
\end{document}